\documentclass[aps,prb,twocolumn,amsmath,amssymb,floatfix,
superscriptaddress]{revtex4-1}
\usepackage{blindtext}
\usepackage{epsfig}
\usepackage{color}
\usepackage{amsmath}
\usepackage{amsfonts}
\usepackage{amssymb}
\usepackage{graphicx}%
\usepackage{url}
\usepackage[breaklinks]{hyperref}
\usepackage[english]{babel}
\usepackage{esvect}
\usepackage[normalem]{ulem}
\usepackage[inline]{enumitem}
\usepackage{xfrac}
\usepackage{dsfont}
\usepackage{bm}
\hypersetup{colorlinks,citecolor=blue,filecolor=blue,linkcolor=blue,urlcolor=blue}

\numberwithin{equation}{section}
\renewcommand{\theequation}{\arabic{section}.\arabic{equation}}

\bibpunct{[}{]}{,}{n}{}{}

\begin{document}

\title{From Topological Superconductivity to Quantum Hall States in Coupled Wires}

\author{Fan Yang}
\affiliation{CPHT, CNRS, Institut Polytechnique de Paris, Route de Saclay, 91128 Palaiseau, France}
\author{Vivien Perrin}
\affiliation{CPHT, CNRS, Institut Polytechnique de Paris, Route de Saclay, 91128 Palaiseau, France}
\affiliation{Institut quantique et D{\'e}partement de physique, Universit{\'e} de Sherbrooke, Sherbrooke, Qu{\'e}bec J1K 2R1, Canada}
\affiliation{Laboratoire de Physique des Solides, CNRS, Universit\' e Paris-Sud, Universit\' e Paris-Saclay, F-91405 Orsay, France}
\author{Alexandru Petrescu}
\affiliation{Institut quantique et D{\'e}partement de physique, Universit{\'e} de Sherbrooke, Sherbrooke, Qu{\'e}bec J1K 2R1, Canada}
\author{Ion Garate}
\affiliation{Institut quantique et D{\'e}partement de physique, Universit{\'e} de Sherbrooke, Sherbrooke, Qu{\'e}bec J1K 2R1, Canada}
\author{Karyn Le Hur}
\affiliation{CPHT, CNRS, Institut Polytechnique de Paris, Route de Saclay, 91128 Palaiseau, France}

\date{\today}

\begin{abstract}
We present a theoretical study of the interplay between topological p-wave superconductivity, orbital magnetic fields and quantum Hall phases in coupled wire systems.  First, we calculate the phase diagram and physical observables of a fermionic ladder made of two coupled Kitaev chains, and discuss the presence of two and four Majorana zero modes. Second, we analyze hybrid systems  consisting of a Kitaev chain coupled to a Luttinger liquid. By tuning the magnetic field and the carrier density, we identify quantum Hall and charge density wave phases, as well as regimes in which superconductivity is induced in the second chain by proximity effect. Finally,  we consider two-dimensional systems made of weakly coupled ladders. There, we engineer a $p+ip$ superconductor and describe a generalization of the $\nu=1/2$ fractional quantum Hall phase. These phases might be realized in solid-state or cold-atom nanowires.
\end{abstract}

\maketitle

\section{Introduction}

The quest of topological phases has attracted a lot of attention in the last decades, starting from the quantum Hall effect \cite{QHE,FQHE} and evolving towards variants on the honeycomb lattice with effectively a zero magnetic flux in a unit cell \cite{Haldane2D}.  These quantum Hall systems are characterized by a robust, unidirectional charge flow at the sample boundaries \cite{Halperin,Buttiker}, as well as by the emergence of fractional charges in the case of Laughlin states \cite{Laughlin,fisherstone,wen2004book,HansReview}. 

More recently, superconducting analogues of quantum Hall phases have also been focus of interest \cite{ReadGreen,MooreRead,Nayak2008Rev}. These are p-wave superconductors hosting chiral (unidirectionally propagating) Majorana modes at the edges. One theoretical approach to study the connections between quantum Hall phases and topological superconductors is to view two-dimensional systems as arrays of  coupled one-dimensional wires \cite{kane2002,kane2014,kane2017}. This so-called wire construction approach comes with the benefit that it allows to investigate interaction effects (such as fractionalization) \cite{Herviou,Fractional,Basel} that go beyond mean-field theory, via the Luttinger liquid paradigm \cite{Haldane,Heinz}. 

Below, we study a multi-wire system in the low-energy subspace, where each wire is described by a topological $p$-wave superconductor \cite{kitaev2001}.  We address the interplay between intrawire and interwire tunneling and superconducting pairing terms, including orbital magnetic field effects. Such orbital magnetic field effects have been realized in coupled nanowires \cite{halperin2008} through the application of a magnetic field perpendicular to the plane of the wires. Similar efforts are underway in cold-atom systems \cite{Monika,Munich,QH2D}, where quantum Hall phases have been observed in ladder geometries \cite{QHEladder1,QHEladder2}. The possibility of engineering fractional quantum Hall phases in these ladders  has stimulated a vigorous research activity, focused on both bosonic and fermionic systems \cite{grusdt2014,alex2015,alex2017,CornfeldSela,Mazza,Michele,Shtengel}. Orbital magnetic field effects have also started to attract some theoretical interest in the case of two coupled Kitaev chains \cite{Wang,Herviou}. Our building block is a two-leg ladder Kitaev system. One scope of our work is to show that coupling such ladder systems together will allow us to realize a $(p+ip)$ topological superconductor, through the engineering of space-dependent magnetic fields. 

We start by studying the phase diagram of two coupled wires in the bonding-antibonding band representation, which allows us to access the strong-tunneling limit \cite{Urs,Review} between the wires. Afterwards, we introduce hybrid (spinless) systems coupling a chain of free fermions with a Kitaev chain and study the effect of Andreev processes. An Andreev process allows the transfer of a Cooper pair from a superconducting system into a normal metal (here the other wire).  Such a process leads to superconducting correlations in the normal wire. In the case of spin-1/2 fermions, Andreev processes have already been shown to influence the properties of cuprate superconductors near the Mott insulating regime \cite{karyn2001}. They also give rise to a fractional quantum Hall phase at filling factor $\nu=1/2$ in the case of hybrid spin-1/2 wire systems  \cite{alex2015}. 

We show that the aforementioned hybrid systems can stabilize various phases: topological superconducting phases with four and two Majorana fermions, Abelian quantum Hall phases at $\nu=1$ and $\nu=1/3$, and a charge density wave.  The charge density wave occurs when the lowest band is completely filled and the upper band is empty. It is analogous to the the rung-Mott phase identified in the two-leg bosonic ladder system \cite{RungMott,RungMottMunich}. Here, the system hosts one particle and one hole per rung, with phase coherence between different particle-hole pairs. We discuss observables in the different phases, the stability of Majorana zero modes, and the robustness towards interaction effects.  We also propose various probes of the phases, such as a Meissner-Majorana current to reveal the presence of four Majorana zero modes in a ladder and a Thouless charge pump geometry to study the bulk response in the quantum Hall phases.

In the later part of this work, we apply the wire construction method \cite{kane2014,kane2017} to engineer, in coupled-ladder geometries, the $p+ip$ superconductor (and more precisely, an $ip_x+p_y$ superconductor) with spinless fermions \cite{ReadGreen} as well as a fractional quantum Hall state at $\nu=1/2$ \cite{alex2015,alex2017} for spin-1/2 fermions \cite{alex2015}. The $p+ip$ superconducting proposal is implemented with two different Peierls phases acting on the intrawire and interwire hopping terms \cite{Monika,Munich} corresponding to a proper choice of space-dependent magnetic field. We identify three phases for our system: a strong-paired phase with no Majorana mode, a quasi-one-dimensional phase with $2N$ Majorana fermions where $N$ is the number of two-leg ladders, and a Moore-Read phase \cite{MooreRead,ReadGreen} with a chiral Majorana fermion flowing along the edge of the two-dimensional sample. A square unit cell in the lattice allows to reveal one Majorana fermion. It has been recently shown that one can also realize a topological $(p_x\pm ip_y)$ superconductor on the honeycomb lattice with Rashba spin-orbit interaction, as a result of the interplay of geometric phase and electron correlation \cite{NaturePhysics}. 

This wire construction could be engineered in solid-state systems. More precisely, a one-dimensional topological p-wave superconductor has been realized through a semiconducting nanowire with spin-orbit coupling in the vicinity of an s-wave superconductor \cite{Delft,Marcus} following theoretical predictions \cite{lutchyn,refael,fukane}.  We assume here that the intrawire and interwire pairing channels come from the vicinity of a common (s-wave) superconducting reservoir. The presence of a Rashba spin-orbit coupling of the form $-i\alpha\sigma_y\partial_x$ \cite{lutchyn,refael,fukane,fisher,pascal} shifts the band structure associated to the $\sigma_y=+1$ state compared to the band structure associated to the $\sigma_y=-1$ state in momentum space. The band structure of a given wire shows four Fermi points. The application of a magnetic field perpendicular to the spin-orbit field, {\it e.g.} in the $z$ direction will then open a gap in the band structure at the crossing region between the two bands, canting slightly the spins at the remaining two Fermi points along the $z$ direction \cite{fisher,alicea}. The spin-polarized nature of the fermions in the low-energy model comes from the fact that the system has now only two remaining gapless (spin-polarized) Fermi points. The coupling with an s-wave superconducting reservoir induces pairing terms, as long as the spins are oriented differently at the two Fermi points.  Coupled-wire spin-1/2 systems with Rashba spin-orbit coupling have been studied, for instance in Ref. \cite{Schrade}, without orbital magnetic field effects. In our work, (chiral) effects will be included directly in the spinless fermion model through Peierls phases, which are also induced by the magnetic field in the $z$ direction.  To realize the $p+ip$ superconductor, we engineer magnetic fields $B_z$ which are staggered along the $y$ direction. In this realization, the magnetic field vanishes at the locations of the wires. Thus, in order to realize spinless fermions, one could resort to an additional uniform magnetic field in the perpendicular (say, $x$) direction.

Similar efforts are realized in ultracold atoms with synthetic spin-orbit interactions \cite{Monika,Galitski,MIT,Nascimbene,spinorbit}. It is also important to mention the current efforts in implementing cold-atom spin-polarized p-wave superfluid states, that would then allow to realize the spinless fermion model without spin-orbit coupling \cite{Petrov,Gurarie}. 

The organization of the paper is as follows. In Sec. \ref{sec:pre}, we introduce the model with the different flux situations to be studied. In Sec. \ref{sec:tw}, we address both analytically and numerically the ladder system comprising two coupled spinless fermionic wires, in the bonding-antibonding band representation, then allowing us to access the strong-tunneling limit \cite{Review}.  In Sec. \ref{sec:hw}, we introduce magnetic flux effects and generalize the analysis to the case of hybrid systems, with one wire being a topological p-wave superconductor and the other wire a free fermion model or a Luttinger liquid, taking into account the physics of Andreev processes \cite{rice1994}. In Sec. \ref{sec:cp}, we design and study two-dimensional topological phases hosting chiral edge modes from coupled ladder geometries. We show how magnetic field effects can turn the one-dimensional topological superconductor (which belongs to the symmetry class BDI) into a two-dimensional topological $p+ip$ superconductor (which belongs to the symmetry class D) by coupling the ladders to the same (s-wave) superconducting reservoir \cite{bernevig2015}. We study the stability of the Majorana zero modes in the different phases in relation with Symmetry Protected Topological (SPT) phases in wire constructions \cite{Fidkowski,Thomale}. We also discuss a realization of a $\nu=1/2$ Laughlin phase in coupled (hybrid) ladders comprising spin-1/2 fermions, in the presence of a uniform magnetic field. The $\nu=1/2$ Laughlin phase belongs to the SPT Class A in terms of the edge K-matrix structure \cite{Thomale}. In Sec. \ref{summary}, we summarize our main findings. In the Appendices, we present a further analysis on the derivation of Andreev processes, and higher-order processes, induced by the interwire tunneling. We also provide additional information on the edge theory and Thouless charge pump geometry \cite{thouless1983} for the quantum Hall phases, and discuss Coulomb interaction effects.

%------------------------------------------------------------------------------------------------------------
\section{Preliminaries}
\label{sec:pre}

\subsection{Model and Definitions}
\label{model}

\begin{figure}[t]
   \begin{center}
      \includegraphics[width=0.75\linewidth]{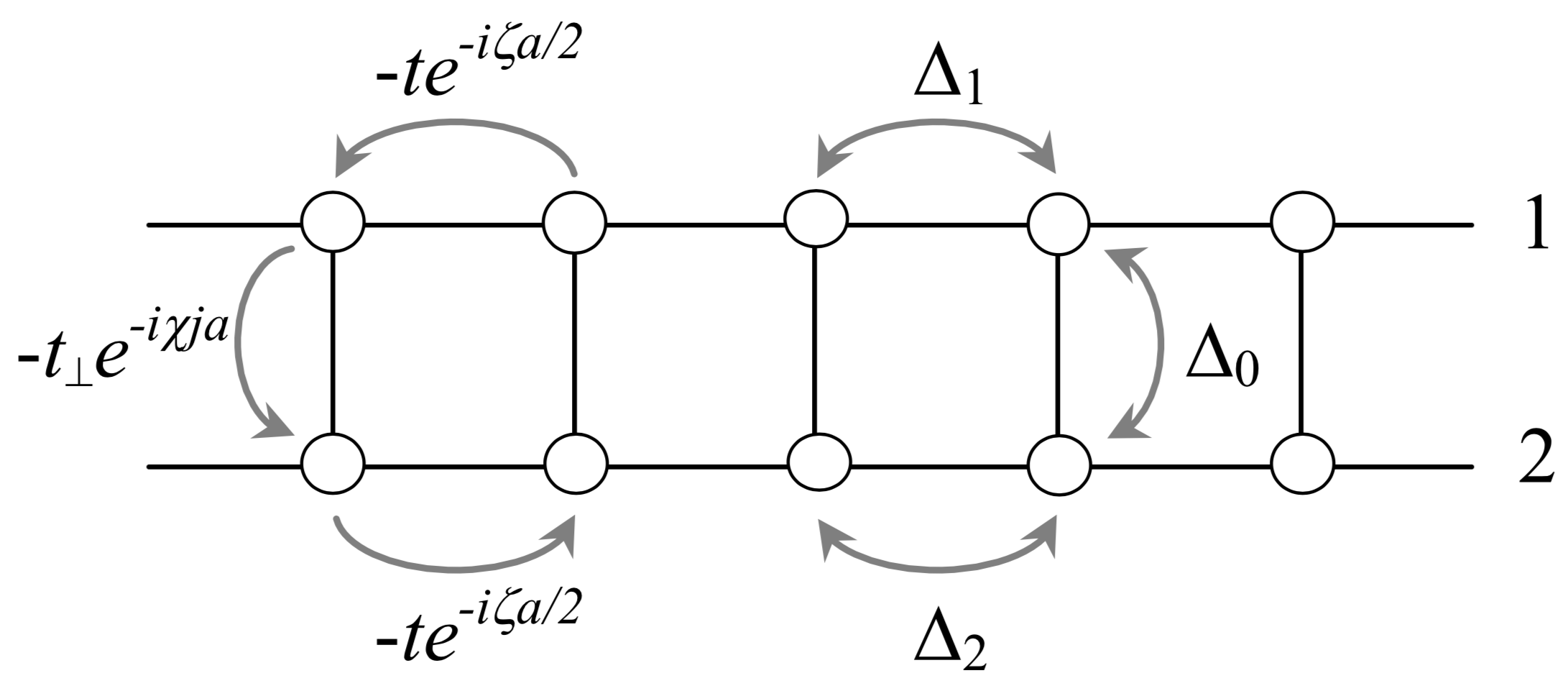} \\
      \vspace{0.2cm}
        \includegraphics[width=0.95\linewidth]{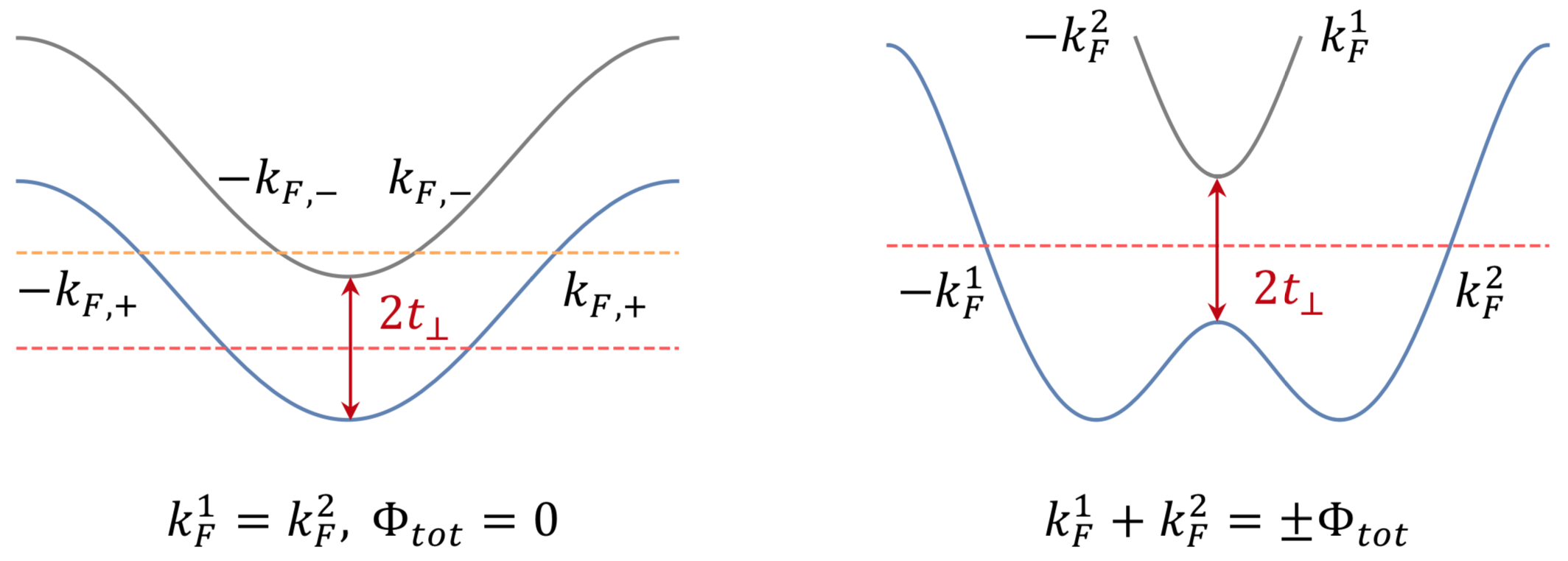}     \end{center}
  \vskip -0.5cm \protect\caption[]
  {(color online) (Top) Two-leg ladder lattice comprising spinless fermions; we introduce the different flux situations discussed in the article through $\chi$ and $\zeta$ such that the total flux per plaquette is $\Phi_{\text{tot}}=(\chi-\zeta)$ assuming $a=a'=1$. (Bottom left) Band structure in the absence of magnetic field and superconductivity. The blue and grey curves correspond to the bonding ($+$) and antibonding ($-$) bands \cite{Review}. A gap of the order of $2t_\perp$ is opened by the interwire tunneling term. Varying the chemical potential, the dashed lines denote the Fermi levels which host different numbers of gapless Fermi points. (Bottom right) Band structure with flux insertion and opening of a gap in the crossing region between the two bands. The lowest bonding band now mixes fermionic states with different chiralities (corresponding to left movers in the first wire with a wavevector $-k_{F}^1$ and to right movers in the second wire with a wavevector $+k_{F}^2$).}
    \label{fig:gladder}
    \vskip -0.5cm
\end{figure}

We begin by illustrating the building block of our coupled wire construction: a flux-assisted two-leg ladder system of spinless fermions, captured by the Hamiltonian (see Fig.~\ref{fig:gladder}, top)
  \begin{align}
    \mathcal{H} = \mathcal{H}_\parallel + \mathcal{H}_\perp + \mathcal{H}_\Delta + \mathcal{H}_{\Delta_0}, \label{eq:h}
  \end{align}
which includes tunneling terms along and between the wires
  \begin{align}
     \mathcal{H}_\parallel &= - \sum_{j}  \sum_{\alpha = 1,2} [\mu c_\alpha^\dagger (j) c_\alpha (j) + t e^ {-i\zeta a/2} c^\dagger_1 (j) c_1 (j+1)  \notag \\
     &\phantom{========} + t e^ {i\zeta a/2} c^\dagger_2 (j) c_2 (j+1)+ \text{H.c.}] , \notag \\
     \mathcal{H}_\perp &= - \sum_j t_\perp e^{i\chi x_j} c_1^\dagger (j) c_2 (j) + \text{H.c.}, \label{eq:h_0}
  \end{align} 
as well as pairing interactions induced by the proximity to the superconducting (superfluid) reservoir 
  \begin{align}
    \mathcal{H}_\Delta &= \sum_{\alpha=1,2} \sum_j \Delta_\alpha c_\alpha^\dagger (j) c_\alpha^\dagger(j+1) + \text{H.c.}, \notag \\
    \mathcal{H}_{\Delta_0} &= \sum_{j} \Delta_0 c_1^\dagger (j) c_2^\dagger(j) + \text{H.c.}. \label{eq:h_1}
  \end{align}
We denote the lattice spacings of the wires as $a$ and $a'$ for the horizontal $x$ and vertical $y$ directions. For a square ladder, $a'=a$. The positions of the sites along each wire are denoted as $x_j = j a$, where $j=1,...,M$.
The total length of one wire becomes $L = Ma$. The operator $c_\alpha^\dagger (j)$ creates a spinless fermion on site $j$ of the wire $\alpha$. Here, $\mu$ stands for the global chemical potential. For symmetric and decoupled (or weakly-coupled) wires, this condition will then ensure that the Fermi wave vectors in the two wires satisfy $k_F^1 = k_F^2$, but we will also address cases (when specified hereafter) with asymmetric wires where $k_{F}^1\neq k_{F}^2$. The intrawire and interwire pairing amplitudes are denoted  $\Delta_{1,2}$ and $\Delta_0$, respectively. Here, the phase associated with the pairing terms is fixed by the properties of the superconducting (superfluid) reservoir. In Sec. \ref{sec:cp}, we will also include a term $\Delta_0 (c_1^\dagger (j) c_2^\dagger(j+1) + c_2^\dagger (j) c_1^\dagger(j+1) +  \text{H.c.})$, which will play an important role for the realization of the $p+ip$ superconducting phase. 

Even though we do not write the Coulomb interaction in the Hamiltonian, we shall comment hereafter on the stability of the physics towards interaction effects. The stability of the topological superconducting phases and the possibility to realize correlated and Mott phases in the presence of Coulomb interactions or nearest-neighbor interactions have been discussed, for instance, for the Kitaev chain in Refs. \cite{Schuricht,Li} and for two coupled wires (when $t_{\perp}=0$) in Ref. \cite{Herviou}.

Below, for simplicity we assume that $t$ and $t_{\perp}$ are real. The orbital effects of magnetic fields are included by multiplying $t$ and $t_\perp$ by  phase factors $e^{\pm i \zeta a/2}$ and $e^{\pm i \chi x_j}$, respectively (see Fig. \ref{fig:gladder}).  We neglect the effect of the magnetic field in the superconducting order parameter of the reservoir. We define the Peierls phases for a unit charge $q=1$ and we set $\hbar=1$.  Furthermore, we introduce the quantity   \begin{align}
    \Phi_{\text{tot}} = \chi-\zeta. \label{eq:flux}
  \end{align}
 This quantity has the dimension of a wave vector.  The flux per plaquette or square unit cell is defined as $(\chi - \zeta)a$.  Therefore, $\Phi_{\text{tot}}$ can also be seen as the total flux in a unit cell or plaquette with $a=a'=1$. 
 
 When $\chi\neq 0$ and $\zeta=0$, this situation will correspond to the case of a uniform magnetic field $B_z=\chi/a'$ applied along $z$ direction and a vector potential gauge $A_y= + x_j \chi/a'$ along $y$. In that case, one can also perform a gauge transformation to absorb the effect of the induced fluxes (or phases) onto a boost of the wave vector $k$ for each wire, resulting in Fig. \ref{fig:gladder} (bottom right).  
 
When $\chi$ and $\zeta$ are both non-zero, this situation will allow us to engineer the $p+ip$ superconductor in Sec. \ref{sec:p+ip}.  In Fig. \ref{fig:gladder}, this corresponds to a vector potential specified by the components $A_x=-(-1)^{y_j/a'}\frac{\zeta}{2}$ and $A_y= x_j \chi/a'$, with $y_j=0$ for the lower wire and $y_j=a'$ for the upper wire. This choice of vector potential corresponds to a magnetic field perpendicular to the plane of the wires, 
\begin{align}
B_z=\partial_x A_y - \partial_y A_x = \frac{\chi}{a'} - \frac{\zeta\pi}{2a'}\sin(\pi y/a'), \label{Bfield}
\end{align}
and reproduces a net flux $\Phi_{\text{tot}}=\chi-\zeta$ in a square unit cell. Formally, to regularize properly the function $A_x$ and ensure that $\partial A_x/\partial y$ is real (as it should be), we can use the form $A_x=-1/2(e^{i\pi y/a'} + e^{-i\pi y/a'})\frac{\zeta}{2}$ which reproduces $A_x=-(-1)^{y/a'}\frac{\zeta}{2}$ for $y=y_j$ and produces the second term $-\zeta\pi/(2a')\sin(\pi y/a')$ in the magnetic field in (\ref{Bfield}). This form of space-dependent magnetic field, with the two distinct geometrical forms coming from $\chi$ and $\zeta$, allows us to break time-reversal symmetry even if $\chi=\zeta\neq 0$ and $\Phi_{\text{tot}}=0$, because $B_z\neq 0$, by analogy to the Haldane model on the honeycomb lattice \cite{Haldane2D}. The situation is physically different compared to the case where one would apply a uniform magnetic field $B_z=(\chi-\zeta)/a'$. In that case, the choice of two independent Peierls phases in Fig. \ref{fig:gladder} would not be justified (for instance, in that case, one could simply choose $A_x=A_y=0$ for $\chi=\zeta$). To realize the $p+ip$ superconductor in weakly coupled-ladder geometries (see Sec. \ref{sec:p+ip}), we require that the vector potentials and the magnetic field(s) are indeed periodic if we change $y\rightarrow y+2a'$. In addition, the magnetic field $B_z$ must be staggered if we change $y\rightarrow y+a'$, therefore we will also assume in Sec. \ref{sec:p+ip} that $\chi$ takes a staggered (periodic) step-like form $\{\chi; -\chi\}$ associated to two successive square cells in the $y$ direction.  

The particular situation in which $\chi=\zeta=\pi/a$, of interest in this work,  admits two physical interpretations in the absence of the superconducting pairing. On the one hand, the system is equivalent to a model of two wires with imaginary hopping terms $\pm i t$ or band dispersions $\mp 2t\sin(k a)$ in Fig. \ref{fig:gladder} (bottom right), with an alternating transverse hopping term $t_{\perp}(-1)^j$. On the other hand, because the total net flux is zero in a given unit (square) cell,  the band structure of the two-wire system is also analogous to the one in Fig. \ref{fig:gladder}  (bottom left) after gauge transformation, with a uniform transverse hopping term $t_{\perp}$ \footnote{The ``local'' gauge transformation on the fermionic operators takes the form $c_1(j)\rightarrow e^{i\chi x_j/2} c_1(j)$ and $c_2(j)\rightarrow e^{-i\chi x_j/2} c_2(j)$ in Eq. (\ref{eq:h_0}). This situation also corresponds to the case where  $A_x=0$ and $A_y = \chi x/a' - \zeta \pi/(2a')\sin(\pi y/a')x$.  We check that transporting one particle from the upper to the lower wire (vertically) produces a (zero) phase $\int_0^{a'=a} A_y dy=0$ accompanying the transverse hopping term $t_{\perp}$ when $\zeta=\chi$.}.

\subsection{Energy Bands and Magnetic Field Effects}
\label{cases}

Below, we introduce four cases of interest, which will be addressed throughout the manuscript. 

\begin{figure}[t]
   \begin{center}
          \includegraphics[width=0.5\linewidth]{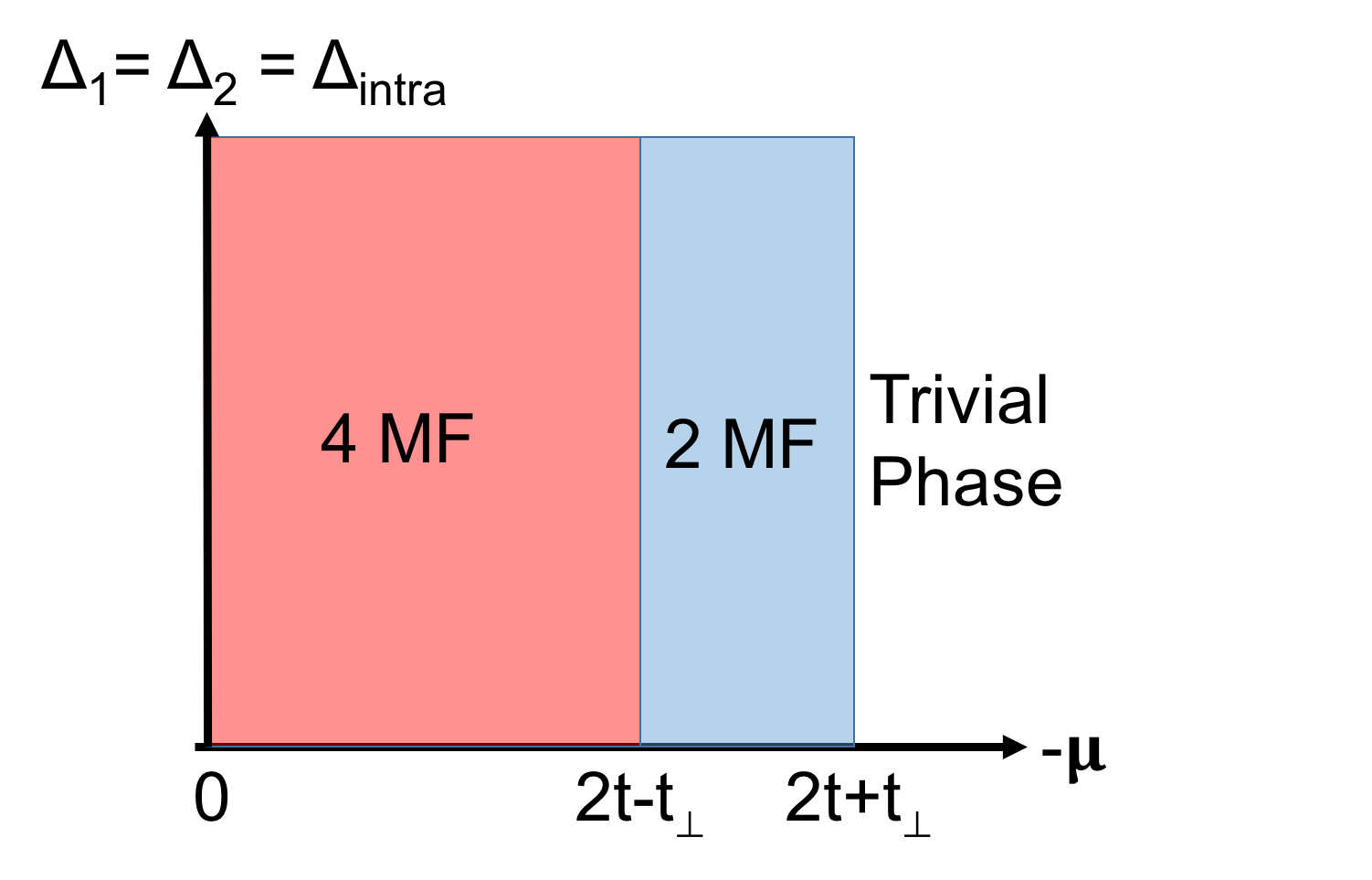}  \includegraphics[height=3cm]{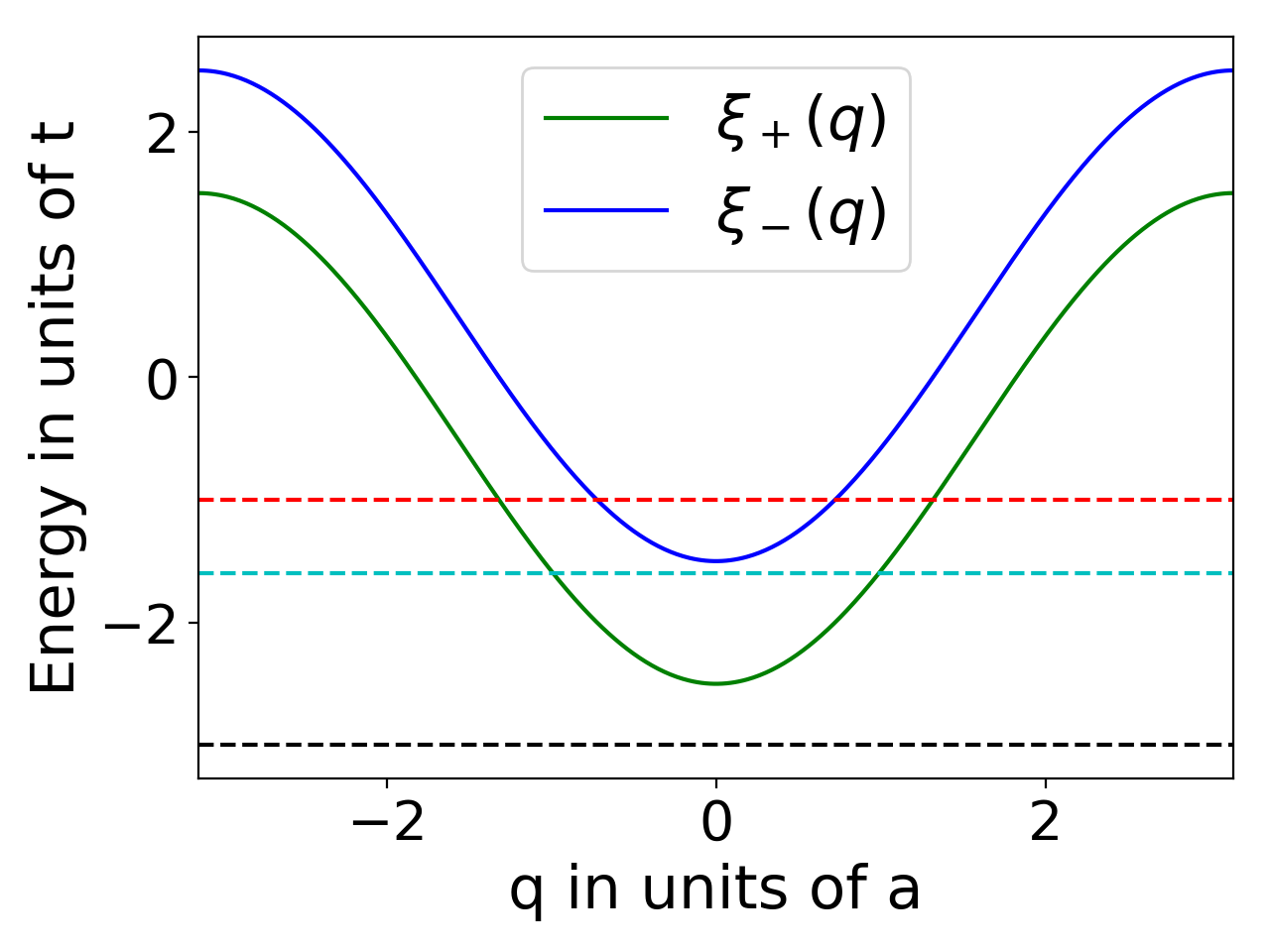}  \\
      \vspace{0.2cm}
        \includegraphics[height=3cm]{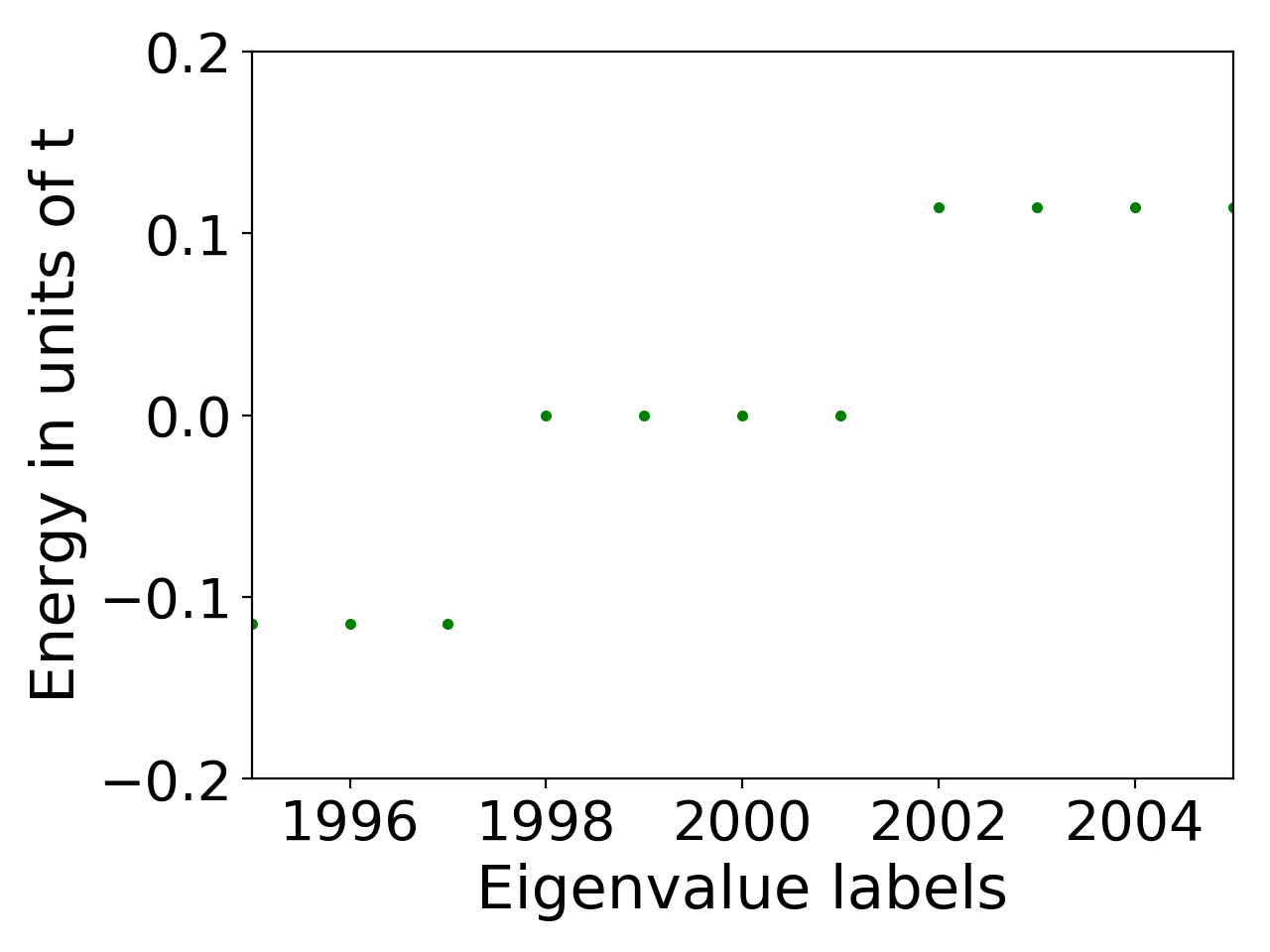}    \includegraphics[height=3cm]{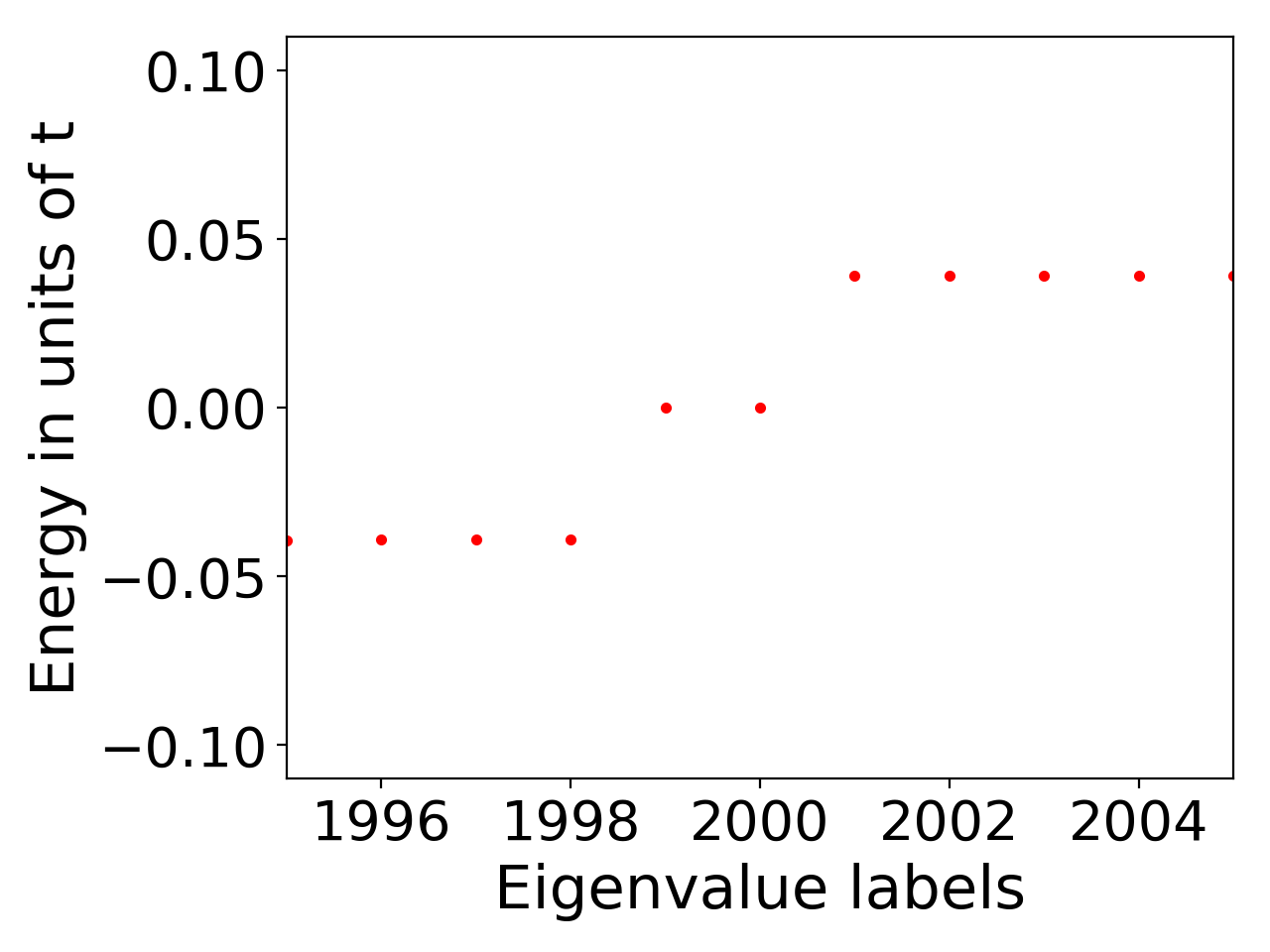}   \includegraphics[height=3cm]{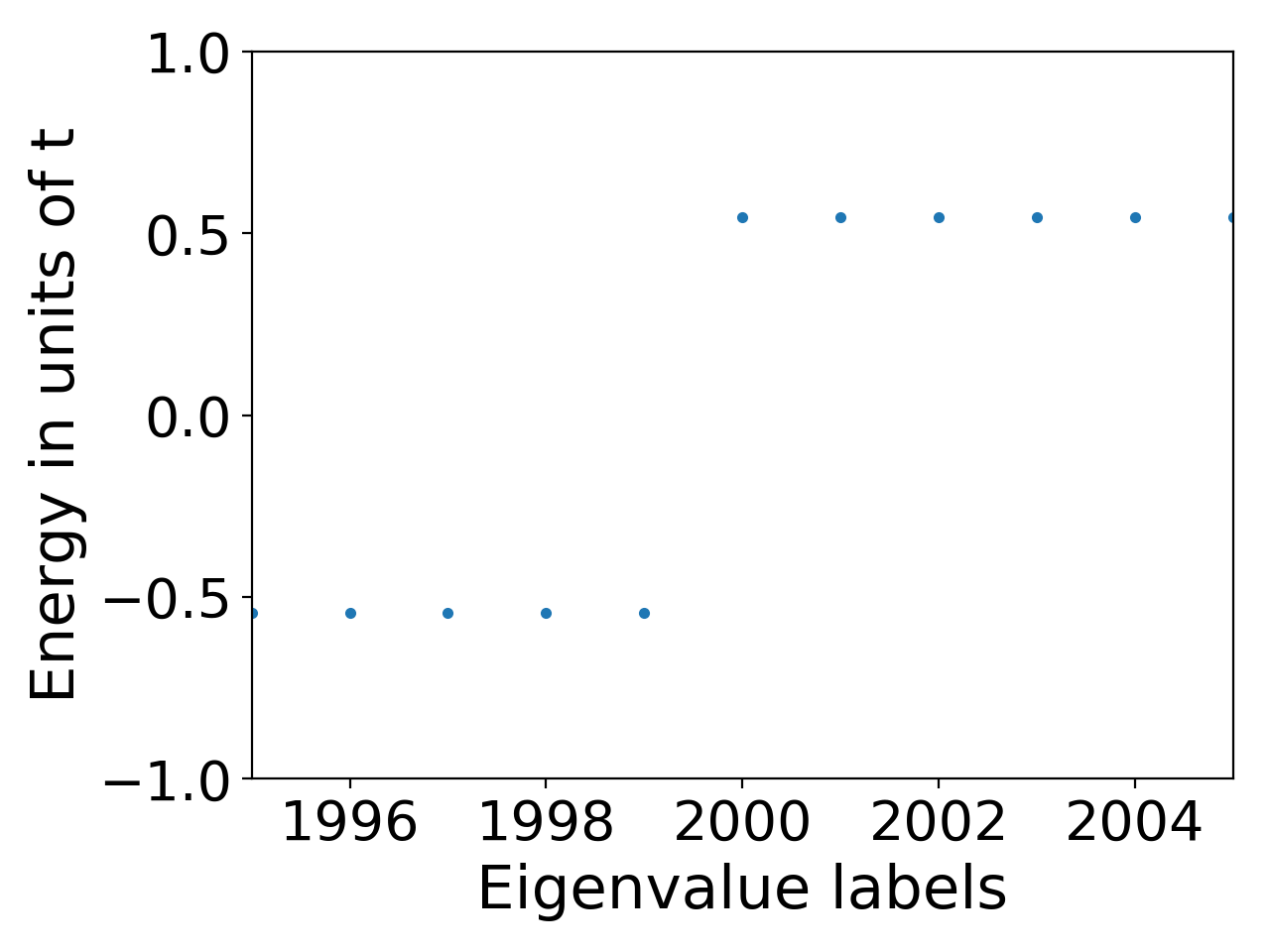}
    \end{center}
  \vskip -0.5cm \protect\caption[]
  {(color online) (Top left) Phase diagram of two coupled wires in the absence of magnetic fields. A Lifshitz phase transition, associated with a change in the topology of the Fermi surface showing respectively 4 and 2 Fermi points, occurs between the topological phases with 4 and 2 Majorana fermions. The trivial phase refers to the strong-paired phase. (Top right) The lowest bonding and antibonding energy bands for $\Delta = 0, t_\perp = 0.5t$ with $\xi_{q,\pm}=-E_{\pm}$. Through a shift of the chemical potential, three possible cases of occupancy are denoted by the dark, red and cyan dashed lines. At the Fermi level, each gapless Fermi point contributes to a Majorana edge mode when $\Delta \ne 0$;  (Bottom) Three energy spectra from exact diagonalization calculations of Eq. (\ref{eq:th}) with a system size $M=1000$ for each wire and open boundary conditions. The parameters are set to $t_\perp = 0.5t$, 
  $\Delta = 0.1t$ and $\Delta_0 = 0.3t$. The different graphs correspond to different values of $\mu = -t, -1.6t, -3t$. In the last graph, the energy gap corresponds roughly to the energy to produce a quasiparticle at the bottom of the lowest bonding band.}
    \label{fig:2kw}
    \vskip -0.5cm
\end{figure}

In the first case, we consider that orbital magnetic fields are vanishing, $\zeta = \chi = 0$. The band structure of two wires is characterized by the bonding (+) and antibonding (-) fermion operators
\cite{Urs} 
   \begin{gather}
       c_\pm(j) = \frac{1}{\sqrt{2}} \left[ c_1(j) \pm c_2(j) \right]. \label{eq:bab}
   \end{gather}
If we neglect the superconducting terms $\Delta_i$, the non-interacting part can be diagonalized as
  \begin{gather}
    \mathcal{H}_0 = \mathcal{H}_\parallel + \mathcal{H}_\perp = \sum_{\lambda = \pm} \sum_k \xi_{k,\lambda}  c_\lambda^\dagger(k) c_\lambda(k), \label{eq:k_0}
  \end{gather}
 with an energy dispersion
 \begin{gather}
   \xi_{k,\pm} = -2t\cos(ka) \mp t_\perp - \mu. \label{eq:dp}
 \end{gather}
 In Eq. (\ref{eq:k_0}), we have used the Fourier transform $c_\alpha (j) = (1/\sqrt{M}) \sum_k e^{ikx_j} c_\alpha (k)$, with $k = 2\pi n /(Ma)$ and $n = -M/2, -M/2 +1, \dots, 0, \dots, M/2-1$ (if we assume $M$ even). The energy spectrum of the two bands is shown in Fig.~\ref{fig:gladder} (bottom left). The Fermi wave vectors for the bonding and antibonding bands are $k_{F,\pm} = (1/a) \arccos[(\pm t_\perp + \mu)/(-2t)]$. We will include the effect of pairing terms at the Fermi points of this band structure, which is justified if the pairing amplitudes satisfy $\Delta_i\ll (t, t_\perp)$ with $i=0,1,2$. In Sec. \ref{sec:hw}, we will use a complementary approach in the wire basis addressing the weak-coupling limit. In Sec.~\ref{sec:tw}, we will turn on the pairing interactions in this strong-tunneling limit and study phase transitions towards SPT phases with 4 and 2 Majorana zero-energy modes (MZM) \cite{Wang}. The transition from 4 MZM to 2 MZM can be viewed as a Lifshitz transition, {\it i.e.} it is induced by a change of the Fermi surface topology or by a change of the number of Fermi points from 4 to 2. The associated Van Hove singularity in the density of states can be observed, for instance, through local compressibility measurements. The trivial phase on the phase diagram of Fig. \ref{fig:2kw} (top left) refers to the strong-paired phase with zero Majorana fermions, which occurs when the chemical potential lies below the bottom of the lowest band or above the top of the upper band.  We study both theoretically and numerically the stability of the Majorana fermions and the structure of induced pairing terms at low-energy in the strong-coupling region. We also study observables and provide complementary results compared to Refs. \cite{Schrade,Nagaosa,Herviou,Wang}.
 
The second case of interest corresponds to $\zeta = 0$ and $\chi  \ne 0$. In Sec.~\ref{sec:hw}, we further set $\Delta_2 = \Delta_0 = 0$ which implies a hybrid system composed of a Kitaev superconducting wire and a free fermion wire. The phase associated to the vertical tunneling terms can be absorbed through the local transformation
  \begin{gather}
      c_1(j) = e^{i\chi x_j/2} \tilde{c}_1(j), \quad
      c_2(j) = e^{-i\chi x_j/2} \tilde{c}_2(j). \label{eq:gt}
  \end{gather}
The effect of the flux insertion $\Phi_{\text{tot}} = \chi$ is then equivalent to a shift (boost) of the momentum of the two energy bands: 
  \begin{gather}
     \xi_{k,\tilde{1}} = -2t\cos[(k+\chi/2)a] - \mu, \notag \\
     \xi_{k,\tilde{2}} = -2t\cos[(k-\chi/2)a] - \mu.
   \end{gather}
  Within these definitions, the two bands cross when $\xi_{k,\tilde{1}}= \xi_{k,\tilde{2}}$, meaning at the wave vector $k=k_0=0$ in the new basis associated to Fig. \ref{fig:gladder} (bottom right), which then makes the effect of a uniform interwire hopping term relevant in this region. In fact, in the strong-coupling limit, $t_\perp$ is only relevant at the band crossing point $k_0$, resulting in \cite{halperin2008}
  \begin{align}
    \mathcal{H}_\perp &= -t_\perp \sum_k \tilde{c}_1^\dagger(k) \tilde{c}_2 (k) + h.c. \notag \\
                                 &\simeq -t_\perp (\tilde{c}_+^\dagger (k_0) \tilde{c}_+ (k_0) - \tilde{c}_-^\dagger(k_0)\tilde{c}_-(k_0)),
  \end{align}
 which then splits the energies of the bonding and antibonding bands at the crossing point $k_0=0$ accordingly. In the original frame, this corresponds to having $k_F^1 + k_F^2 = \pm \Phi_{\text{tot}}$ at the crossing point such that a gap $2t_\perp$ can be opened, as shown in Fig.~\ref{fig:gladder} (bottom right). When the Fermi level is between the upper and lower bands, the two gapless modes form an edge state falling into the category of an Abelian quantum Hall phase provided that $\Delta_i\ll t_{\perp}$.

In the weak-coupling limit, on the other hand, we find a superconducting topological phase at flux $\Phi_{\text{tot}} = \pi$. When adjusting the densities of the wires such that $(k_F^1+k_F^2)=\pi/a$, Andreev processes between the two wires stabilise superconductivity in the two wires similarly to the case of zero magnetic flux. The system shows 4 Majorana fermions via proximity effect. At $\pi$ flux, in the strong-tunneling limit, the lowest (bonding) band which now mixes the two wire states at the Fermi points can be fully filled and can therefore open a gap in the single-particle spectrum when the upper band is empty. This situation, which is analogous to a band insulator, gives rise to a charge density wave in the wire basis associated to long-range order with one particle on each rung. We show below that this charge density wave state forms phase-coherent particle-hole pairs along the wires which then survive even if the transverse hopping term $t_{\perp}$ becomes comparable to (slightly larger than) the pairing channel amplitudes. 

The third case is $\zeta = \chi=\pi/a$ with the vector potential and magnetic field described in Sec. \ref{model}. It is our starting point towards implementing a $p + ip$ superconductor in coupled-ladder geometries via the $\Delta_0$ (interwire) pairing channel. Due to the $\pm it$ hopping terms in the two wires, the time-reversal symmetry and the chiral symmetry are not present, but particle-hole symmetry is preserved.  The transverse hopping term $t_{\perp}(-1)^j$ now becomes real, and this will then ensure the stability of Majorana modes at zero energy even if $t_{\perp}\neq 0$. In Sec.~\ref{sec:pxy}, first we perform the local gauge transformation to map the band structure onto the one of Fig. \ref{fig:gladder} (bottom left), similar to the one in the absence of magnetic fields, and then we show how the interwire pairing term $\Delta_0$ can give rise to a purely imaginary $ip_x$ channel for the bonding fermions in a two-leg ladder architecture. Coupling pairs of wires or ladders together, we realize a $p_y$ channel through the same superconducting (s-wave) reservoir, assuming that the $B_z$ magnetic field in Eq. (\ref{Bfield}) is uniform in the $x$ direction and staggered in the $y$ direction (which implies that $\chi$ generates a step function changing of sign on each successive plaquette in $y$ direction).  Then, we build a low-energy model in a two-dimensional representation of Majorana fermions to study the existence of boundary edge modes. We check that similar results can be obtained with the band structure of Fig. \ref{fig:gladder} (bottom right), corresponding to the same value of the magnetic field $B_z$. We also discuss the stability of the results when fixing $\Phi_{\text{tot}}=0$ when decreasing progressively the value of $\zeta = \chi$.

The fourth case is analogous to the second  $\zeta = 0, \chi \ne 0$ with spinful fermions. In Ref. \cite{alex2015}, two of us have shown the possibility to realize a fractional quantum Hall phase at filling factor $\nu = 1/2$ in hybrid systems. In Sec. \ref{sec:det}, we generalize the analysis in coupled-ladder geometries with a uniform magnetic field showing how the chiral edge mode becomes more protected towards backscattering effects (when the bulk becomes larger).

The last two cases place us on track to search for topological phases in quasi-one-dimensional systems by coupling flux-assisted two-leg ladders through vertical tunnelings. This is the subject of the analysis performed in Sec. \ref{sec:cp}.

%------------------------------------------------------------------------------------------------------------
\section{Two-Band Model in the Strong-Coupling Limit}
\label{sec:tw}

We start from a system of two coupled Kitaev superconducting wires with $\Delta_1=\Delta_2=\Delta$ and $\chi=\zeta=0$ (corresponding to the first case mentioned above).  We consider the limit $(t,t_{\perp})\gg (|\Delta_0|, |\Delta|)$, where one can apply the bonding-antibonding band representation. We take $t>0$ and $t_\perp>0$ without loss of generality. Hereafter, we build a Bardeen-Cooper-Schrieffer (BCS) model in this representation, and to discuss interaction as well as interband/interwire pairing effects we also introduce the Luttinger liquid formalism \cite{Haldane,Heinz,thierry2004,Tsvelik}. In Sec. \ref{IIIA}, we study the physics and observables associated to the three phases, namely the two topological superconducting phases with 4 and 2 MZM and the trivial phase corresponding to the strong-paired phase. In Sec. \ref{IIIB}, we also discuss the stability of MZM modes in relation with SPT phases and check the results numerically.

A key property of the system is that for arbitrary values of $t_{\perp}$, the interwire pairing term $\Delta_0 c^{\dagger}_1(j)c^{\dagger}_2(j) + \text{H.c.}$ is irrelevant from renormalization group arguments. To prove this important point, we resort to the bosonization formalism or Luttinger liquid description, which will be also useful when discussing proximity effects and magnetic field-induced phases in Sec. \ref{sec:hw}. We assume the continuum limit $\psi_{\alpha} (x) =  c_{\alpha}(j)/{\sqrt{a}}$ with $x = x_j= ja$ \cite{Haldane,Heinz,thierry2004,Tsvelik}. A fermionic operator can be written in terms of bosonic fields $\phi_{\alpha}$ and
 $\theta_{\alpha}$:
  \begin{gather}
    \psi_{\alpha} (x) = \psi_{R}^{\alpha}(x) +  \psi_{L}^{\alpha}(x), \notag \\
    \psi_{r}^{\alpha} (x) = \frac{U_{r}^\alpha}{\sqrt{2\pi a}} e^{irk_{F}^{\alpha}x} e^{-i \left[ r \phi_{\alpha} (x) - \theta_{\alpha} (x)\right]}. \label{eq:b1}
  \end{gather}
The index $r = +1(-1)$ is taken for $r = R(L)$, {\it i.e.} for the right (left) moving particle. Formally, $\alpha = 1,2 $ or $+, -$ can embody the wire or band basis. In this subsection, we switch to the bonding and anti-bonding basis (\ref{eq:bab}). The Klein factors $U_{r}^{\alpha}$ enforce the Fermi statistics  and satisfy the relations: $U_{r,\alpha}^\dagger = U_{r,\alpha}$, $\{U_{R,\alpha}, U_{L, \alpha} \} = 0$. For convenience, it is sufficient to set  $U_{R, \alpha}U_{L, \alpha}= i$ and all others to unity $U_{r, \alpha} U_{r', \alpha'}= 1$. The bosonic fields satisfy the commutation relation
  \begin{gather}
    \left[ \phi_{\alpha} (x), \theta_{\beta}(x')\right] = i\frac{\pi}{2} \delta_{\alpha \beta} \text{Sign} (x'-x). \label{eq:b2}
  \end{gather}   
  
For $\Delta=0$, each band gives rise to a Luttinger quadratic Hamiltonian in real space, characterized by the Luttinger parameter $K$ that is equal to one for free fermions \cite{Haldane,thierry2004} (see Eq. (\ref{eq:h0pm})).  We then notice that for general values of $t_{\perp}$ the interwire pairing contribution coupling the two bonding and antibonding bands now oscillates rapidly in the real space domain:
  \begin{align}
    \mathcal{H}_{\Delta_0}  &= -\Delta_0 \sum_j  c_+^\dagger(j) c_-^\dagger (j) + \text{H.c.}   \notag
   \\  &= -\frac{4|\Delta_0|}{\pi a}   \int dx \cos (k_{F,+} x - \phi_{+} ) \cos(k_{F,-} x - \phi_-) \notag
   \\ & \phantom{ = -\frac{4|\Delta_0|}{\pi a}   \int dx}  \times \cos(\delta - \theta_+ - \theta_-),    \label{eq:iwp}
   \end{align}
where $\Delta_0 = |\Delta_0|e^{i\delta}$. This contribution which oscillates rapidly is irrelevant from renormalization point of view and can be neglected in the effective BCS model even when the chemical potential lies above the bottom of the antibonding band. Below, we check numerically that the interband pairing term $\Delta_0$ in Eq. (\ref{eq:iwp}) does not modify the low-energy properties. It should be noted that the phase $\delta$ coming from the fact that the pairing term is not gauge-invariant, should not influence the physical properties. For instance, one could re-absorb it through a redefinition of the phases $\theta_+ \rightarrow \theta_+ +\delta/2$ and $\theta_- \rightarrow \theta_- +\delta/2$, which would then only modify slightly the fermion operators in Eq. (\ref{eq:b1}) through a global phase. The system thus behaves as if $\delta=0$ in this section, and similarly we can take the order parameter $\Delta$ to be real.

Assuming $K=1$, the Hamiltonian $\mathcal{H}$ can be rewritten as BCS Hamiltonians in the two subspaces $\mathcal{H} = \mathcal{H}(\zeta=\chi=0, \Delta_1=\Delta_2=\Delta) = \mathcal{H}_+ \oplus \mathcal{H}_-$, each sharing the form of a Kitaev wire Hamiltonian \cite{kitaev2001}. We obtain the model
  \begin{gather}
    \mathcal{H}_\pm 
      =  \frac{1}{2} \sum_k \Psi_{\pm}^\dagger (k) h_{\pm}(k) \Psi_{\pm}(k), \notag \\
      h_{\pm} (k) = \begin{pmatrix}
                              \xi_{k,\pm} & -\Delta_k \\
                              -\Delta_k^* & - \xi_{k,\pm}
 			  \end{pmatrix}, \quad
     \Psi_{\pm}(k) = \begin{pmatrix}
                      c_\pm(k) \\
                      c_\pm^\dagger (-k)
     	           \end{pmatrix}.  \label{eq:th}
  \end{gather}
 $\xi_{k, \pm}$ is given in Eq.~(\ref{eq:dp}) and the off-diagonal elements read $\Delta_k = -2i\Delta \sin (ka)$. 
  
 The $\Delta_0$ term, as written in Eq. (\ref{eq:iwp}), does not contribute to the intraband pairing term as a result of the Pauli principle. Adding a contribution of the form $\Delta_0 c^{\dagger}_1(j)c^{\dagger}_2(j+1)$ simply renormalizes the intra-band $\Delta$ term in $\Delta_k$ in an additive manner. 
 
The total band structure resulting from the diagonalization of the Bogoliubov-de-Gennes Hamiltonian (\ref{eq:th}) consists of four branches ($\pm E_{\pm}$) with 
 \begin{gather}
E_{\pm} (k) = \sqrt{(2t\cos(ka) \pm t_\perp + \mu)^2 + 4\sin^2 (ka) |\Delta|^2}. 
  \end{gather}
If $\Delta = 0$, as depicted in Fig.~\ref{fig:2kw} (top right), there are three possibilities for the occupancy of two lowest bonding and antibonding bands ($-E_{\pm}$) depending on the chemical potential: (i)  
4 MZM phase: $2t+t_{\perp}> \mu > -(2t - t_\perp)$;  (ii) 2 MZM phase: $- (2t + t_\perp)< \mu < -(2t - t_\perp)$; (iii) Trivial phase: $|\mu | > (2t + t_\perp)$. Once $\Delta \ne 0$, by analogy with the Kitaev chain \cite{kitaev2001}, a superconducting gap is opened at the Fermi level,  giving rise to a protected Majorana mode associated to each gapless Fermi point. We thus confirm the presence of two topological phases and one trivial phase hosting 4, 2 and 0 MZM in the phase diagram of Fig.~\ref{fig:2kw} (top left) \cite{Nagaosa,Wang}. This phase diagram holds as long as $t_{\perp}<2t$. 

Let us comment briefly on the stability of the phases against the Coulomb interaction in each wire. Following Refs. \cite{Herviou,Review,Urs}, we observe that interactions would have two effects. First, they would modify the Luttinger parameter $K<1$ in each band, therefore decreasing slightly the size of the intraband pairing amplitude as $\Delta_{\text{eff}} \sim \Lambda_c (\Delta /\Lambda_c)^{1/(2-1/K)}$ due to renormalization effects when integrating modes at high energy (with $\Lambda_c\sim t,t_{\perp}$ being a high-energy cutoff in the model). Second, they would produce a Josephson term or interband Cooper pair channel between the band pairs which corresponds then to locking the phase difference between the superconducting gaps in each band. As long as Coulomb interactions are moderate and the pairing amplitude $\Delta_{\text{eff}}$ remains appreciable, implying that $K>1/2$, then the phase with four Majorana fermions subsists, and equally for the phase with two Majorana fermions.  We then conclude that as long as $K$ is not too far from one, the topological phase diagram remains unchanged. The stability of the Majorana edge structure towards interactions in the phase with four Majorana fermions was discussed thoroughly in Ref. \cite{Herviou} for $t_{\perp}=0$. 
 
Additionally, we perform numerical calculations on Eq. (\ref{eq:th}) through exact diagonalization (ED) in a finite-size system ($M=1000$) with open boundary conditions (OBCs) using the Hamiltonian $\mathcal{H}_{\pm}$ above (assuming here that $K=1$), including the effect of $\Delta_0$. Fig.~\ref{fig:2kw} (bottom) shows the energy spectra in the strong coupling regime $(t, t_\perp) \gg (\Delta, \Delta_0)$. In particular, we observe that the 4 and 2 MZM are both robust against relatively strong interwire pairing interactions, even when $\Delta_0 > \Delta$, in agreement with the theory above and in Sec. \ref{IIIB}. In bosonization, the pairing term in the bonding (antibonding) band gives a contribution of $\Delta\cos(2\theta_+)$ $(\Delta\cos(2\theta_-))$ in the Hamiltonian. The relevance of these intraband pairing channels then ensures that higher-order contributions of the $\Delta_0$ term in Eq. (\ref{eq:iwp}) will not affect the low-energy physics.

%---------------------------------------------------------------------------------------------------------
\subsection{Observables}
\label{IIIA}

From Eq. (\ref{eq:th}), one can extract physical quantities such as superconducting correlation functions and the local compressibility.

\begin{figure}[t]
   \begin{center}
         \includegraphics[height=3cm]{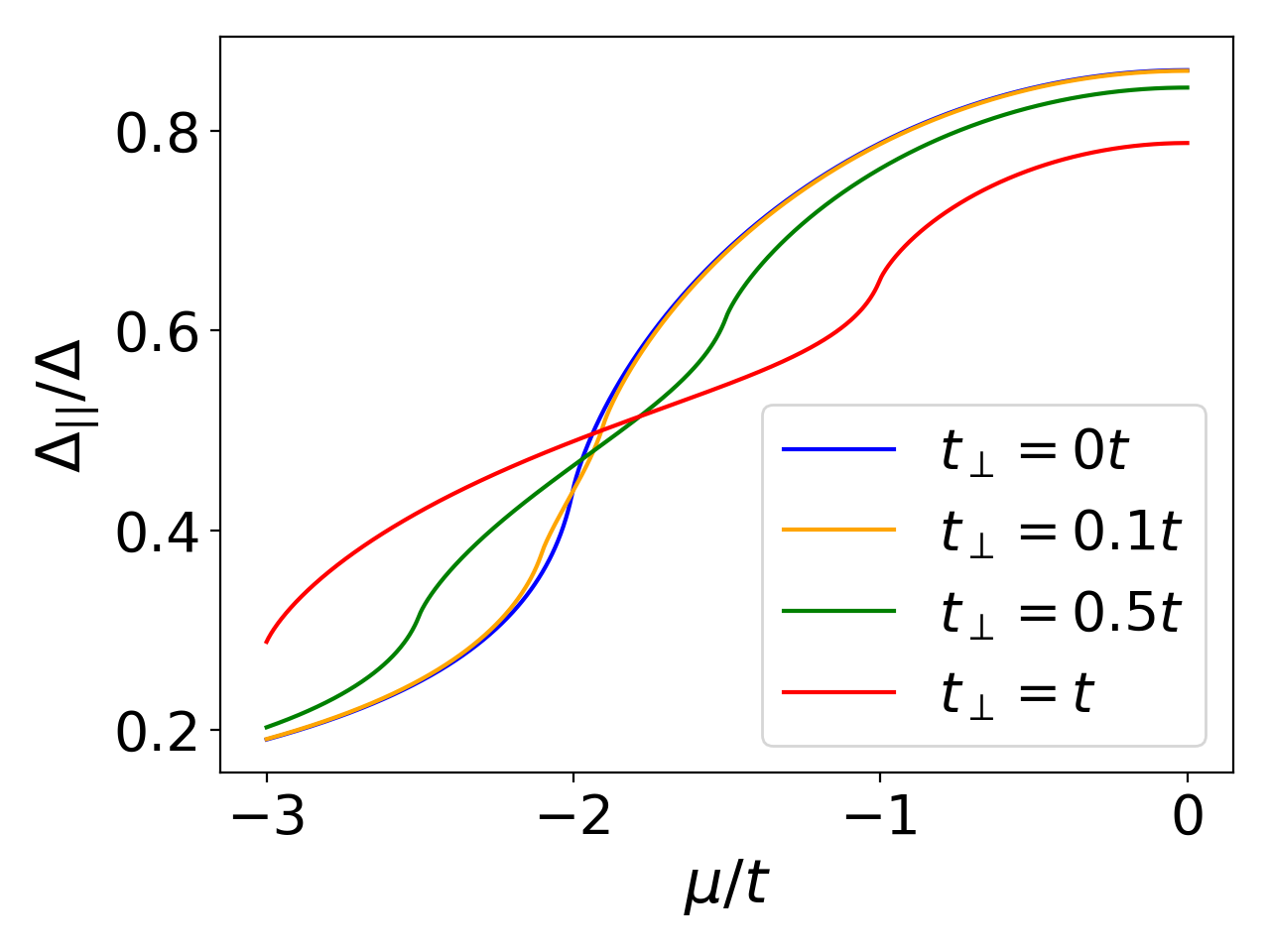} \hspace{0.15cm }\includegraphics[height=3cm]{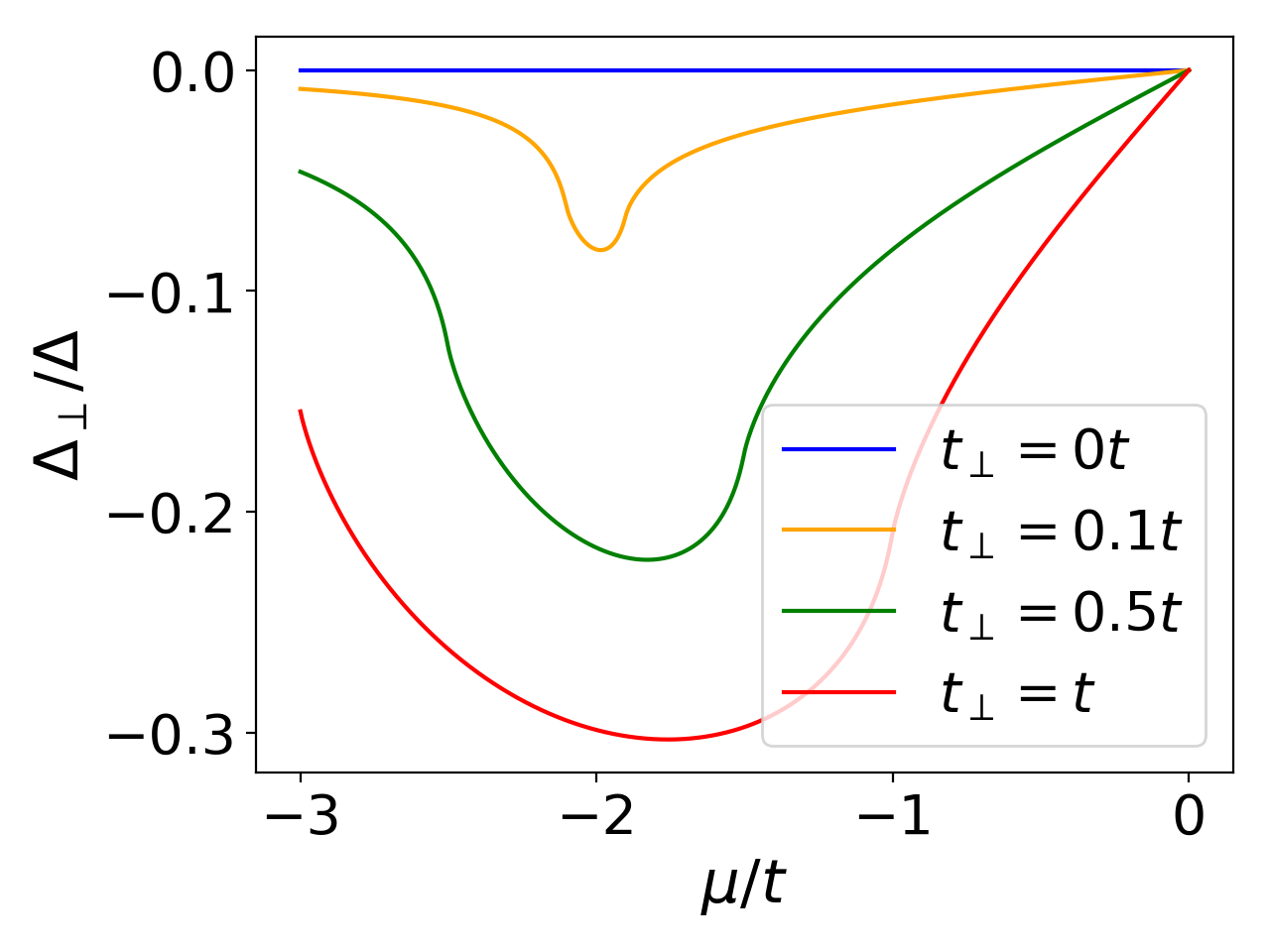} \hspace{0.15cm} \\
        \includegraphics[height=3cm]{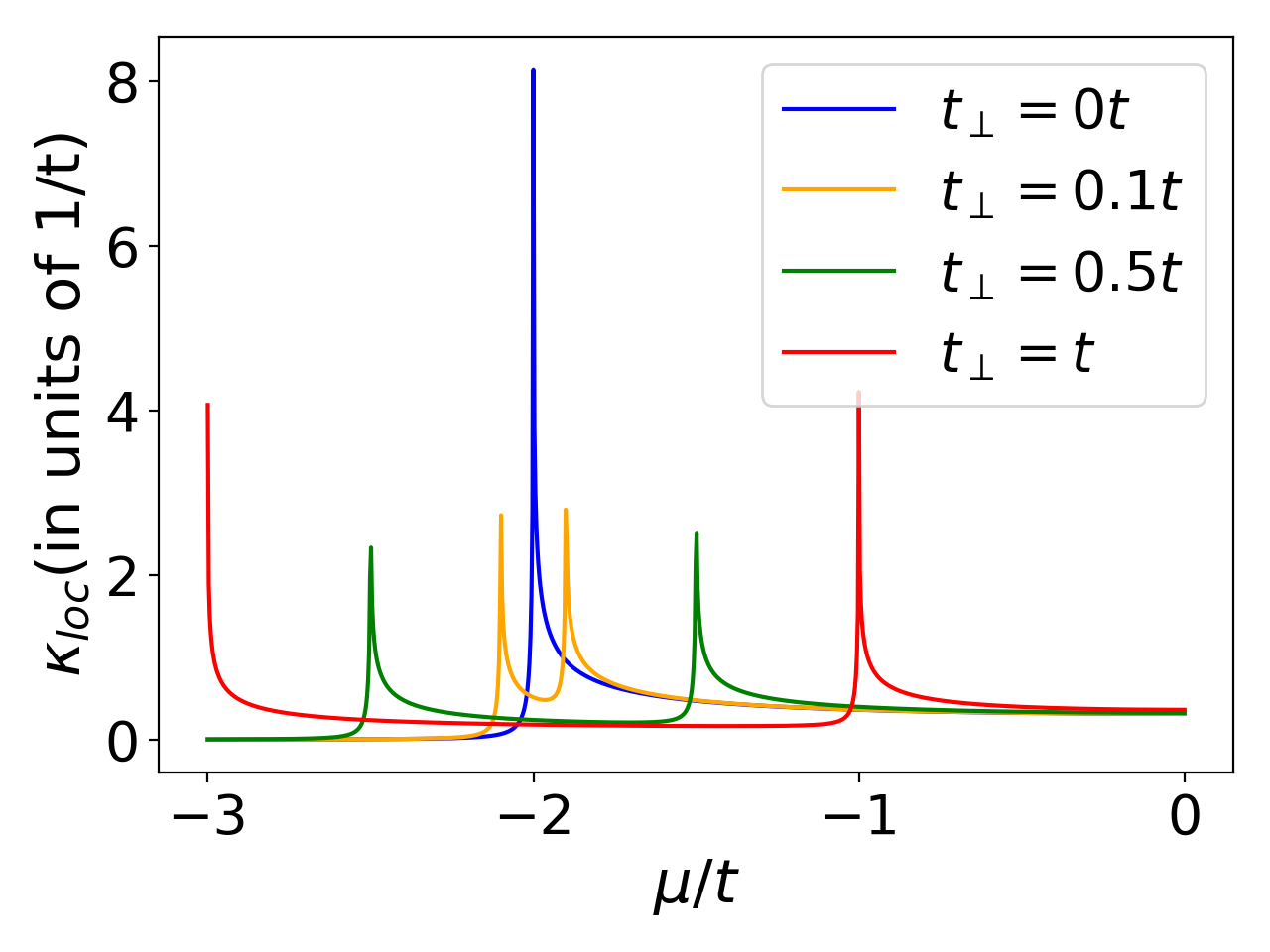}    \includegraphics[height=3cm]{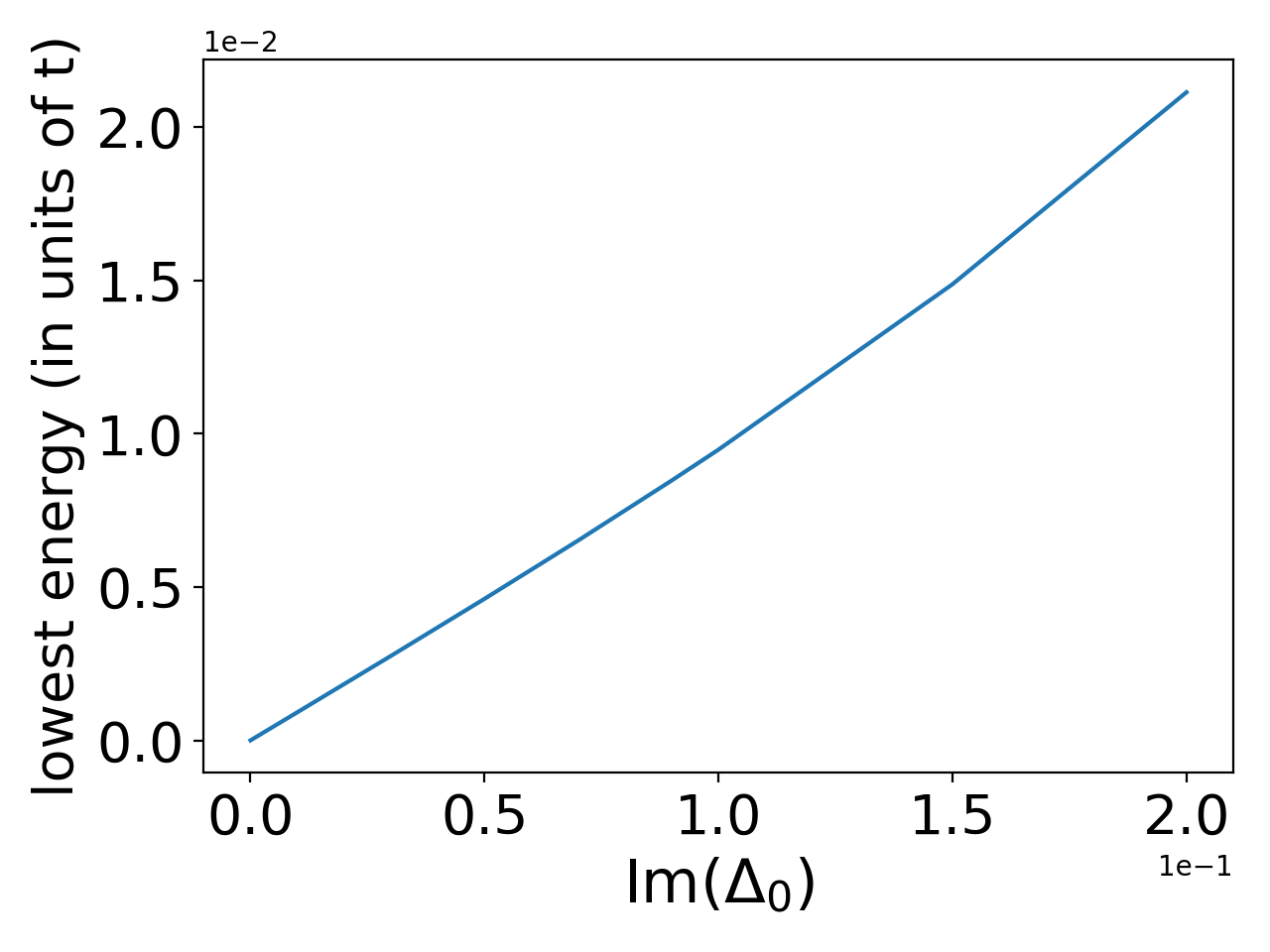}
    \end{center}
  \vskip -0.5cm \protect\caption[]
  {(color online) (Top) Formation of the intrawire gap $\Delta_\parallel$ and the interwire gap $\Delta_\perp$ as a function of the chemical potential $\mu$ for various tunneling strengths $t_\perp$. The pairings induced by the superfluid reservoir are set to $\Delta = 0.1t$ and $\Delta_0 = 0$; (Bottom left) Local compressibility. The parameters are kept the same as previous two plots; (Bottom right) Linear growth of the lowest single-particle excitation energy as a function of $\mathfrak{Im}{\Delta_0}$. In ED, the parameters are the same as Fig.~\ref{fig:2kw} (bottom) with  $\mu = -t$ corresponding to the 4 MZM phase.}
    \label{fig:2kob}
    \vskip -0.5cm
\end{figure}

We define two types of ``gap'' (pairing correlation functions) in the system, one intrawire gap as 
\begin{align}
\Delta_{\parallel} = (1/M) \sum_j \langle c_1(j) c_1(j+1)\rangle,
\end{align}
and the other interwire gap as 
\begin{align}
\Delta_{\perp} = (1/M) \sum_j \langle c_1(j) c_2(j+1) \rangle,
\end{align}
which is induced by the band contribution $\Delta_k c^{\dagger}_+(k)c^{\dagger}_+(-k)$ where $\Delta_k=-2i\Delta\sin(ka)$. This induced pairing term between the two wires corresponds to the transfer of one particle forming a Cooper pair in one wire into the other wire as a result of $t_{\perp}\neq 0$. The interwire quantity $(1/M) \sum_j \langle c_1(j) c_2(j) \rangle$ is equal to zero. By symmetry $\Delta_{\parallel}$ can also be defined with the fermions $c_2$. Both observables can be extracted from the Hamiltonian (\ref{eq:th}) through the Bogoliubov transformation. One finds
  \begin{align}
    \Delta_\parallel  &= \int \frac{dk}{2\pi} \Delta \sin^2(ka) \left( \frac{1}{2E_+(k)} + \frac{1}{2E_-(k)} \right), \notag \\
     \Delta_\perp  &= \int \frac{dk}{2\pi} \Delta \sin^2(ka) \left( \frac{1}{2E_+(k)} - \frac{1}{2E_-(k)} \right),\label{Delta}
   \end{align}
where $k \in [ -\pi/a, \pi/a [$. We plot them by varying $t_{\perp}$ and $\mu$ in Fig.~\ref{fig:2kob} (top). When $t_{\perp}=0$, we check that $\Delta_{\perp}=0$. We also check that $\Delta_{\perp}\rightarrow 0$ in the strong-paired phase, for instance when $\mu\sim -3t$, showing that in this regime the system behaves as if the kinetic terms were almost zero and the intrawire pairing terms $\Delta_{\parallel}$ approach smoothly their uncoupled-wire values. We also find that $\Delta_{\perp}=0$ when $\mu=0$ independently of the value of $t_{\perp}$. This can be explained by the fact that $E_+(k+\pi) = E_-(k)$ when $\mu=0$.

As the tunneling effect between the two wires increases, the splitting of the bonding and antibonding bands becomes more significant, and $\Delta_{\perp}\neq 0$ results from the transport of one electron forming a Cooper pair in a wire, into the other wire. Indeed, we observe an enhancement in $\Delta_\perp$ with larger $t_\perp$. We study the behaviors of $\Delta_{\parallel}$ and $\Delta_{\perp}$ in the vicinity of the two quantum phase transition points occurring at $\mu = -(2t \pm t_\perp)$; here, $\Delta_\parallel$ shows a combination of smooth steps and inflection points.

To locate the phase transitions accurately, we study the local compressibility $\kappa_{\text{loc}}$, which can be measured via scanning electron transistors in nanowire systems \cite{david2016}. One could also 
introduce bipartite fluctuations to localize the quantum phase transitions and study some aspects of the topology in the bulk \cite{Herviou,fluctuations}. By definition, $\kappa_{\text{loc}} = \partial{n_{\text{loc}}} / \partial \mu$ with $n_\text{loc}$ the local number of particles in two wires. Using $(1/M) \sum_{j,\alpha} \langle c_\alpha^\dagger (j) c_\alpha (j) \rangle$ for $n_{\text{loc}}$ we obtain
  \begin{gather}
    \kappa_{\text{loc}} = - \int \frac{dk}{2\pi} \frac{\partial}{\partial\mu} \left( \frac{\xi_{k,+}}{2E_+(k)} + \frac{\xi_{k,-}}{2E_-(k)} \right).
  \end{gather}
 Originating from Van Hove singularities in the density of states of each band ($\pm$), in Fig.~\ref{fig:2kob} (bottom left), $\kappa_{\text{loc}}$ shows a clear divergence as a function of  $\mu$ at the quantum phase transitions towards the 
 phases with 2 and 4 MZM.
 
%------------------------------------------------------------------------------------------------------------
\subsection{Stability of Majorana zero modes}
\label{IIIB}

Here, we study the stability of the Majorana edge modes against the interwire pairing term $\Delta_0$, using symmetry arguments. 
The 4 MZM phase belongs to the class BDI \cite{bernevig2015}, which respects the time-reversal symmetry $\mathcal{T}$, chiral symmetry $\mathcal{C}$ 
and particle-hole symmetry $\mathcal{P}$. To gap four Majorana zero-energy modes, we need a term which breaks $\mathcal{T}$. Since complex spinless fermion operators are left invariant under $\mathcal{T}$,
 the interwire pairing transforms as
  \begin{gather}
   \mathcal{T } \Delta_0 c_+^\dagger(j) c_-^\dagger (j) \mathcal{T}^{-1} =  \Delta_0^* c_+^\dagger(j) c_-^\dagger (j).
 \end{gather}
Only the imaginary part of $\Delta_0$ could break $\mathcal{T}$ and couple the edge modes. We have justified above that for $\chi=\zeta=0$, the pairing terms can be taken to be real (see discussions below Eq. (\ref{eq:iwp})), then ensuring the stability of the 4 MF edge states as numerically checked in Fig. \ref{fig:2kw} (bottom).  The phase of $\Delta_0$ must be related to the superconducting phase difference between the two chains. Such a difference is gauge-invariant, hence observable.
Below, for completeness, we study briefly the effect of  $\mathfrak{Im}{\Delta_0}\neq 0$, which can occur in certain geometries when applying magnetic fluxes per plaquette different from $0$ and $\pi$ (see {\it e.g.} Sec. \ref{sec:p+ip}). 

To be more precise, we introduce local Majorana fermion operators
     $\gamma_A^\lambda (j) = c_\lambda^\dagger(j) + c_\lambda(j)$ and
     $ \gamma_B^\lambda (j) = i(c_\lambda^\dagger(j) - c_\lambda(j))$
with $\lambda = \pm$ for the bonding and anti-bonding bands. In terms of these, the Hamiltonian of each subspace reads
   \begin{align}
     \mathcal{H}_{\lambda} &= \frac{i}{2} \sum_j \left[ -(t +\Delta) \gamma_A^\lambda(j) \gamma_B^\lambda (j+1) \right. \notag \\
     &\left. + (t -\Delta)\gamma_B^\lambda (j) \gamma_A^\lambda (j+1) -(\mu + \lambda t_\perp ) \gamma_A^\lambda (j) \gamma_B^\lambda (j) \right].
  \end{align} 
For $(\mu+\lambda t_{\perp})\ll t$ and $\Delta\sim t$, we obtain the Majorana pairs of $\gamma_A^\lambda(j) \gamma_B^\lambda (j+1)$. If we relabel the site index as $j = 0, \cdots, M-1$, then free particles on the two boundaries constitute a 4-Majorana edge state $\{\gamma_B^+(0), \gamma_B^-(0), \gamma_A^+(M-1), \gamma_A^-(M-1)\}$.  At a boundary, either the Majorana fermion $\gamma_A^\lambda$ or $\gamma_B^\lambda$ is gapped confirming the protection of Majorana modes towards large values of $t_{\perp}$ in the band basis;  for $\Delta\sim t$ and $(\mu+\lambda t_{\perp})\ll t$  one can neglect the effect of the term $-(\mu + \lambda t_\perp ) \gamma_A^\lambda (j) \gamma_B^\lambda (j)$. We can now rewrite the interwire pairing term at the boundaries as:
 \begin{align}
    \delta \mathcal{H}_{\Delta_0} &= - \frac{\mathfrak{Im}{\Delta_0}}{2} \left[ i \gamma_B^+(0) \gamma_B^-(0)+ i \gamma_A^+(M-1) \gamma_A^-(M-1) \right].
    \end{align}
    The fermion parity operators $i \gamma_B^+(0) \gamma_B^-(0)$ and $i \gamma_A^+(M-1) + \gamma_A^-(M-1)$ commute with the Hamiltonian and take eigenvalues $\pm 1$ at zero temperature. 
    In general, a gap is induced by the  interwire pairing with the behavior:
  \begin{gather}
    \Delta E \propto \mathfrak{Im}{\Delta_0}. \label{Imaginary}
  \end{gather}
This behavior is verified by ED calculations (see Fig.~\ref{fig:2kob}, bottom right).

The arguments above confirm the stability of the topological phases with 4 and 2 MZM when $\Delta\sim t$ assuming that $(\mu+\lambda t_{\perp})\ll t$. We check numerically the stability of the topological phases and MZM for other choices of parameters $\Delta/t$ and $\mu$; see Fig. \ref{fig:2kw}.

%------------------------------------------------------------------------------------------------------------
\section{Magnetic Flux Effects and Andreev Mechanism}

\begin{figure}[t]
   \begin{center} 
      \includegraphics[width=0.75\linewidth]{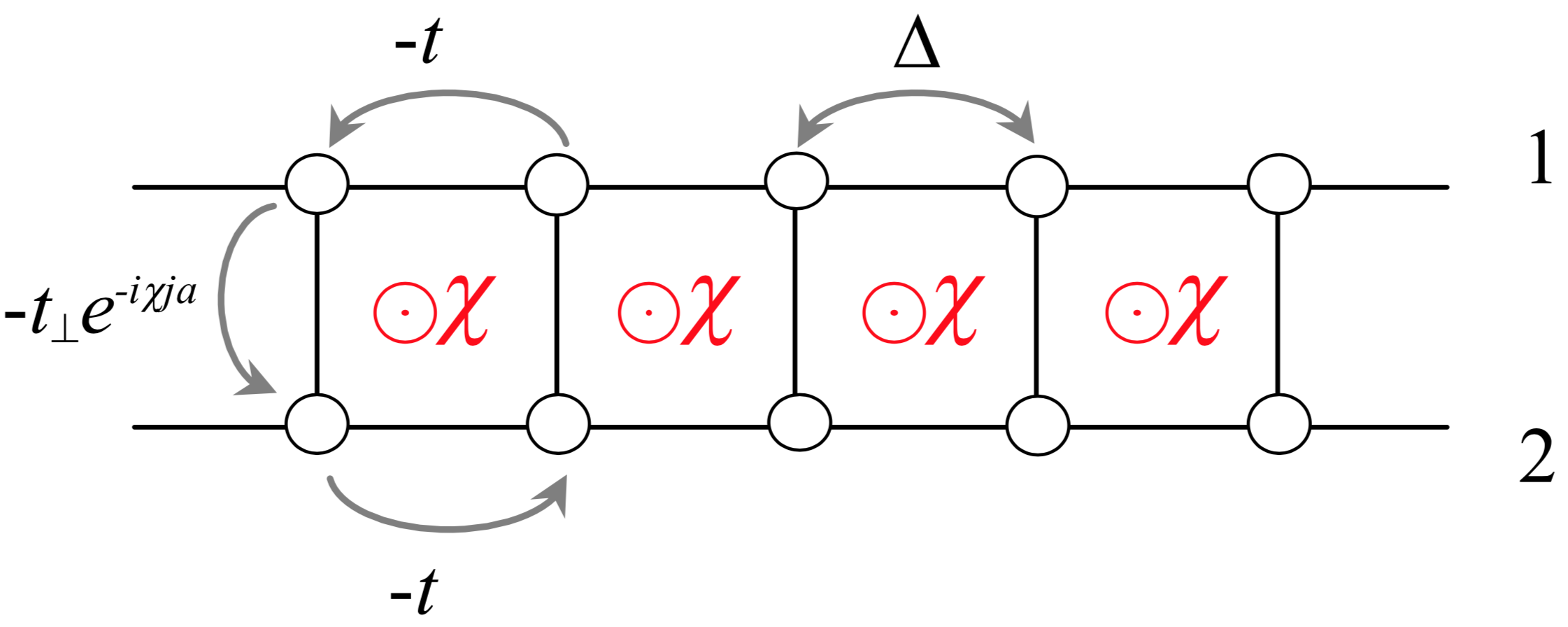} \\ \vspace{0.3cm}
      \includegraphics[width=0.618\linewidth]{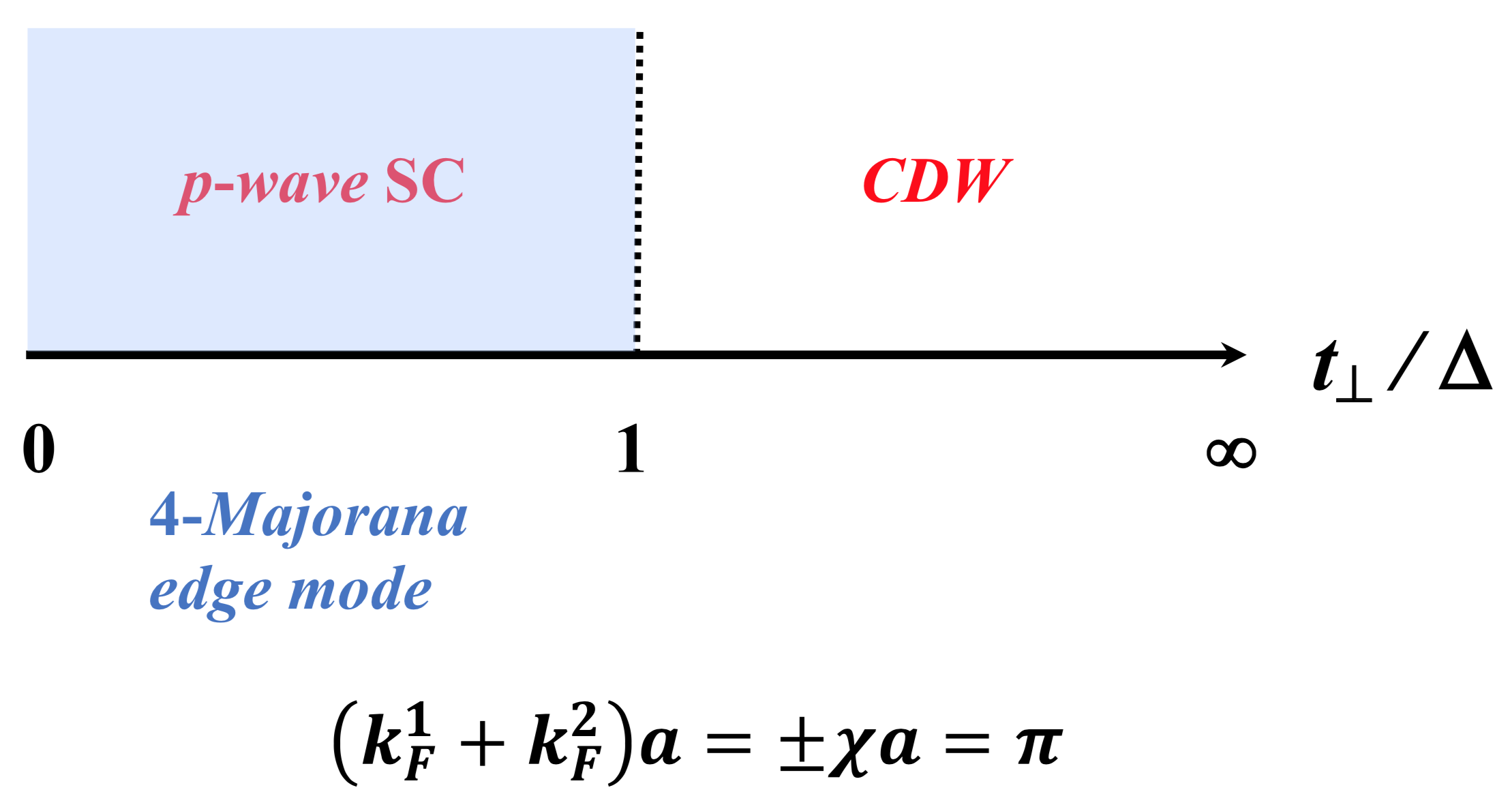}    \\ \vspace{0.5cm}
      \includegraphics[width=0.618\linewidth]{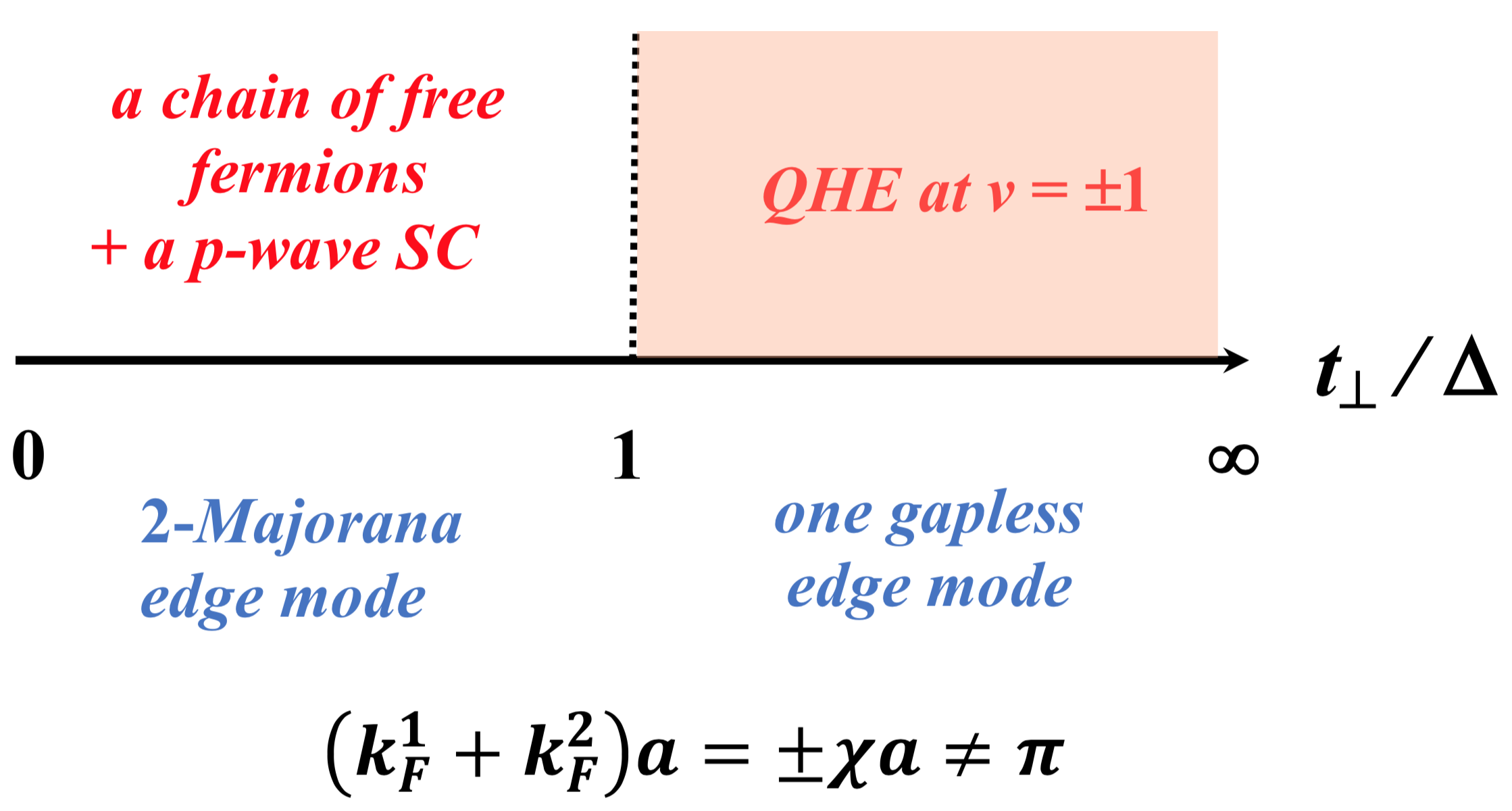}    
    \end{center}
  \vskip -0.5cm \protect\caption[]
  {(color online)   (Top) Hybrid system comprising a topological superconducting Kitaev p-wave chain and a chain of free fermions. (Middle) and (Bottom) Phase diagrams for $K=1$ and for densities in the wires such that $(k_F^1+k_F^2)a=\pm \chi a=\pi$ and
$(k_F^1+k_F^2)a=\pm \chi a \neq\pi$.  For the superconducting phase in blue, the condition on $k_{F}^i$ ensures the occurrence of 4 MZM until the charge density wave transition. The dashed lines refer to quantum phase transitions.}
    \label{fig:hladder}
    \vskip -0.5cm
\end{figure}

Now, we generalize the results of two wires to the situation of a hybrid system comprising a Kitaev superconducting chain and a metallic chain, and study the interplay between Andreev processes \cite{rice1994,karyn2001} and orbital magnetic-field effects. In the following, we concentrate on the case of spinless or spin-polarized fermions and study the proximity effect between the superconducting wire and the Luttinger wire through the bosonization formalism. In Sec. \ref{sec:spinful}, we address a comparison to the
case of spin-1/2 fermions. In Fig.~\ref{fig:hladder} (top), we depict this situation (corresponding to the second case discussed in Sec. \ref{cases}). The model is described by
  \begin{gather}
    \mathcal{H}_{\text{hybrid}} = \mathcal{H} (\zeta = 0, \chi\neq 0, \Delta_2 = \Delta_0 = 0, \Delta_1 = \Delta). \label{eq:hh}
  \end{gather}
The precise goals below are as follows: first, we study the evolution of the topological superconducting phases in the presence of a gap anisotropy and study the weak-coupling limit. 
In addition, we study possible quantum phase transitions under a uniform magnetic flux. The schematic phase diagram is plotted in  Fig.~\ref{fig:hladder} (middle) and (bottom) for different flux conditions. When $\pm \chi a = \pi$, in the weak coupling limit $p$-wave superconductivity is induced in the free fermion wire. Deviating slightly from the $\pm \chi a = \pi$ situation, we analyze the Meissner and Majorana currents in the bulk and at the boundaries, respectively. We also  address Abelian quantum Hall phases for particular relations between magnetic flux and densities in the wires and describe properties of the charge density wave state in real space when the bonding band is fully filled, occuring at $\pm \chi a = \pi$ in the strong-tunneling limit. 

\label{sec:hw}
 
To acquire a physical understanding of the system's properties in real space (such as induced currents) when applying a magnetic field and to generalize the results when $t_{\perp}\ll \Delta$, we now switch to the bosonization picture in the wire basis. This allows us to include higher-order processes in the tunnel-coupling description. The approach below specifically addresses the limit $t\gg t_{\perp}\gg \Delta$, such that we can apply the continuum limit in the $x$ direction parallel to the wires. In the bosonization picture (\ref{eq:b1}), the kinetic term along the wires in $\mathcal{H}_{\text{hybrid}}$ takes the form
   \begin{align}
     \mathcal{H}_\parallel &=  \mathcal{H}_0^+ +   \mathcal{H}_0^-, \notag \\
     \mathcal{H}_0^\pm &= \frac{v^{\pm}}{2\pi} \int dx \left[  K^{\pm} ( \nabla \theta^\pm)^2 + \frac{1}{K^{\pm}} (\nabla \phi^\pm)^2 \right], \label{eq:h0pm}
   \end{align}  
 with the rotated fields $\theta^{\pm} = (\theta_1 \pm \theta_2)/\sqrt{2}$ and $\phi^{\pm} = (\phi_1 \pm \phi_2)/\sqrt{2}$.  It is important to emphasize here that the modes $\phi^{\pm}$ and $\theta^{\pm}$ are not the same as the ones $\phi_{\pm}$ and $\theta_{\pm}$ in the preceding section. Here,  these modes must be understood as symmetric and anti-symmetric superpositions of bosonic fields in the wire basis, whereas in the preceding section the label $\pm$ referred to the two band indices. The sound velocities and Luttinger liquid parameters satisfy $v^{\pm}K^{\pm} = vK=v_F$, where the Fermi velocity takes the form $v_F = 2ta \sin(k_Fa)$ if we linearize the band structure of each wire close to each Fermi point, and in this formula $k_F$ refers either to $k_F^1$ or $k_F^2$. For simplicity, we assume that the two wires have (almost) the same velocity. If we neglect the effect of the Coulomb interaction or nearest-neighbor interaction parallel to the chain $V_{\parallel}$, then formally $K=1$. If we take into account this interaction, as discussed in Sec. \ref{IIIA} then this would adiabatically renormalize $K$ to a value smaller than $1$, according to $v^{\pm}/K^{\pm}=v/K\approx v+2V_{\parallel}a/\pi$. 
   
  The interwire hopping term takes the form
  \begin{align}
    \mathcal{H}_{\perp} = & - \frac{2t_{\perp}}{\pi a} \int dx    \cos (\sqrt{2} \theta^{-} - \chi x)  \notag \\
                               &  \times [\cos((k_F^1+k_F^2)x - \sqrt{2} \phi^+) \notag \\
                                    & + \cos[(k_F^1-k_F^2)x-\sqrt{2} \phi^-) ]. \label{eq:hp}
  \end{align}
 The conservation of the total number of particles in the system at $t_{\perp}=0$ and $t_{\perp}\neq 0$ implies the equality $k_F^1+k_F^2=k_{F,+} + k_{F,-}$ in accordance with Luttinger's theorem. The pairing term in wire 1 takes the form
  \begin{gather}
    \mathcal{H}_\Delta = -\frac{2\Delta}{\pi a} \int dx \sin (k_F^1 a)\cos(2\theta_1). \label{eq:pd}
   \end{gather}
Furthermore, we choose densities in the wires $k_{F}^i\neq \pi/(2a)$, such that intrawire insulating transitions do not occur, since they have been well studied \cite{Herviou,Schuricht,Li}. Still, we will show below that a charge density wave transition can occur in the system, giving rise to an analogue of the rung-Mott insulator for which the total charge $\phi^+$ and the superfluid phase difference $\theta^-$ are pinned \cite{RungMott}.  The physics is adiabatically linked to the half-filled situation where the lowest band is fully occupied and the upper band is empty, implying $k_{F,-}=0$ and $k_{F,+}=\pi/a$ when $t_{\perp}>2t$. In the wire basis, the half-filled condition at $t_{\perp}=0$ refers accordingly to $(k_{F}^1+k_{F}^2)=\pi/a$.

It should be noted here that in principle the effect of Coulomb interactions between the wires \cite{Herviou} could give rise to an additional term proportional to $V_{\perp}\int dx \cos[2(\phi_1-\phi_2)(x)-2(k_F^1-k_F^2)x]$ in the wire basis. To be able to describe the topological superconducting proximity effect induced by $t_{\perp}$, we assume below that $k_{F}^1\neq k_{F}^2$ such that the $V_{\perp}$ term averages to zero. Since $V_{\perp}$ is already a four-fermion operator, higher-order contributions will likewise not be relevant. 
 
%------------------------------------------------------------------------------------------------------------
\subsection{Topological proximity effect when $\chi=0$}
\label{chi0}

Let us start with the case without flux insertion, $\chi = 0$. For general situations where $k_{F}^1\neq k_{F}^2$, all the terms in Eq. (\ref{eq:hp}) oscillate rapidly and average to zero. This implies that one must include higher-order effects in $t_{\perp}$ in the weak-coupling regime \cite{rice1994,karyn2001,alex2015}. Appendix~\ref{app:pert} presents a perturbative approach developed to build the possible non-oscillating terms to higher order in $t_{\perp}$. 
In the weak-coupling limit $t_\perp \ll \Delta\ll t$, to show the emergence of 4 MZM as soon as $t_{\perp}\neq 0$, we expand the partition function to second order and 
identify the effective Hamiltonian
  \begin{align}
    \mathcal{H}_\perp^{(2)} &= -\frac{2t^2_\perp}{ \pi a \Delta} \int dx  \cos(2\sqrt{2}\theta^-) \notag \\
     &=   -\frac{2\bar{\Delta}_2}{ \pi a} \int dx  \cos(2\theta_2).
  \end{align}
 This corresponds to an Andreev process where a Cooper pair is transported from wire 1 to wire 2. For the second equality, we use the fact that at energies smaller than $\Delta$, the charge field of  wire 1 is pinned to a classical value $\langle \cos(2\theta_1) \rangle \sim 1$ implying $\theta_1\sim 0$, due to the strong pairing $\mathcal{H}_\Delta$ term  (\ref{eq:pd}). Classically, to minimize energy (at zero temperature) the superfluid phase $\theta_1$ will be pinned at one minimum of the (periodic) cosine potential $-\cos(2\theta_1)$. Each minimum $\theta_1=2n\pi$ with $n\in\mathbb{Z}$ is equivalent and thus we can assume that the phase is pinned at $\theta_1\sim 0$, neglecting instanton effects from one minimum to another. We observe an effective induced superconducting gap in the second wire: $\bar{\Delta}_2 \sim t_\perp^2/\Delta$. Therefore in the weak tunneling region, the superconducting phase shows 4 MZM for a large range of chemical potentials, when $-2t<\mu<2t$. This conclusion is in agreement with the results of Sec. \ref{sec:tw} in the band basis, since the 2 MZM region shrinks to zero when $t_{\perp}\rightarrow 0$.
 
Formally, minima of the form $\theta_1=(2n+1)\pi$ are also allowed and correspond to a twist of $\pi$ in the definition of the fermionic operator in Eq. (\ref{eq:b1}). In the Kitaev model for the wire $1$, this $\pi$ phase shift is equivalent to the $\mathbb{Z}_2$ symmetry $c_1\rightarrow -c_1$ in the BCS Hamiltonian. A redefinition of the global phase by $\pi$ of the fermionic operator associated with the wire 1 also corresponds to change $t_{\perp}\rightarrow -t_{\perp}$ and $\Delta_0\rightarrow -\Delta_0$ in Eq. (\ref{eq:h}). This corresponds to the transformation $c_1\leftrightarrow c_2$ in the Hamiltonian. In the band picture, this $\pi$-phase shift simply inverts the bonding and anti-bonding bands. It is however important to recognize that the intra- and interwire pairing terms would have a $\pi$-phase shift of difference, which seems difficult to realize with a unique (three-dimensional) superfluid reservoir. 

 \begin{figure}[t]
   \begin{center}
        \includegraphics[width=0.5\linewidth]{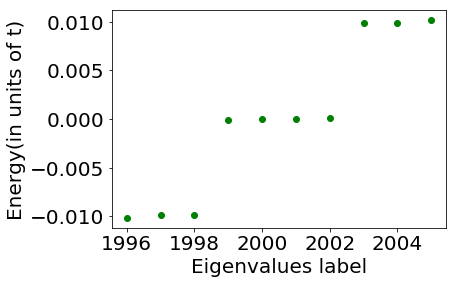}    
         \includegraphics[width=0.5\linewidth]{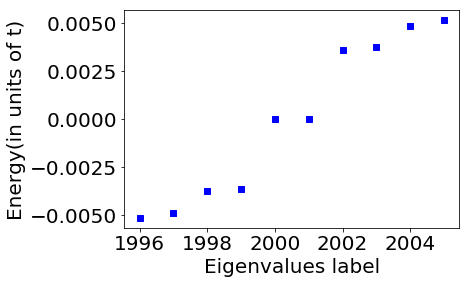}  
           \includegraphics[width=0.5\linewidth]{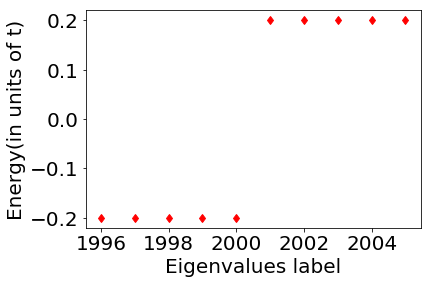}  
    \end{center}
  \vskip -0.5cm \protect\caption[]
 {(color online) Energy spectra from ED with the same eigenvalue labels as in Fig. \ref{fig:2kw}, with $\Delta_1=\Delta=0.01t$, $\Delta_2=0$ and $t_{\perp}=0.1t$ for $t=1$. This corresponds to the limit where $t\gg t_{\perp}\gg \Delta$ with anisotropic bare values of the superconducting gap. The three graphs correspond to $\mu=-1t$, $-2t$ and $-2.3t$ respectively. In the strong-paired phase with $\mu=-2.3t$, there is no MZM (red dots), in the case with two Fermi points where $\mu=-2t$, there are 2 MZM (blue dots), and in the case with four Fermi points where $\mu=-t$, there are 4 MZM (green dots). Starting with $t_{\perp}=0.1t$, fixing $\mu=-2t$ corresponds to having the Fermi level close to the bottom of the lower band, showing the fragility of the proximity effect close to the strong-paired phase since formally $\Delta\sin(k_F^1 a)$ becomes very small, resulting in very small gaps in (bulk) energy spectra.}
    \label{fig:gap2}
    \vskip -0.5cm
\end{figure}

In the regime $t \gg t_\perp \gg \Delta$, one can also generate a similar term
  \begin{gather}
    \mathcal{H}_\perp^{(2)} = -\frac{2t^2_\perp}{\pi a \Lambda} \int dx  \cos(2\sqrt{2}\theta^-). \label{eq:htv}
  \end{gather}
The energy cutoff $\Lambda$ depends on the short time and short distance considered in the virtual processes (see more details in Appendix~\ref{app:pert}). The original pairing term in wire 1 can be estimated as $\langle \cos(2\theta_1) \rangle \sim \Delta/\Lambda$, when evaluating the effect of the pairing term at low energy perturbatively in $\Delta$. More precisely, we calculate $\langle \cos(2\theta_1(x)) \rangle=\hbox{Tr}[e^{-\beta H}\cos(2\theta_1(x))]$ to first order in $\Delta$ leading to $\langle \cos(2\theta_1(x)) \rangle \sim \Delta\int_{1/\Lambda}^{\beta\rightarrow +\infty} d\tau \langle\cos(2\theta_1(x,\tau)) \cos(2\theta_1(x,0))\rangle$, where $\tau$ represents the imaginary time and $\langle ... \rangle$ means an average on the quadratic Luttinger theory of the correlation function at the position $x$. Remembering that $\langle \cos(2\theta_1(x,\tau))\cos(2\theta_1(x,0))\rangle\sim (\Lambda\tau)^{-2}$, we check that $\langle \cos(2\theta_1) \rangle \sim \Delta/\Lambda$. The $\mathcal{H}_\perp^{(2)}$ term then gives a contribution in $\sim (t_{\perp}^2/\Lambda)\langle\cos(2\theta_1)\rangle\int dx \cos(2\theta_2)$. Starting from a Gaussian Luttinger theory, we have the equality $\langle e^{2i\theta_1(x)}\rangle = \langle e^{-2i\theta_1(x)}\rangle=e^{-2\langle \theta_1^2(x)\rangle}$, implying that only the product of cosine functions contributes. A gap is induced in wire 2 and its amplitude now evolves as $\bar{\Delta}_{2} \sim t_\perp^2\Delta/\Lambda^2 \sim \Delta$. This argument can also be checked applying the arguments in the bonding - antibonding representation.  The term $\Delta_1$ in the band basis gives a term such as $\Delta_1 \sin(k_F^1 a)(c^{\dagger}_{+}(k)c^{\dagger}_+(-k)+c^{\dagger}_{-}(k)c^{\dagger}_-(k))$, which induces a contribution $\Delta_1 \sin(k_F^1 a)c^{\dagger}_{2}(k)c^{\dagger}_2(-k)$ with $\Delta_1=\Delta$. This confirms the preceding argument that the induced superconducting gap in wire $2$ becomes equal to the superconducting gap in wire $1$ in the strong-coupling limit. Bosonization arguments apply as long as the energy spectrum is linear, implying that $k_F^1$ is not too close to zero and therefore that $\sin(k_F^1 a)\sim 1$, which then validates the equivalence with Eq. (\ref{eq:htv}). If we consider the regime close to the bottom of a band, then one must rely on
the band basis arguments showing that the induced gap in wire $2$ is $\Delta \sin(k_F^1 a)$. 

To identify the number of Majorana fermions in this regime, we resort to the band-structure arguments of the preceding section assuming that $K\rightarrow 1$. The proximity effect gives rise to 4 MZM, 2 MZM or 0 MZM depending on the value of $\mu$, as checked numerically using Eq. (\ref{eq:th}) (see Fig. \ref{fig:gap2}). We also check numerically that when approaching the bottom of the lowest band, close to the strong-paired phase transition, the proximity effect becomes fragile as $\Delta\sin(k_F^1 a)$ becomes very small.

%------------------------------------------------------------------------------------------------------------
\subsection{$\pi$-Flux per plaquette at half-filling}
\label{piflux}

For non-zero values of the magnetic field with $\chi\neq 0$, one must adjust the densities or the Fermi wave-vectors in the two wires in Eq. (\ref{eq:hp}) to produce a proximity effect, {\it e.g.} to make the tunneling term $t_{\perp}$ or higher-order contributions relevant. Here, we study the situation with $\pi$-flux per plaquette. We assume that $(k_{F}^1+k_{F}^2)a=\pi$ such that the following commensuration relation is satisfied
  \begin{gather}
    a\left[ (k_F^1 + k_F^2)  \mp \chi  \right] = 0\mod 2\pi, \label{eq:fc}
  \end{gather}
 with $\chi a = \pm  \pi$. This corresponds to a half-filled ladder with one particle per rung.  The interwire hopping term (\ref{eq:hp}) becomes 
   \begin{gather}
     \mathcal{H}_{\perp} = -\frac{2t_{\perp}}{\pi a} \int dx    \cos (\sqrt{2} \theta^{-}) \cos(\sqrt{2} \phi^+), \label{eq:hp_pi}
   \end{gather}   
   modulo an oscillatory term. In the strong-coupling limit, both modes $\theta^{-}$ and $\phi^+$ are pinned to the classical values: $\theta^{-}\sim 0$, $\phi^+\sim 0$. The pinning of the mode $\phi^+$ suppresses fluctuations in the total density (or total charge) on a given rung.  The system shows $a(k_{F}^1+k_{F}^2)/\pi = 1$ particle and 1 hole per rung. Adding a particle or a hole at a given rung costs an energy, and the system shows a long-range charge order associated to the channel $\phi^+$. Each rung is equivalent with one another, leading to a uniform charge density wave order with an effective wave vector $q=(k_F^1+k_F^2\mp \chi) = 0$.

Essentially, when $t_{\perp}\gg \Delta$ and more precisely $t_{\perp}>2t$, in the bonding and antibonding representation of Fig. \ref{fig:gladder} bottom right, the lowest band becomes filled and adding one more particle in the antibonding band costs an energy of the order of $t_{\perp}$. The bosonization argument above suggests that this conclusion remains in fact correct even when $t_{\perp}<t$, but with $t_{\perp} \geq \Delta$, due to renormalization group arguments. Indeed, $t_{\perp}$ is a relevant perturbation associated to the kinetic terms ${\cal H}_0^{\pm}$, and therefore will grow under the renormalization scheme to values larger than $t$.  More precisely, defining the dimensionless quantities $\tilde{\Delta}=\Delta/\Lambda^*$ and $\tilde{t}_{\perp}=t_{\perp}/\Lambda^*$ with $\Lambda^*$ being the high-energy cutoff (which can be taken to be larger than $t$ since the total bandwidth for a given wire is $4t$), the invariance of the partition function to second-order in $\tilde{t}_{\perp}\ll 1$ and $\tilde{\Delta}\ll 1$ leads to the following two equations:
\begin{eqnarray}
\frac{d\tilde{t}_{\perp}}{dl} &=& \left(2-\frac{K_+}{2} - \frac{1}{2 K_-}\right)\tilde{t}_{\perp}, \nonumber \\
\frac{d\tilde{\Delta}}{dl} &=& \left(2-\frac{1}{2 K_+} - \frac{1}{2 K_-}\right)\tilde{\Delta},
\label{flow}
\end{eqnarray}
where $l=-\log(\Lambda^*/E)$ and $E$ corresponds to the energy scale of interest. This renormalization procedure can also be seen as an integration of modes at short distances with typical lengths between $\hbar v/\Lambda^*\sim a$ (with $\hbar=1$) and ${\cal L}=\hbar v/E$, corresponding to an integration of blocks in real space and a redefinition of the lattice spacing as ${\cal L}$. For free fermions, setting $K_+=K_-=1$, we check that both $\tilde{t}_{\perp}$ and $\tilde{\Delta}$ flow to strong couplings. If we assume free fermions $(K_+=K_-=1)$ or weakly-interacting fermions ($K_+<1$, $K_-<1$) and if we consider the limit where $t_{\perp}\gg \Delta$, then we confirm that the term $\tilde{t}_{\perp}$ will flow to strong coupling faster than the term $\tilde{\Delta}$. In this case, the low-energy physics will be strongly associated with the properties of Eq. (\ref{eq:hp_pi}) and the ground state is a charge density wave. 

For free fermions, our results are also in agreement with a filled lowest bonding band. Excitations above the charge gap corresponding to transferring a particle in the anti-bonding band then are accompanied with a phase change of $\pi$ associated to the fermion operator $c_2$ (from Eq. (\ref{eq:bab})), which is equivalent to $\theta_1-\theta_2+\chi a=\pm \pi$ in the presence of the magnetic field. Another manner to understand the pinning of the phase $\theta^-$ is through the condition that at each site $j$ (or each wave-vector $k$) the charge density wave formation implies $\langle \tilde{c}^{\dagger}_1 \tilde{c}_1\rangle +\langle \tilde{c}^{\dagger}_2 \tilde{c}_2\rangle=1$ or  $k_{F}^1+k_{F}^2=\pi/a = k_{F,+}+k_{F,-}$. If we consider the strong-coupling fixed point of the renormalization group arguments then this corresponds to a situation with a large $\tilde{t}_{\perp}\sim 1$ and with a fully occupied
lowest band, {\it i.e.} with $k_{F,+}=\pi/a$ and $k_{F,-}=0$. From the equality $\langle \tilde{c}^{\dagger}_+ \tilde{c}_+ \rangle =1 = \frac{1}{2}(\langle \tilde{c}^{\dagger}_1 \tilde{c}_1\rangle +\langle \tilde{c}^{\dagger}_2 \tilde{c}_2\rangle) + \frac{1}{2}(\langle \tilde{c}^{\dagger}_1 \tilde{c}_2\rangle + \langle \tilde{c}^{\dagger}_2 \tilde{c}_1\rangle)$, we then infer $\langle \tilde{c}^{\dagger}_1 \tilde{c}_2\rangle + \langle \tilde{c}^{\dagger}_2 \tilde{c}_1\rangle = 1$ on each rung of the ladder system. In the continuum limit, phase coherence takes place between a particle $({\psi^1_R})^{\dagger}$ and a hole $\psi^2_L$ .  This constraint is naturally fulfilled through the pinning conditions of $\phi^+$ and $\theta^-$ in $\mathcal{H}_{\perp}$.  Here, $\theta^-$ can be seen as the phase associated to the bosonic particle-hole pair wavefunction, and there is then a global phase coherence for the particle-hole pairs. 

We also observe that the $\pm \pi$ magnetic flux suppresses the effect of the superconducting term $\Delta_1$ when $t_{\perp}\gg \Delta$, since at low energy $\cos(2\theta_1)=\cos(\sqrt{2}(\theta^+ +\theta^-))$ and $\langle e^{i\sqrt{2}\theta^+}\rangle\sim 0$ due to the pinning of the dual mode $\phi^+$, and the quantum uncertainty principle resulting from commutation rules between $\phi^+$ and $\theta^+$. 

In the wire or chain representation, the system is analogous to the rung-Mott insulator \cite{RungMott,RungMottMunich} of the ladder system.  In the rung-Mott state of hard-core bosons, the Josephson effect produced the pinning of the phase $\theta^-$. We name the phase found here Charge Density Wave (CDW) in the phase diagram of Fig. \ref{fig:hladder} associated to the long-range correlations of the field $\phi^+$ in the wire basis. This state of matter is driven here by the $t_{\perp}$ term (rather than the interaction term), and is also related to a filled band insulator when $t_{\perp}\sim 2t$. 

Now, we study the opposite limit where $t_\perp \leq \Delta$, referred to as the weak-coupling regime. The parameter $\tilde{\Delta}$ flows first to strong-coupling values in Eq. (\ref{flow}) assuming moderate repulsive interactions $(K_+,K_->1/2)$. Therefore, we have $\theta_1\sim 0$ and its dual mode $\phi_1$ becomes fast oscillating inside $\cos(\sqrt{2}\phi^+) = \cos(\phi_1 + \phi_2)$. $\mathcal{H}_\perp$ is then irrelevant to the first order. To the second order in perturbation, however,  we still find an effective term reminiscent of an Andreev process between wires where a Cooper pair is transported from wire 1 to wire 2, then triggering a superconducting gap in the free fermion wire  (see Fig.~\ref{fig:hpi} and more details in Appendix~\ref{app:pert})
    \begin{align}
     \mathcal{H}_\perp^{(2)} 
        &= -\frac{2t^2_\perp}{ \pi a \Delta} \int dx  \cos(2\sqrt{2}\theta^- - 2\chi x) \notag  \\
     &=  -\frac{2t^2_\perp}{ \pi a \Delta} \int dx  \cos(2\theta_2). \label{eq:2nd}
    \end{align}
 In the last equality, we regard $2\chi x$ as multiples of $2\pi$ and $\theta_1$ is pinned to zero for large $\Delta$. Similar to the $0$-flux case, the induced gap takes the form $\bar{\Delta}_2 = t_\perp^2/\Delta$. To minimize classically the energy in $\mathcal{H}_\perp^{(2)}$, we obtain the pinning condition
    \begin{gather}
      \langle \cos(2\sqrt{2} \theta^- - 2\chi x) \rangle = 1. \label{eq:ccf}
   \end{gather} 
 This phase is referred to as the p-wave SC phase in the phase diagram of Fig. \ref{fig:hladder}.  Here, the phase reveals 4 MZM until the occurrence of the CDW order transition.  Together, the conditions $k_F^1+k_F^2=\pi/a$ and $\chi=\pi$ imply that the chemical potential will take a value where the 4 MZM phase occurs.  At the transition between the CDW and the 4 MZM topological phase, the modes $\phi^+$ and $\theta^+$ should become gapless, since in the CDW phase the system tends to favor the pinning of the mode $\phi^+$ and in the superconducting phase, both $\theta_1$ and $\theta_2$ are pinned, implying consequently that both $\theta^-$ and $\theta^+$ are pinned. Hence, a Luttinger liquid is expected in the vicinity of the transition line (dashed line in Fig. \ref{fig:hladder}) separating the CDW and the 4 MZM phases.
 
 Below, we study the Meissner and Majorana currents that originate when $a\chi$ deviates slightly from $\pm \pi$. It is interesting to comment that the argument below would also be applicable for the 4 MZM topological superconducting phase close to the zero flux situation. We define and measure the Meissner-Majorana current at the edge by a small disturbance in fluxes $\Delta \chi$, as in Fig. \ref{fig:hpi}. From Eq. (\ref{eq:ccf}), we obtain 
  \begin{gather}
    \widetilde{\chi} = \chi + \Delta \chi, \quad  \langle \nabla \theta^-(x)\rangle = \Delta \chi/\sqrt{2}. \label{theta}
  \end{gather}
It is relevant to observe that $\Delta\chi$ acting on the $\theta^-$ mode in Eq. (\ref{eq:h0pm}) plays a similar role as a chemical potential $-\tilde{\mu}\nabla\theta^-(x)$ on a band insulator with $\tilde{\mu}=\Delta\chi v^- K^-/(\sqrt{2}\pi)$. Therefore, the topological proximity effect takes place as long as $|\tilde{\mu}|<\bar{\Delta}_2$. For $|\tilde{\mu}|=\bar{\Delta}_2$, by analogy to the commensurate-incommensurate transition \cite{HeinzMott}, the pinning of $\theta^-$ should be suppressed, and the system effectively behaves as if there is a chain of free fermions and a topological superconducting wire with 2MZM in Fig. \ref{fig:hladder}. Next, we develop a linear-response analysis in $\Delta\chi$ for the ladder system along the lines of Ref. \cite{Nagaosaladder}.  In particular, we address the Majorana particle current at the boundaries of the system with 4 MZM, building an analogy with the topological Josephson junction \cite{fukane,Rokhinson}. 
 
   \subsection{Superconducting response close to $\pi$-flux}
   \label{MMSC}

To evaluate the Meissner and Majorana currents, we employ the Heisenberg equation of motion for the density operator $d(n_1-n_2)/dt = i[{\cal H}_{\text{hybrid}},n_1-n_2]$ with $\hbar=1$ and with $n_1(x)-n_2(x) =- \sqrt{2}\partial_x\phi_-/\pi$ (modulo a global background charge). Integrating the continuity equation $\int dx [\partial (n_1-n_2)/\partial t +\mathbb{\nabla}.{\bf j}(x)]=0$, then we identify the parallel (intrawire) current (density) $j_{\parallel}$ and the interwire (vertical) current $j_{\perp}$ associated respectively to $\mathcal{H}_0^-$ and $\mathcal{H}_{\perp}$. 

\begin{figure}[t]
   \begin{center}
       \includegraphics[width=1\linewidth]{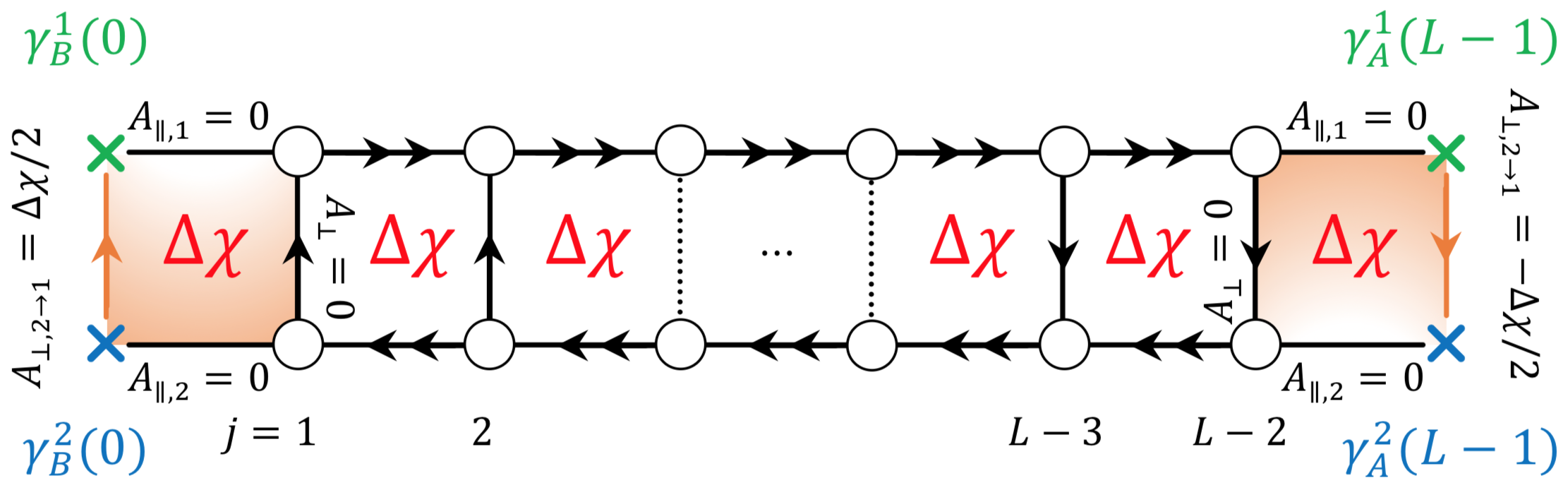}
    \end{center}
  \vskip -0.5cm \protect\caption[]
  {(color online) Weakly coupled hybrid wires in the presence of a uniform magnetic field with $\Phi_{\text{tot}} = \chi a = \pm \pi$. The condition $k_{F}^1+k_{F}^2\mp\chi=0$ allows us to make Andreev terms flowing to strong couplings close to $\Phi_{\text{tot}}=\mp \pi$, similarly as the situation at zero net flux. A Meissner current formed by the Cooper pairs and proportional to $\Delta\chi$ is formed within each plaquette under a small flux deviation from $\Phi_{\text{tot}} = \chi a = \pm \pi$, and the bulk transverse current effectively averages to zero as described through the vertical dashed lines. The induced parallel flow screens the effect of the perturbation $\Delta \chi$. The conservation of the Meissner current close to the boundaries is ensured through Andreev processes shown at rungs $(1,2)$ and $(L-3,L-2)$ which have flown to the strong-coupling limit. In addition, at the sites $0$ and $L-1$, similar to the Josephson junction with Majorana fermions \cite{fukane,Rokhinson,alicea}, the presence of gapless Majorana modes allows for a (perturbative) current proportional to $t_{\perp}$.  We then find a Majorana current in orange satisfying the property $\langle j_{M,2\rightarrow 1}(0)\rangle=-\langle j_{M,2\rightarrow 1}(L-1)\rangle$, where $2$ and $1$ refer to the wire indices. The choice of vector potentials at the first and last rungs are fixed such that the Peierls phase for an electron or Majorana fermion is half of that of a Cooper pair.}
    \label{fig:hpi}
    \vskip -0.5cm
\end{figure}

The perpendicular current associated with a unit charge $q=1$ takes the form:
     \begin{align}
       j_\perp (x) = -\frac{4t_\perp}{\pi} \sin (\sqrt{2}\theta^-) \cos(\sqrt{2}\phi^+).
       \end{align}
If we apply the pinning condition found above $\langle\theta^-(x)\rangle=\Delta\chi x/\sqrt{2}$, we observe that the vertical current which was formally zero when $\Delta\chi=0$ now gives an oscillatory response with the space variable $x$ in the continuum limit.  We check that $\int dx \langle j_{\perp}(x) \rangle = 0$, as an indication that circulating Meissner currents still take place if we assume a small deviation from $\pi$ flux. Below, we assume that the left and right boundaries of the sample are located at $x=0$ and $x=L-1$ as in Fig. \ref{fig:hpi}.  Since we have fixed $a'=a=1$, the variable $L-1$ here will also denote the position of the last site. 

To evaluate the parallel current from Eq. (\ref{eq:h0pm}), it is important to mention that the charge operator has been defined as $n_1(x)-n_2(x) =- \sqrt{2}\partial_x\phi_-/\pi$ for a unit charge $q=1$. To define the current associated to Cooper pairs, we shall then multiply this operator by $q=2$, which then results in:
\begin{align}
 j_\parallel (x) = -\frac{2\sqrt{2} vK}{\pi}  \nabla \theta^-(x).
\end{align}
In the bulk, a Meissner current with charge $q=2$ is then formed. From Eq. (\ref{theta}), we obtain:
  \begin{align}
     \langle j_\parallel (x) \rangle = -\frac{2\sqrt{2} vK}{\pi} \langle \nabla\theta^-(x) \rangle  =  -\frac{2vK\Delta \chi}{\pi}, \label{currentM}
  \end{align}
which screens the effect of the magnetic flux variation; see Fig.~\ref{fig:hpi}. The continuity of the bulk Meissner current $\langle j_{\parallel}\rangle=-2vK\Delta\chi/\pi$ close to the two boundaries in the vertical direction (at sites $j=1,2$ and sites $j=L-3,L-2$ formally) is ensured by the Andreev processes, which have flowed to the strong-coupling limit.  Once $\Delta \chi a \ne 0$, in addition to the normal Meissner current formed by the bulk Cooper pairs, the hopping term at the two edges also induces an edge Majorana fermion current. The Majorana responses at sites $0$ and $L-1$ will be proportional to the perturbation $t_{\perp}$.

Based on the analysis of Sec.~\ref{IIIB},  we introduce the four MZM operators $\{\gamma_B^1(0), \gamma_B^2(0), \gamma_A^1(L-1), \gamma_A^2(L-1)\}$. To describe the effect of $\Delta\chi$ symmetrically on the Majorana fermions in the first and last (orange) unit cells in Fig. \ref{fig:hpi}, we define a vector potential of the form $A_{\perp}(0)=A_{\perp,2\rightarrow 1}=(\Delta\chi a)/(2a')$ at the first rung (and effectively zero vector potentials in the other segments of the first cell) associated to a unit charge $q=1$; the symbol $2\rightarrow 1$ means going from (lower) wire $2$ to (upper) wire $1$. The associated Peierls phase at $j=0$ for the Majorana fermions then takes the form $A_{\perp}(0) a'=\Delta\chi a/2$. In the last unit cell, we may define $A_{\perp}(L-1)=A_{\perp,2\rightarrow 1}=-A_{\perp}(0)$ (and effectively zero vector potentials in the other segments of the last cell). In this geometry of parallel wires, the operator $c^{\dagger}_1 c_2$ then turns into $\gamma_B^1(0)\gamma_B^2(0)/4$, which results in   
\begin{align}
  \mathcal{H}_\perp (0) &=  \frac{t_\perp}{2}\sin\left(\frac{\Delta \chi a}{2}\right) i\gamma_B^1(0) \gamma_B^2(0) \notag \\ 
  &\approx \frac{t_{\perp} a}{4}\Delta \chi  i\gamma_B^1(0) \gamma_B^2(0). \label{eq:em}
  \end{align}
When $\Delta \chi a = 0$, $\mathcal{H}_\perp (0)$ is zero, reflecting that the phase with 4-Majorana edge modes is protected against a $\pi$-flux since the hopping term $t_{\perp}e^{i\chi x_j}=t_{\perp}(-1)^j$ becomes real and since the arguments of Sec. \ref{IIIB} show that in this case $t_{\perp}$ should not hybridize the MZM. Similarly to Eq. (\ref{Imaginary}),  we check that Majorana fermions move adiabatically (linearly with $\Delta\chi$) from zero energy when deviating from $\pi$ flux per plaquette.
  
 At the left boundary of the system, the Majorana current is denoted as $\langle j_M (0)\rangle = \langle j_{M,2\rightarrow 1}(0)\rangle$ and is defined as $\langle j_M (0)\rangle = \partial \langle \mathcal{H}_\perp (0)\rangle/{\partial (a A_{\perp,2\rightarrow 1}(0))}=2\partial \langle \mathcal{H}_\perp (0)\rangle/\partial (a\Delta\chi)$, which results in:
  \begin{align}
      \langle j_M (0)\rangle &=   ({t_\perp}/{2}) \langle  i\gamma_B^1(0) \gamma_B^2(0)\rangle. 
 \end{align} 
 In the $\pi$-flux configuration, $[i\gamma_B^1(0) \gamma_B^2(0), \mathcal{H}_{\text{hybrid}}] = 0$, the parity operator $i\gamma_B^1(0) \gamma_B^2(0)$ associated to these two MZM is equally likely to take an expectation value $\pm 1$. Then, 
 $\langle i\gamma_B^1(0) \gamma_B^2(0) \rangle = 0$ on average and accordingly $\langle j_M (0)\rangle$. As soon as one switches on $\Delta \chi a = 0^+$, to minimize the energy, the parity operator $\langle i\gamma_B^1(0) \gamma_B^2(0) \rangle$ is locked to  $-1$, producing a
 response $\langle j_M (0)\rangle =  ({t_\perp}/{2})$. Here, the Majorana current satisfies $\langle j_M(0)\rangle = -\langle j_M(L-1)\rangle$ where $j_{M}(L-1) = j_{M,2\rightarrow 1}$. This relation can be derived from the boundary term
  \begin{gather}
     \mathcal{H}_\perp (L-1) \approx  -\frac{t_\perp a}{4} \Delta \chi i\gamma_A^1(L-1) \gamma_A^2(L-1). \label{HMajor}
  \end{gather}
 By symmetry between the two wires' boundaries, we have the relations $\gamma_A^1(L-1)\leftrightarrow \gamma_B^2(0)$ and $\gamma_A^2(L-1)\leftrightarrow \gamma_B^1(0)$, ensuring
 that $\langle \mathcal{H}_{\perp}(0) \rangle = \langle \mathcal{H}_{\perp}(L-1) \rangle$. The identification $\langle j_M(0)\rangle =-\langle j_M(L-1)\rangle$ comes from the vector potential $A_{\perp,2\rightarrow 1}(L-1)=-A_{\perp,2\rightarrow 1}(0)=-(\Delta\chi a)/2a'$, as shown in Fig. \ref{fig:hpi}. When writing Eq. (\ref{HMajor}), we have assumed that $t_{\perp}(-1)^j$ takes the same values at $j=0$ and $j=L-1$ implying that $L-1$ is even; an odd value of $L-1$ would lead to the same physical result and would just flip the value of $\langle i\gamma_A^2(L-1) \gamma_A^1(L-1) \rangle$ at the boundary to satisfy $\langle \mathcal{H}_{\perp}(0) \rangle = \langle \mathcal{H}_{\perp}(L-1) \rangle$. This shows that the physics remains the same independently of the relative parity between the two pairs of Majorana fermions at the boundaries. 
  
 If we change $\Delta\chi \rightarrow -\Delta \chi$, all the currents should flip their signs. The energy conservation of $\langle {\cal H}_{\perp} \rangle$ is related to a flip of the parity operator formed by the two Majorana fermions at one boundary. There is then a jump of size $t_{\perp}$ associated with the edge MZM current when changing the sign of $\Delta\chi$ close to $\pi$-flux per plaquette. This situation is therefore very similar to a Josephson junction with a resonant level which can be realized with a double-dot charge qubit and which shows a similar `jump' in the superconducting Josephson response for a $\pi$ phase shift \cite{JJ}. The resonant level here is formed at $\pi$ flux per plaquette through the two values of the expectation value of the parity operator $ i\gamma_B^1(0) \gamma_B^2(0) =\pm 1$ at the boundary.  Fig.~\ref{fig:hpi} illustrates the formation of the composite Meissner-Majorana current. It enables us to detect the 4-Majorana edge mode through the measurement of the vertical current at the two boundaries. Physically, to measure the Majorana current, one can resort to a setting similar to the one suggested in Ref. \cite{teleLF}, where an analogy between Majorana fermions and a resonant level was also addressed.

%------------------------------------------------------------------------------------------------------------
\subsection{Abelian Quantum Hall phases}
\label{edge}

We can also consider, in general, any configuration allowed by Eq.~(\ref{eq:fc}):
  \begin{gather}
    a\left[ (k_F^1 + k_F^2)  - m  \chi  \right] = 0\mod 2\pi, \label{eq:gfa}
  \end{gather}
with $m$ odd and $\chi a \ne \pm  \pi$. Adjusting the densities in the two wires, one can then reach quantum Hall plateaux for specific values of the magnetic field.  As before, one can safely drop out fast oscillating terms, such as  $\propto \int dx \cos (\cdots \mp 2\chi x)$ in $\mathcal{H}_\perp$ (\ref{eq:hp}), and reach for $m=1$
   \begin{gather}
    \mathcal{H}_{\perp} = -\frac{t_{\perp}}{\pi a} \int dx \cos [\sqrt{2} (\theta^- - m\phi^+)]. \label{eq:hp1}
   \end{gather}
 In the strong-coupling limit, $t_{\perp}\gg \Delta$, the form of the interwire hopping term $\mathcal{H}_{\perp}$ (\ref{eq:hp1}) satisfies the classification of a $\nu = 1/m$ Laughlin state (with $m$ odd) and integer quantum Hall effect when $\nu=1$ \cite{kane2002, kane2014}. We show below that in that case the intrawire pairing term $\Delta$ flows to zero according to renormalization group arguments.  In the case of $m=1$, the quantum Hall phase can be achieved for free fermions, as experimentally confirmed in ultra-cold atoms \cite{QHEladder1}, and we check that the hopping term is relevant in that case for $K=1$ under the renormalization group procedure. Long-range Coulomb forces resulting in $K\ll 1/2$ can also stabilize the fractional quantum Hall state  at $\nu=1/3$ by making the tunneling term in Eq. (\ref{eq:hp1}) relevant for $m=3$, as numerically observed \cite{CornfeldSela,Mazza}. In that case, one must include higher harmonics contributions to the definition of the fermion operator in Eq. (\ref{eq:b1}) to obtain Eq. (\ref{eq:hp1}) for the same density-flux constraint as in Eq. (\ref{eq:gfa}) \cite{CornfeldSela}. 
 
 The relevance of the hopping term $t_{\perp}$ here pins the mode $(\theta^- - m\phi^+)$ to zero, which can be interpreted as a (bulk) gapped mode by analogy to a two-dimensional system. A ladder system is described by four bosonic fields, which implies that two modes are still gapless. The latter describe the two chiral edge modes of the ladder system. To properly describe the system and the edge states, it is convenient to introduce four chiral fields \cite{alex2015,alex2017}
  \begin{gather}
    \phi_r^\alpha = \frac{\theta_\alpha}{m} + r\phi_\alpha, \label{eq:cm0}
  \end{gather}
  with $\alpha = 1,2$ representing the wire index and now $r = + 1/ -1$ denoting the left/right moving particles in agreement with Eq. (\ref{eq:b1}). The chiral bosonic fields satisfy the commutation relation
  \begin{gather}
    [ \phi_r^\alpha (x), \phi_p^\beta (x') ] = i r\frac{\pi}{m} \delta_{rp}\delta_{\alpha \beta} \text{Sign} (x' - x).
  \end{gather}
New modes can be constructed from $\phi_r^\alpha$ which capture the properties of  the gapped bulk states ($\theta, \phi$) 
  \begin{gather}
    \begin{cases}
       \phi = (-\phi_{-1}^1 + \phi_{+1}^2)/2, \\
       \theta = (\phi_{-1}^1 + \phi_{+1}^2)/2, 
    \end{cases}\label{eq:cm}
  \end{gather}
and gapless edge states ($\theta', \phi'$)  
  \begin{gather}
      \begin{cases}
        \phi' = (-\phi_{-1}^2 + \phi_{+1}^1)/2, \\
        \theta' = (\phi_{-1}^2 + \phi_{+1}^1)/2.  \label{eq:cm1}
      \end{cases}
   \end{gather}
 The commutation relation reads: 
   \begin{gather}
     [\phi(x), m\theta(x') ] = i(\pi/2) \text{Sign} (x'-x),
   \end{gather}
 and the same for the gapless modes  ($\phi', \theta'$). The bulk gapped mode is now related to the field $\phi$. When $t_\perp \gg \Delta$, correspondingly, the bulk mode $\phi$ is pinned to a classical value since 
   \begin{gather}
     \mathcal{H}_{\perp}  = -\frac{t_{\perp}}{\pi a} \int dx \cos(2m\phi). \label{eq:hpt}
    \end{gather}
 A gap is opened in the bulk as previously shown in Fig.~\ref{fig:gladder} (bottom right). In Appendix \ref{app:et} and Eq. (\ref{B11}), we check that the two gapless modes which produce a chiral Luttinger theory are $L(x)=\phi_{+1}^1(x)=\theta'(x)+\phi'(x)$ and $R(x)=\phi_{-1}^2(x)=\theta'(x)-\phi'(x)$. Starting from free fermions and $K=1$, the edge Luttinger parameters are $K^e = 1$, $v^e = v$. We thus realize an Abelian quantum Hall phase at filling factor $\nu  = 1/ m = 1$ in the presence of an arbitrary uniform flux. If we change the direction of the magnetic field, which is equivalent to change $m\rightarrow -m$ in Eq. (\ref{eq:gfa}), we obtain similar results by inverting the left and right-moving particles' definitions in Eq. (\ref{eq:cm0}). 
  
 The two gapless modes are, furthermore, protected against a small intrawire pairing. Indeed, we can rewrite $\mathcal{H}_\Delta$ as
  \begin{align}
     \mathcal{H}_\Delta = -\frac{\Delta}{\pi a} \int dx e^{i[2m\theta + \sqrt{2}(\theta^- + m\phi^-)]} + \text{H.c.}.
  \end{align}
Since $\phi$ is pinned to a classical value minimizing the cosine potential term, then the dual mode $\theta$ oscillates rapidly: $\langle e^{i\theta(x)} e^{-i\theta(0)} \rangle \propto e^{-x/\xi}$ with a correlation length $\xi$ proportional to $1/t_\perp$.  Therefore, $\mathcal{H}_{\Delta}$ flows to zero for large $t_\perp$. For the observables, the integer Laughlin state is revealed in the edge current. A direct calculation of the parallel and perpendicular currents defined in Sec.~\ref{MMSC} for a unit charge $q=1$ leads to
  \begin{align}
     \langle j_\perp (x) \rangle &= -\frac{2t_\perp}{\pi} \langle \sin (2m\phi) \rangle = 0, \notag \\
    \langle j_\parallel (x) \rangle &=  -\frac{\sqrt{2} vK}{\pi}  \langle \nabla \theta^- (x) \rangle.
  \end{align}
 Choosing an alternative gauge for the magnetic vector potential
   \begin{gather}
     A_\perp = 0, \qquad \oint \vec{A} \cdot d\vec{l} = \left(A_\parallel^1 - A_\parallel^2 \right)a = \chi a,
   \end{gather}
 we rewrite the quadratic contribution to the action for the $-$ bosonic field, as
  \begin{gather}
    S [\phi^-, \theta^-] = \frac{v}{2\pi} \int dx d\tau \left[ \frac{1}{K} \left( \nabla \phi^-\right)^2 + K \left( \nabla \theta^- + A_\parallel^- \right)^2\right], \label{eq:ac}
  \end{gather}
 where $A_\parallel^- = (A_\parallel^1 - A_\parallel^2)/\sqrt{2}$.
To extremize the action $\partial S/ \partial \theta^- = 0$, it requires 
    $\langle \nabla \theta^- (x) \rangle = -A_\parallel^-$.
The edge current of the integer Laughlin state for a unit charge $q=1$ and for free electrons $K=1$ hence becomes
  \begin{gather}
    \langle j_\parallel (x) \rangle = \frac{vK\chi}{\pi}. \label{current}
  \end{gather}
It follows the same direction as the vector potential produced by the magnetic flux (shown in Fig.~\ref{fig:har}).  
   
 In the presence of Coulomb interactions, as shown in Appendix \ref{app:et}, backscattering effects occur for $K\neq 1$ when integrating out the bulk (gapped) mode on the two-leg ladder system, and as a result the charge at the edges will be adiabatically deformed  \cite{halperin2008,alex2015,alex2017}. In contrast, the bulk polarization in a Thouless pump geometry \cite{thouless1983} is stable under Coulomb interactions as shown in Appendix~\ref{app:tp}, reflecting the Zak phase, and measuring $\nu$. Similar arguments apply to the Laughlin phase $\nu=1/3$ stabilized with long-range Coulomb interactions \cite{CornfeldSela,Mazza}. The charge at the edges in these ladder systems could be measured accurately \cite{halperin2008,alex2017,Glattli,Ramsey,kim,Ion}.
      
 \begin{figure}[t]
   \begin{center}
     \includegraphics[height=1.5cm]{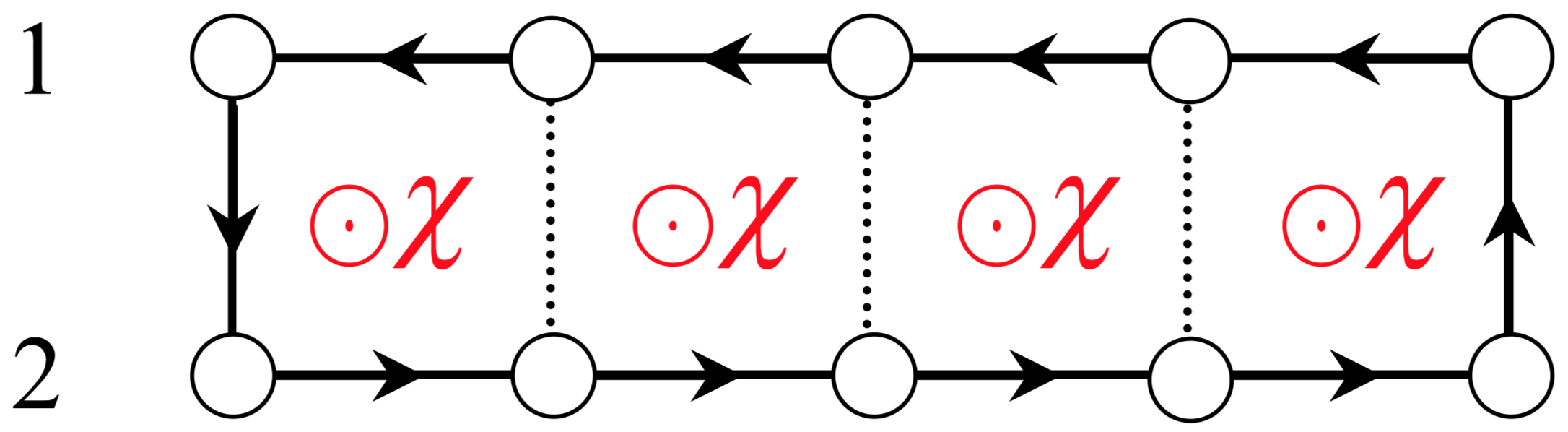} 
    \end{center}
  \vskip -0.5cm \protect\caption[]
  {(color online) Edge current for a unit charge $q=1$ in the Abelian quantum Hall state ($\nu = 1$) when $t_\perp \gg \Delta$. An arbitrary uniform magnetic flux is applied on each plaquette.}
    \label{fig:har}
    \vskip -0.5cm
\end{figure}
  
The quantum Hall phase at $\nu=1$ occurs when $t_{\perp}\geq \Delta$.  In the weak-coupling limit $t_{\perp}\leq \Delta$, for an arbitrary flux, both the first order (\ref{eq:hp1}) and the second order (\ref{eq:2nd}) contributions of $t_\perp$ are irrelevant from renormalization group arguments since the superfluid mode $\theta_1$ is now pinned. In Appendix~\ref{app:pert}, we find that the most relevant term is
  \begin{align}
    &\mathcal{H}_\perp^{(4)} = - \frac{t_\perp^4}{3\pi a \Delta^3} \int dx \cos [ 2 (\theta_2(x+2a) - \theta_2(x)) + 4\chi a],  \notag \\
                                          &\propto e^{4i\chi a}c_2^\dagger(x-a) c_2^\dagger(x)c_2(x+a)c_2(x+2a) + H.c.. \label{eq:4th}
  \end{align}
  The constraint on the phases
  \begin{gather}
    \theta_2 (x+2a) - \theta_2(x) = - 2\chi a,
  \end{gather}
helps to form a local current formed by the Cooper pairs within three adjacent plaquettes and using Eq. (\ref{currentM}), this leads to
   \begin{align}
        \langle j_\parallel (x) \rangle_{\text{plaquette}} = - \frac{2vK\chi}{\pi}.
   \end{align}
Deviating from the $\pi$-flux substantially in the weak-coupling region, an arbitrary flux breaks the symmetry $\mathcal{T}$ and destabilizes the proximity effect (see related discussion in \ref{piflux}). Consequently, there is no induced superconducting gap in the free fermion wire. Still, two Majorana edge modes persist in the original Kitaev superconducting wire.  The protection against orbital magnetic effects in this weak-coupling region could be useful for applications in quantum computation and engineering of Majorana fermions  with magnetic fluxes \cite{Beenakker}. 

Fig.~\ref{fig:hladder} shows the phase diagrams for the cases of $\pi$ flux per plaquette at half filling and arbitrary fluxes, when adjusting the densities in the wires such that the proximity effect can effectively takes place. In the strong-coupling limit, on the other hand, one is able to distinguish the CDW and Abelian quantum Hall states from the responses of Thouless pump \cite{thouless1983}. We refer the reader to Appendix~\ref{app:tp}, where a proposal for the measurement is raised and the stability of the bulk polarization under Coulomb interactions is also discussed.

%------------------------------------------------------------------------------------------------------------
\subsection{$\nu=1/2$ Quantum Hall phase for a spin-1/2 ladder}
\label{sec:spinful}

\begin{figure}[t]
   \begin{center}
     \includegraphics[height=2.5cm]{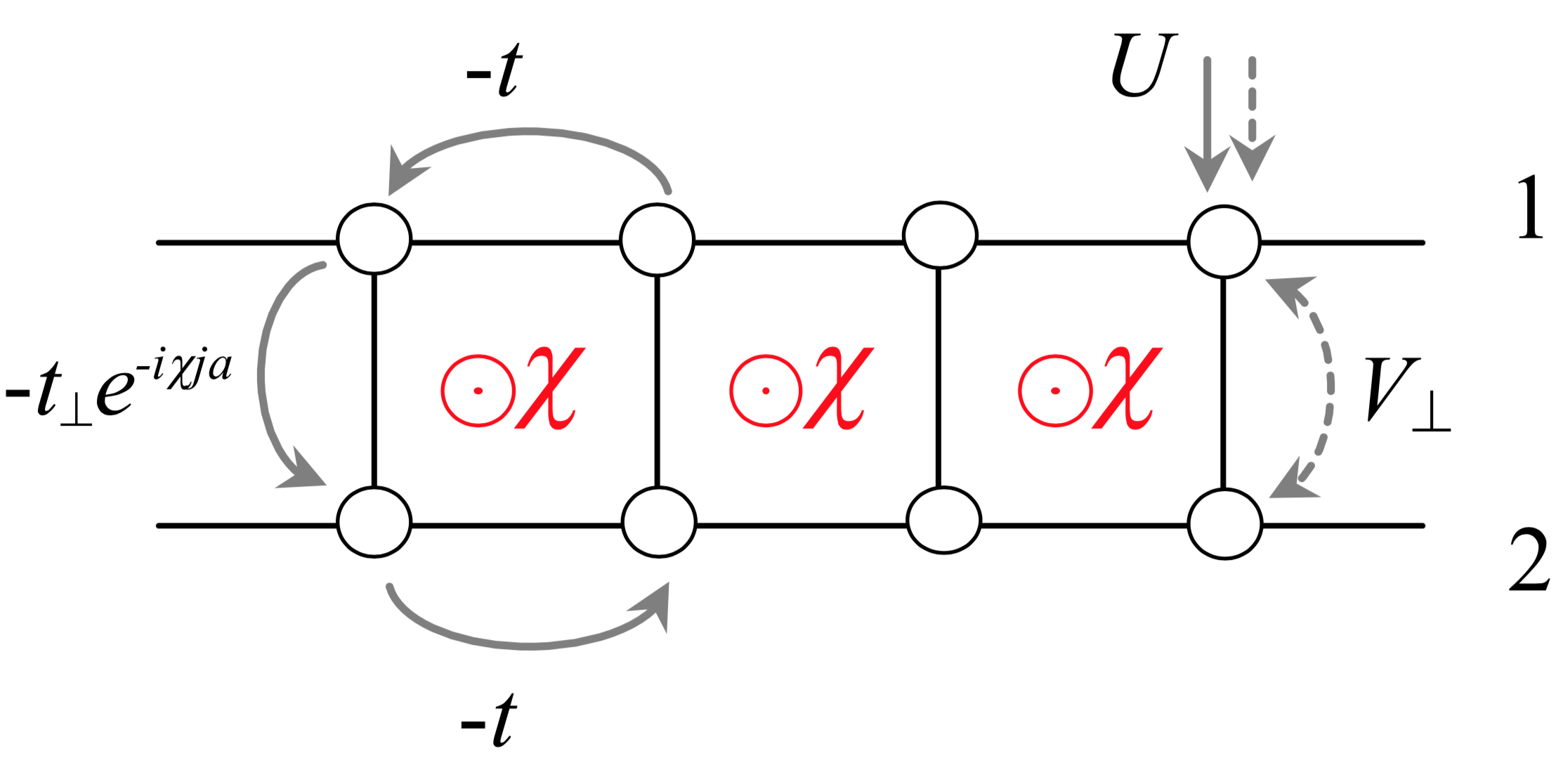}  \includegraphics[height=2.5cm]{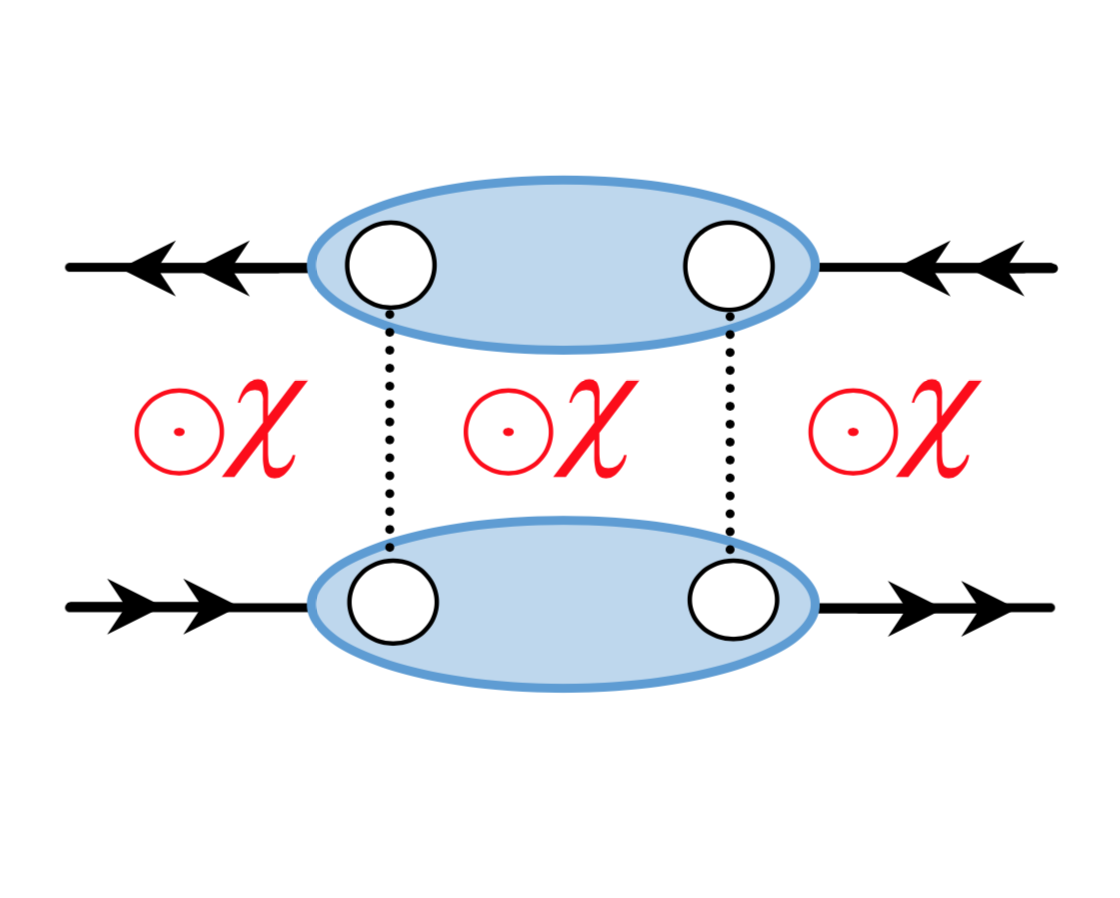} 
    \end{center}
  \vskip -0.5cm \protect\caption[]
  {(color online) (Left) Weakly coupled spinful (spin-1/2) wires with a flux configuration $\Phi_{\text{tot}} = \chi a = (k_F^1+ k_F^2)a$. The dashed arrows represent the interactions between fermions with different spins; (Right) Formation of Cooper pair Laughlin state at $\nu =  1/2$, where the chiral edge currents move in opposite direction compared to those in the Meissner effect.}
    \label{fig:sladder}
    \vskip -0.5cm
\end{figure}

With arbitrary fluxes,  another interesting comparison can be made with the case of spinful or spin-1/2 fermions.  Two of us have previously shown in Ref.~\cite{alex2015}, for a spin-1/2 system, that the attractive Hubbard interaction and the long-range repulsive interwire interaction help stabilize a ladder generalization of a Cooper pair Laughlin state at $\nu = 1/2$. The attractive channel allows us to realize a Luther-Emery model \cite{LutherEmery} with Cooper pairs or bosons, and the long-range repulsive interaction then realizes the hard-core boson regime.  An essential difference with the spinless situation is that here we take into account the spin-charge separation phenomenon of the Luttinger liquid. Therefore, as for a Luther-Emery liquid, the Cooper pairs are realized through a gap in the spin channel in the two wires. The two charge fields are still free in the wires, allowing more tunability to realize a Laughlin state at $\nu=1/2$ of hard-core bosons from the charge sector.  The filling factor $\nu=1/2$ reflecting the charges of quasiparticles in the bulk can be accessed through Thouless pump measurements. The main difference with the Meissner effect is that the edge current now flows in opposite direction, as shown in Fig. \ref{fig:sladder} compared to Fig. \ref{fig:hpi}. Below, we first briefly review the main results and later in Sec.~\ref{sec:det} we will extend the building block to coupled-ladder systems referring then to the fourth case of study.

The key ingredients of the spinful ladder can be found in Fig.~\ref{fig:sladder} (left). Four parts enter into the Hamiltonian
  \begin{gather}
    \mathcal{H}_{\text{spinful}} = \mathcal{H}_{\parallel} + \mathcal{H}_\perp + \mathcal{U} +\mathcal{V}.
  \end{gather}
The intra- and interwire hopping terms $\mathcal{H}_{\parallel}$ and $\mathcal{H}_\perp$ keep the form of the spinless case with the addition of the spin flavor $s = \uparrow, \downarrow$: $c_{\alpha}(j) \to c_{\alpha, s}(j)$. 
The flux configuration also satisfies $\zeta = 0$ and the constraint (\ref{eq:gfa}) for an arbitrary value of $\chi$. For the interactions, we take into account the Hubbard and interwire Coulomb repulsion:
  \begin{align}
    \mathcal{U} &= \sum_j \sum_{\alpha=1,2} U_\alpha n_{\alpha, \uparrow} (j) n_{\alpha, \downarrow} (j), \notag \\
    \mathcal{V} &= V_\perp \sum_j \sum_{s,s' = \uparrow, \downarrow}  n_{1,s}(j) n_{2,s'} (j),
  \end{align}
where $n_{\alpha, s}(j) = c_{\alpha,s}^\dagger (j)c_{\alpha,s} (j)$ denotes the particle number operator. For simplicity, we consider two identical wires sharing the same attractive Hubbard interaction $U = U_1 = U_2 < 0$ and a repulsive long-range interaction $V_\perp > 0$ is added between the wires. We consider the weak-coupling regime $t_\perp \ll (|U|, |V_\perp|)$, where $\mathcal{H}_\perp$ acts as a small perturbation. 

 Applying bosonization techniques \cite{Haldane,thierry2004}, we introduce the spin ($\sigma$) and  charge ($\rho$)  degrees of freedom
  through the transformation
   \begin{align}
     \phi_\sigma (x) &= \frac{1}{\sqrt{2}} (\phi_\uparrow (x) - \phi_\downarrow (x)),\notag \\
     \phi_\rho (x) &= \frac{1}{\sqrt{2}} (\phi_\uparrow (x) + \phi_\downarrow (x)),
    \end{align}
 and the same for the dual modes $\theta_{\sigma, \rho}$. The Hamiltonian then takes the form:
   \begin{align}
    \mathcal{H}_{\text{spinful}} &= \mathcal{H}_\sigma + \mathcal{H}_\rho + \mathcal{H}_\perp, \notag \\
    \mathcal{H}_{\rho} & = \mathcal{H}_{0,\rho}^+ + \mathcal{H}_{0,\rho}^-, \notag \\
    \mathcal{H}_{\sigma} &= \sum_{\alpha = 1,2} \mathcal{H}_{0,\sigma}^\alpha  + \frac{U}{2\pi^2 a}\int dx \cos(2\sqrt{2} \phi_\sigma^\alpha), \notag \\
     \mathcal{H}_\perp & = -\frac{t_\perp}{2\pi a} \int dx \sum_{s = \pm 1} \sum_{r,r' = \pm 1} e^{i\chi a} e^{i r' (k_F^1+rk_F^2 )x} \notag \\
                                   & \times   e^{-ir'(\phi_\rho^r + r' \theta_\rho^- + s(\phi_\sigma^r + r'\theta_\sigma^-))} + \text{H.c.}.
   \end{align}
 Here, $\mathcal{H}_{0,\rho}^{\pm}$ and $\mathcal{H}_{0,\sigma}^\alpha$ with $\alpha=1,2$ take the Luttinger form. The Luttinger parameters are defined as
      $K_\sigma = K_\sigma^1 = K_\sigma^2= (1+u)^{-1/2}$, 
      $K_\rho^\pm = (1-u \pm v_\perp)^{-1/2}$, where $u = |U|a/(\pi v)$, $v_{\perp} = 2V_\perp a/(\pi v)$.
 In the spin part, the attractive Hubbard interaction $\mathcal{U}$ plays the same role as the pairing term $\mathcal{H}_{\Delta}$ for the spinless particles, resulting in the pinning conditions for the spin modes (modulo $\pi/\sqrt{2}$):
   \begin{align}
     \phi_\sigma   = \phi_\sigma^1  =  \phi_\sigma^2  \sim 0.
   \end{align}
The Cooper pairs formed along each wire are protected by a spin gap $\Delta_\sigma$ which grows exponentially fast at small $U$ and then $\Delta_\sigma \propto |U|$ for strong interactions. 

Below, we assume the flux condition $2(k_{F}^1+k_F^2) \pm 2\chi=0$ following Ref. \cite{alex2015}.  For the energy scales smaller than $\Delta_\sigma$,  the dual mode $\theta_\sigma^- = (\theta_\sigma^1 - \theta_\sigma^2)/\sqrt{2}$, meanwhile, oscillates rapidly in $\mathcal{H}_\perp$ making the coupling term $t_{\perp}$ irrelevant to first order in perturbation theory. The second order term gives the charge contribution
   \begin{gather}
     \mathcal{H}_\perp^{(2)} = -\frac{t_\perp^2}{\Delta_\sigma \pi a} \int dx \cos[2(\theta_\rho^- \pm \phi_\rho^+)].
   \end{gather}
To study the motion of Cooper pairs,  we notice that the corresponding creation operator satisfies $(\psi_{-r,\uparrow}^\alpha \psi_{r,\downarrow}^\alpha)^\dagger \sim e^{-ir\sqrt{2}\theta_\rho^\alpha}$. This implies a canonical transformation for Cooper pairs on each wire:
   \begin{gather}
     \Theta_\rho^\alpha = \sqrt{2} \theta_\rho^\alpha, \quad \Phi_\rho^\alpha = \phi_\rho^\alpha /\sqrt{2}, \notag \\
      \mathcal{H}_\perp^{(2)} = -\frac{t_\perp^2}{\Delta_\sigma \pi a} \int dx \cos[\sqrt{2}(\Theta_\rho^- \pm 2\Phi_\rho^+)]. \label{eq:sg}
   \end{gather}
By analogy with spinless fermions [see Eq. (\ref{eq:hp1})], a fractional Cooper pair Laughlin state is observed at $\nu = 1/(2m) = \pm 1/2$ on the spinful ladder, as shown in Fig.~\ref{fig:sladder} (right). 
 
In the end, we comment that a relatively strong long-range repulsive interwire interaction $V_{\perp}$ plays a vital role in driving the system towards a fractional quantum Hall state \cite{alex2015}. This  will be studied in more detail in Sec. \ref{sec:det}.

%------------------------------------------------------------------------------------------------------------
\section{Topological phases in coupled ladders}
\label{sec:cp}

In this section, we present two proposals to realize two-dimensional topological phases using multiple coupled ladders (corresponding to the third and fourth cases of Sec. \ref{cases}). 
First, we design a $p+ip$ superconductor \cite{ReadGreen} starting from the bonding-antibonding band representation of a pair of wires --- ``ladder geometry'' --- and generalizing the analysis to coupled ladder systems. Here, a pair of wires with appropriate Peierls phases $\zeta$ and  $\chi$ will generate an $ip_x$ superconducting channel, and coupling weakly the ladders together will provide the $p_y$ channel. If we would consider instead only strongly coupled wires or symmetrically coupled wires, then the projection onto the lowest band which is assumed below would result in long-range pairing terms in the direction perpendicular to the wires, through the tunneling term $t_{\perp}$ and a generalization of Eq. (\ref{Delta}).  Other proposals coupling quantum wires have suggested the possible engineering of $ip$ channels through bath or reservoir engineering \cite{kane2017,Asahi,Seroussi,Bardyn} by analogy to the two-dimensional case \cite{ReadGreen}. It is also important to mention that for spins-1/2 fermions, similar ladder constructions allow us to reproduce d-wave superconductivity, D-Mott and pseudogap physics, relevant to cuprate superconductors \cite{Lin, Ladders, Review}. Second, we address the case of coupled spinful or spin-1/2 ladder systems where, as obtained in Sec.~\ref{sec:spinful}, each ladder hosts a Cooper pair Laughlin state at filling $\nu=1/2$. In the presence of a uniform magnetic field and long-range repulsive interactions, we show how the coupled wire system can form a single bulk fluid with chiral modes at the edges now carrying a charge in agreement with the two-dimensional bulk-edge correspondence \cite{fisherstone}. 

 %-----------------------------------------------------------------------------------------------------------
\subsection{$p + ip$ superconductor from spinless fermions}
\label{sec:p+ip}

\begin{figure}[t]
   \begin{center}
     \includegraphics[width=1\linewidth]{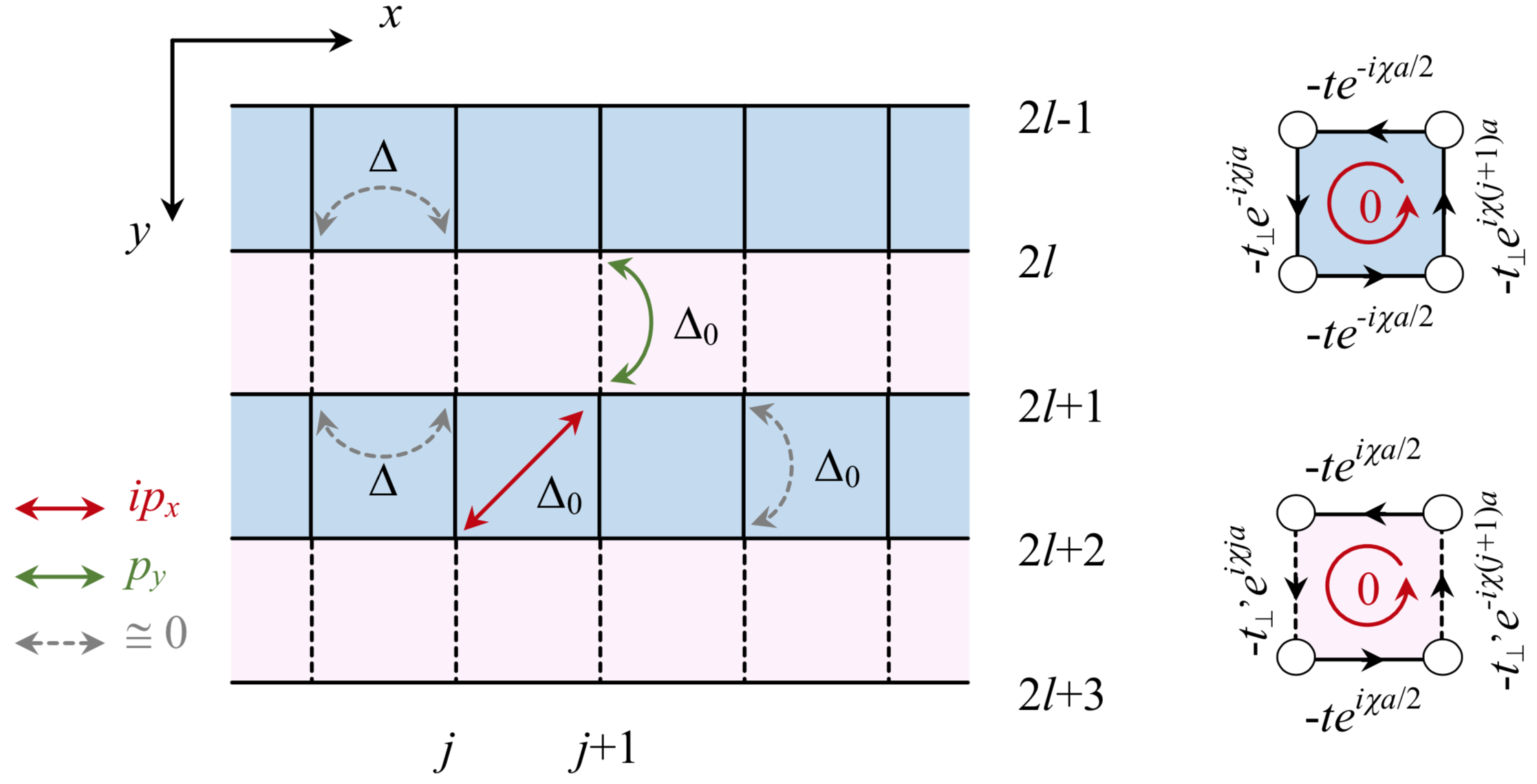}  
    \end{center}
  \vskip -0.5cm \protect\caption[]
  {(color online) (Left) Coupled ladder construction for a $p + ip$ superconductor.  A building block is formed in blue with two strongly-coupled wires. The blue ladders then couple together via the pink regions. 
  While the pairing terms denoted by grey arrows (both intra- and interwire)  are suppressed in the strong coupling limit, the pairings denoted by red and green arrows (interwire only) are responsible for the $ip_x$ and $p_y$ channels; (Right) Flux attachment in each square plaquette with $\zeta a=\chi a = \pi$. In each square unit cell, the total net flux is zero.  Peierls phases take opposite values within two successive plaquettes in $y$ direction for the same hopping process.}
    \label{fig:pxy}
    \vskip -0.5cm
\end{figure}

Here, we present our proposal to engineer the $p+ip$ superconductor in coupled ladder geometries (referring to the third case of study). Depicted in Fig.~\ref{fig:pxy} (left), our building block is a blue ladder comprising two strongly-coupled wires labeled by 
$\alpha = 2l-1, 2l$ where $l$ is an integer. The flux attachment which will be responsible for the $ip_x$ channel is shown in Fig.~\ref{fig:pxy} (right top). 

First, we re-analyze a block, say with $\alpha = 1, 2$, for the case with $\zeta=\chi$. The kinetic part of the Hamiltonian is described by $\mathcal{H}_\parallel$ and $\mathcal{H}_\perp$ in Eq.~(\ref{eq:h_0}) with
  \begin{gather}
    \zeta = \chi, \quad \Phi_{\text{total}} = 0.
  \end{gather}
  As discussed in Sec. \ref{sec:pre}, we assume here that the vector potentials related to $\zeta$ and $\chi$ are engineered, for instance, in ultra-cold atom systems.  In solid-state nanowires, this can also be realized through a space-dependent magnetic field $B_z=\chi/a' - \zeta\pi/(2a')\sin(\pi y/a')$. Physically, the two pieces related to $\chi$ and $\zeta$ correspond to the superposition of two distinct forms of magnetic fields, as emphasized in Sec. \ref{sec:pre}. Since the total flux per plaquette is zero, we can rewrite the band structure as the one of Fig. \ref{fig:gladder} (bottom left), in the absence of magnetic flux, performing a proper gauge transformation similar to Eq. (\ref{eq:gt}). The pairing terms will be modified accordingly, then providing the required physics, namely an $ip_x$ channel, after fixing the value of $\chi=\zeta=\pi/a$. Coupling weakly the ladders together, we realize a $p_y$ channel. There are two key properties to this proposal. First, the magnetic field $B_z$ will be staggered in $y$ direction implying that $\chi$ also takes opposite values in two successive plaquettes in $y$ direction with the requirement that $\Phi_{\text{tot}}=0$ in all plaquettes. Here, $\chi$ represents a periodic staggered step-like function in $y$ direction. Therefore, induced Zeeman effects at the position of the wires will be small and controllable through another magnetic field $B_x$ along $x$ direction (which is important to induce the proximitized p-wave pairing potentials \cite{fisher}). Second, the choice $\chi=\zeta=\pi/a$ leads to imaginary hopping terms $\pm it$ along the wires, which break chiral symmetry and time-reversal symmetry while preserving particle-hole symmetry. This then allows for a wire-construction of a topological phase of class D, {\it e.g.} a two-dimensional $p+ip$ superconductor \cite{bernevig2015}. The transverse hopping term of the form $t_{\perp}(-1)^j$ is real, therefore ensuring that Majorana modes can occur at zero energy. Below, we provide two  physical interpretations of the results when using equivalent low-energy theories related to Figs. \ref{fig:gladder} (bottom left and right), which implies that the results are gauge independent. 
  
Performing the gauge transformation $(\ref{eq:gt})$, 
 \begin{gather}
      c_1(j) = e^{i\chi x_j/2} \widetilde{c}_1(j), \quad
      c_2(j) = e^{-i\chi x_j/2} \widetilde{c}_2(j), \label{eq:gtt}
  \end{gather}
we can define the new basis for the bonding ($+$) and anti-bonding ($-$) fermions 
  \begin{gather}
    \widetilde{c}_{\pm} (j) = [\widetilde{c}_1(j) \pm \widetilde{c}_2(j)]/\sqrt{2}.
  \end{gather}
The two-band model in Eqs.~(\ref{eq:k_0})-(\ref{eq:dp}) is recovered with the band structure shown in Fig.~\ref{fig:gladder} (bottom left). A gap of scale $(2t_\perp)$ is opened between ``$+/-$" bands. Below, we fix the chemical potential $\mu$
such that the $+$ band becomes partially filled and the $-$ band remains empty. We then project a pair of wires, the blue ladder system, onto the lowest bonding band pair basis:
  \begin{gather}
    \mathcal{H}_{0,+} = - \sum_{k_x} [2t\cos(k_xa) + t_\perp +\mu] \widetilde{c}_+^\dagger(k_x) \widetilde{c}_+(k_x).
  \end{gather}
  
    \label{sec:pxy}
  
 In the strong coupling limit $t_\perp \gg \Delta_i$,  the phase $\zeta$ is important to suppress the intrawire pairing. 
The intrawire pairing term $\Delta_1 = \Delta_2 = \Delta$ becomes
  \begin{align}
      \mathcal{H}_\Delta = i\Delta \sum_{k_x} \sin(k_x a)\cos(\zeta/2a)\widetilde{c}_+^\dagger (k_x) \widetilde{c}_{+}^\dagger (-k_x) + \text{H.c.}. \label{HDelta}
   \end{align}
 Below, we set $\chi=\zeta=\pi a$, such that this contribution becomes zero. As mentioned above in Sec. \ref{IIIA}, the interwire contribution $\Delta_0$ involving a given rung also vanishes between nearest (N) neighbors when projecting
 onto the lowest band:
  \begin{align}
   0 = \mathcal{H}_{\Delta_0, \text{N}} &= \Delta_0 \sum_j c_1^\dagger(j) c_2^\dagger(j)+\text{H.c.} \notag \\
                                                       &= \frac{\Delta_0}{4} \sum_{k_x} \{\widetilde{c}_+^\dagger(k_x), \widetilde{c}_+^\dagger(-k_x)\} + \text{H.c.},  \end{align}
  but survives between next-nearest (NN) neighbors
  \begin{align}
   \mathcal{H}_{\Delta_0, \text{NN}} &= \Delta_0 \sum_j c_1^\dagger(j) c_2^\dagger(j+1)+\text{H.c.} \notag \\
                                                       &= \widetilde{\Delta} \sum_{k_x} i\sin(k_x a) \widetilde{c}_+^\dagger(k_x) \widetilde{c}_+^\dagger(-k_x) + \text{H.c.}, \label{Delta0}                                     
  \end{align}
 with $\widetilde{\Delta} = (\Delta_0 e^{i\chi a/2}) /2$. Adding a term $\Delta_0\sum_j c_2^\dagger(j) c_1^\dagger(j+1)+\text{H.c.}$ would result in an identical contribution in the bonding band basis.  In the equations above, we have ignored all irrelevant terms involving the $``-"$ band. 
 Once we fix the flux value
   \begin{gather}
     \chi a = \pi, \quad \widetilde{\Delta} = i \Delta_0/2,
    \end{gather}
   we engineer a purely imaginary $ip_x$ channel for the bonding fermions in each blue building block $\alpha = 2l-1, 2l$. The pair of strongly-coupled wires then effectively behaves as a topological $ip_x$ superconductor with two Majorana zero modes, one at each boundary.  The projection onto the lowest bonding band (of each ladder) is important to produce a purely $ip_x$ channel.
           
 Now, we couple the blue building blocks or ladders along the $y$ direction by the tunneling amplitude $t'_\perp$ (corresponding to the pink blocks in Fig.~\ref{fig:pxy}, left) with a reversed flux unit $\chi \to -\chi$ depicted in Fig.~\ref{fig:pxy} (right bottom).  The magnetic field  $B_z=\chi/a' - \zeta \pi/(2a') \sin(\pi y/a')$ then becomes staggered in the $y$ direction.  To realize the $ip_x+p_y$ superconductor (or $p_x\pm ip_y$ superconductor modulo gauge redefinitions), the magnetic field $B_z$ must be uniform in the $x$ direction but staggered in the $y$ direction. 
 
 Formally, this construction is correct as long as $t'_{\perp}\ll t_{\perp}$, such that the bonding-antibonding band representation of a two-coupled wire model remains valid. Through a pink region, we obtain the Hamiltonian:
    \begin{gather}
     \mathcal{H}'_{\perp} = -t'_\perp \sum_{l=1}^{M'/2} e^{-i\chi x} c^\dagger_{2l}(x) c_{2l+1} (x) + \text{H.c.}, \notag \\
     \mathcal{H}'_{\Delta_0, \text{N}} = \Delta_0 \sum_{l=1}^{M'/2} c_{2l}^\dagger(x) c_{2l+1}^\dagger (x) + \text{H.c.}
   \end{gather} 
 For the coupled wire system,  the gauge transformation (\ref{eq:gtt}) can be generalized as  
   \begin{gather}
         c_{2l-1}(j) = e^{i\chi x_j/2} \widetilde{c}_{2l-1}(j), \quad
         c_{2l}(j) = e^{-i\chi x_j/2} \widetilde{c}_{2l}(j), \label{gauge}
  \end{gather}
  which leads to
     \begin{align}
       \mathcal{H}'_\perp &= -t'_\perp \sum_{k_y} \cos(k_y a') \widetilde{c}_+^\dagger(k_y)\widetilde{c}_+(k_y), \notag \\
       \mathcal{H}'_{\Delta_0,\text{N}} &= \frac{\Delta_0}{2} \sum_{k_y} i \sin(k_y a') \widetilde{c}_+^\dagger(k_y)\widetilde{c}_+^\dagger(-k_y) + \text{H.c.}. \label{structure}
    \end{align}  
 Thus we can implement a purely real $p_y$ channel for the bonding fermions, assuming that $t_{\perp}' \ll (t_{\perp}, t)$. The relative phase between the $ip_x$ and $p_y$ channel is locked through the gauge transformation (\ref{gauge}). 
 
 Next, we define two-dimensional fermion operators $\widetilde{c}_+({\bf k})=\widetilde{c}_+(k_x,k_y)$. The goal is to study the phase diagram and edge state properties associated with the system. To that end, it is convenient to switch to the Majorana fermion representation,
   \begin{gather}
     \widetilde{c}_+({\bf k}) = \frac{1}{2} \left(\gamma_1({\bf k}) + i\gamma_2({\bf k})\right).
  \end{gather}
 Similar to Ref. \cite{kane2017}, for simplicity we set the lattice spacing $a=a'=1$ and we concentrate on the regime with low densities around ${\bf k} \simeq (0, 0)$, where we have
   \begin{align}
     \mathcal{H}_{+} &\simeq \mathcal{H}_{0,+} +  \mathcal{H}_{\Delta_0, \text{NN}} +  \mathcal{H}'_\perp + \mathcal{H}'_{\Delta_0,\text{N}} \notag \\
     &= - \frac{1}{4} \sum_{\bf k} \gamma^{T}(-{\bf k}) \mathcal{H}_{+}({\bf k})\gamma({\bf k}), 
    \end{align}
    with $\gamma^{T}(-{\bf k}) = (\gamma_1(-{\bf k}), \gamma_2(-{\bf k}))$ and
    \begin{align}     
      \mathcal{H}_{+}(k) = u k_x  \tau^z + [ \epsilon_0 + \frac{k_x^2}{2m} + T_2 \cos(k_y) ] \tau^y +  R_2 \sin(k_y) \tau^x. \label{eq:hplus}
    \end{align}
 Here $\tau^i \ (i=x,y,z)$ denote Pauli matrices and effective parameters are given by $u = \Delta_0$, $m=1/(2t)$. The three parameters locating the phase transitions read
  \begin{align}
    \epsilon_0 = -2t-t_\perp-\mu, \quad T_2 = -t'_\perp, \quad R_2 = -\Delta_0.
  \end{align}
The phase diagram of $\mathcal{H}_+$ (\ref{eq:hplus}) is carefully studied in Ref.~\cite{kane2017}, and we adjust the discussion for the Majorana zero modes at a boundary to our situation. By tuning $\mu$, $\epsilon_0$ goes from a large negative value to a large positive value (with respect to $\pm|T_2|$), and this produces two phase transitions from an anisotropic topological superconducting phase to the trivial strong paired state through the occurrence of an intermediate topological two-dimensional Moore-Read Particle-Hole phase \cite{MooreRead}. 

A large (positive) $\epsilon_0\gg |T_2|$ value hinders the occurrence of MZM from energetics point of view in the trivial strong-paired phase. The chemical potential is located below the bottom of the lowest band. When $\epsilon_0 = + t_{\perp}' = +|T_2|$, we enter into the intermediate topological phase. The energy spectrum becomes gapless in the two-dimensional sense with the lowest energy eigenvalue $-\Delta_0\sqrt{k_x^2+k_y^2}$ close to $k_x=k_y=0$. A two-dimensional gapless Majorana mode flows around the sample, as a chiral edge mode. The physics is then related to the neutral sector of a Moore-Read Particle-Hole Pfaffian phase \cite{MooreRead} with an effective $p_x-ip_y$ channel for $R_2 < 0$. By changing the sign of $\chi=\zeta$ by $-\pi$, we flip the sign of $R_2$, and the intermediate region now becomes the Pfaffian phase \cite{MooreRead}. Changing the sign of $R_2$ will change the velocity (direction) of propagation of the Majorana mode. In fact, as long as $|\epsilon_0| \leq |T_2|$ \cite{kane2017}, the system stabilizes one gapless chiral Majorana fermion. Increasing the chemical potential further, this results in (very) negative values of $\epsilon_0\leq -|T_2|$ and therefore low-energy modes move away from $k_x=k_y=0$. Modes in the $y$ direction associated to the $R_2\sin(k_y)\tau^x$ term now cost a finite energy and the system becomes anisotropic. This parameter regime is adiabatically linked to $N$ decoupled $ip_x$ topological superconductors with $2N$ MZM. From Sec. \ref{IIIB}, in our geometry, the $2N$ gapless Majorana modes should be protected against {\it real} $\Delta_0$ and $t_{\perp}'$ terms in Eq. (\ref{structure}), meaning that the Majorana fermions of each chain should remain decoupled from those in other chains. 

For completeness, we provide an alternative understanding of the emergence of a purely $ip_x$ channel using the band structure of Fig. \ref{fig:gladder} (bottom right). As mentioned in Sec. \ref{sec:pre}, the $\zeta a=\pm \pi$ phase on the two wires produces equivalently purely imaginary hopping terms $\pm it$ for the two wires, then modifying the band structure of free fermions as $\mp 2t\sin(k_x a)$. When $t_{\perp}=0$, these two bands cross at $k_x=k_0=0$ with the indices $1$ and $2$ in Fig. \ref{fig:gladder} switched (bottom right). Furthermore, in Eq. (\ref{eq:h_0}), the perpendicular tunneling term $t_{\perp}$ becomes modified as $t_{\perp} (-1)^j$. By Fourier transform, we then conclude that for this situation, the tunneling term $t_{\perp}$ does not open a gap at the crossing point, {\it i.e.} at $k_0=0$. If we also Fourier transform the $\Delta_0$ channel in Eq. (\ref{Delta0}) including the effect of the phases $\zeta a=\pm \pi$ for the two wires, then the wave-vectors (of the two wires) in Fig. \ref{fig:gladder} (bottom right) are modified as $k_x\rightarrow k_x\pm \zeta/2$, and around the crossing point $k_0=0$, the $\Delta_0$ NN-neighbor channel can indeed produce a term, as:
 \begin{equation}
\frac{ \Delta_0}{M} \sum_j \sum_{k_x,k'_x} \widetilde{c}_1^{\dagger}(k'_x) \widetilde{c}_2^{\dagger}(k_x) e^{i(k_x+k_x')ja} e^{i k_x' a} e^{i\frac{a\zeta}{2}} + \text{H.c.}. \label{Deltanew}
 \end{equation}
When $k_x=-k_x' =0$, the pairing term vanishes. Then, we confirm that the induced gap at the Fermi energy takes the form 
 $\widetilde{\Delta} = (\Delta_0 e^{i\zeta a/2}) /2$ with here $\zeta=\chi=\pi/a$. We also check that the N-neighbor $\Delta_0$ channel vanishes in the low-energy subspace due to anti-commutation rules between fermionic operators $\Delta_0 \widetilde{c}_1^{\dagger}(0)\widetilde{c}_2^{\dagger}(0)+\text{H.c}\approx 0$. To proceed and understand the correspondence with Eq. (\ref{eq:hplus}),  we remind that the mapping onto Fig. \ref{fig:gladder} (bottom right) is applicable as long as one assumes to be close to the band crossing point. Taking into account the momentum boost $k\rightarrow k\pm \zeta/2$ with $\zeta=\pi/a$, this corresponds to the case where each wire is half-filled. In the corresponding Fig. \ref{fig:gladder} (bottom left) this implies that the lowest band is now close to the full filling, which corresponds to change $-2t\rightarrow +2t$ for the lowest bonding band compared to the case studied before where the (lowest) band is almost empty.  One must therefore re-adapt in the arguments that $\epsilon_0$ now becomes $+2t-\mu-t_{\perp}$; when the lowest band is filled then this means that $k_{F,+} = \pi/a$. Then, to make the correspondence with Eq. (\ref{eq:hplus}) complete, one can then re-identify the fermionic operators $\widetilde{c}_1(k)$ and $\widetilde{c}_2(k)$ close to the crossing point $k_0$ with the left-moving and right-moving branches of the lowest band operator $\widetilde{c}_+(k)$. From Fig. \ref{fig:gladder},  one can then apply the same arguments as in Eq. (\ref{eq:hplus}), modulo the fact that one re-defines $\epsilon_0=2t-t_{\perp}-\mu$.   
 
 It is perhaps important to mention a duality in the system when exchanging the blue and pink ladders in Fig. \ref{fig:pxy} (with open boundary conditions in the $y$ direction), showing that the above two-dimensional Majorana fermion analysis remains applicable when $t_{\perp}'\gg t_{\perp}$.

When fixing the total flux per plaquette $\Phi_{\text{tot}}=\chi-\zeta=0$ and changing (decreasing) adiabatically the value of $\chi=\zeta$ from $\pi/a$, {\it i.e.} decreasing slightly the value of the magnetic field $B_z$, we observe that the transformation (\ref{gauge}) remains applicable. Then, this produces additional real channels to the $ip_x$ superconducting channel coming from Eqs. (\ref{HDelta}) and (\ref{Delta0}), resulting in a $i(p_x-ip_x')$ channel, whereas the $p_y$ channel remains identical. Through the arguments of Sec. \ref{IIIB}, a superconducting channel of the form $(p_x-ip_x')$ (defined modulo the gobal phase $\pi/2$) should then slightly move progressively the $2N$ Majorana fermions in the anisotropic phase away from zero energy. On the other hand, the two-dimensional chiral gapless edge mode seems to survive in the Moore-Read phase. More precisely, entering this phase from the strong-paired phase which means that $\epsilon_0 = + t_{\perp}' = +|T_2|$, the lowest energy eigenvalue turns slightly into $-(u^2 k_x^2 + (R_2 k_y +  v k_x)^2)^{1/2}$, where $v$ is proportional to $\delta\chi\Delta_0$ and $\delta\chi=\pi/a-\chi$ (with $a=1$). Therefore, when fixing either $k_x\neq 0, k_y=0$ or $k_x = 0, k_y\neq 0$, the energy spectrum still allows for a linear gapless Majorana mode. 

It is instructive to briefly address the limiting case where $B_z=0$. In that case, all the channels $p_x$ and $p_y$ are real. In the low-energy description of Eq. (\ref{eq:hplus}) we have $u=0$ and the last term $R_2$ describes all the superconducting terms $\tilde{\Delta}\sin k_y +\tilde{\tilde{\Delta}}\sin k_x$, where $\tilde{\Delta}=-\Delta_0$, and now $\tilde{\tilde{\Delta}}$ takes into account all the intra-ladder pairing terms $\tilde{\tilde{\Delta}}\sim -\Delta$. Assuming that $\tilde{\Delta}\sim  \tilde{\tilde{\Delta}}$, we can redefine the $R_2$ contribution as $2\tilde{\Delta}\sin((k_x+k_y)/2)\cos((k_y-k_x)/2)$.  The superconducting pairing term can be then re-written in terms of the wave-vectors $k'=(k_x+k_y)/2$ and $k''=(k_y-k_x)/2$, which implies that in that case gapless excitations defined around $k'=0$ will correspond to nodal quasiparticles in the bulk propagating between a $+$ and $-$ $p$-wave lobe, associated with zeroes of the superconducting term.  
   
%------------------------------------------------------------------------------------------------------------
\subsection{Fractional quantum Hall state at $\nu = 1/2$ with spinful fermions}
\label{sec:det}

Here, we would like to address the coupled spinful ladders based on the building block in Sec.~\ref{sec:spinful}, referring to the fourth case of analysis. 
A two-dimensional fractional quantum Hall state at $\nu = 1/2$ can be built from the Cooper pair Laughlin states formed on each ladder.

\begin{figure}[t]
   \begin{center}
     \includegraphics[width=1\linewidth]{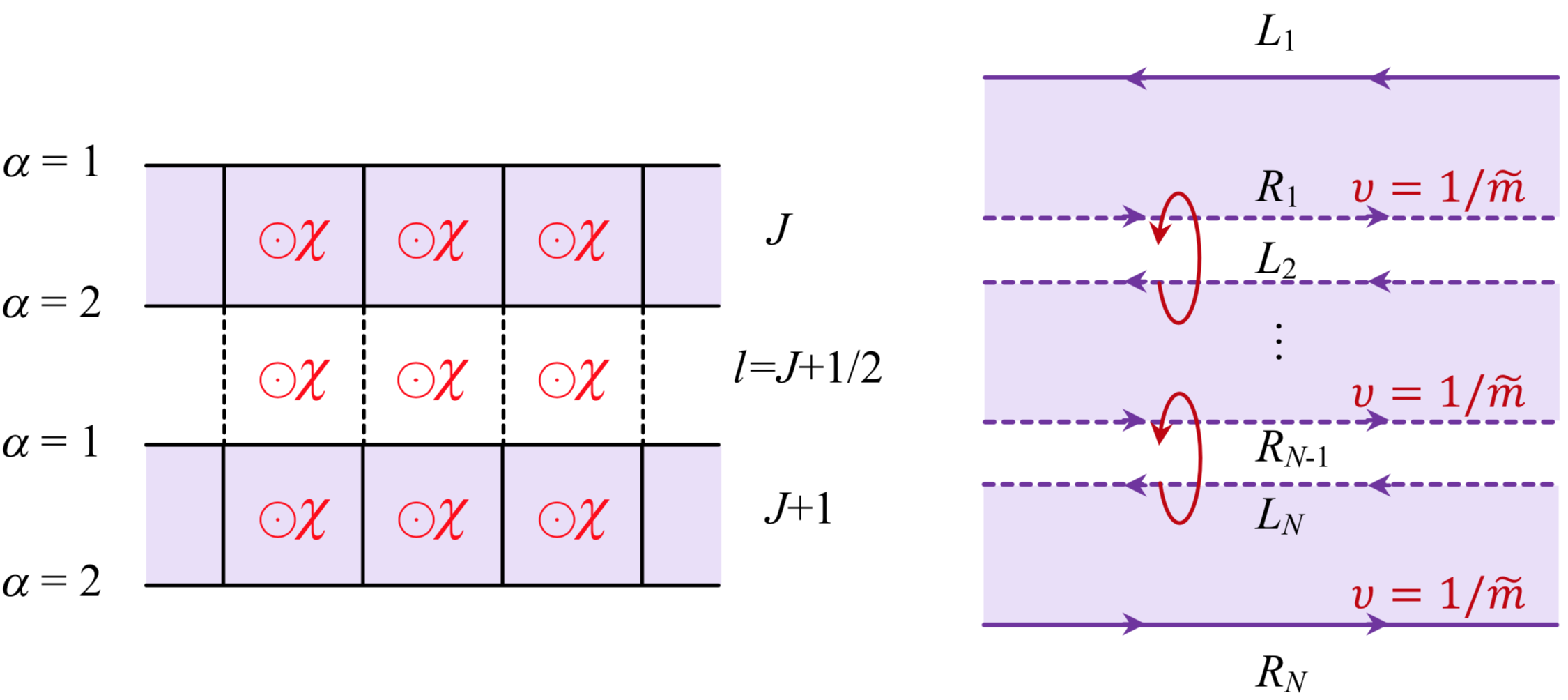}  
    \end{center}
  \vskip -0.5cm \protect\caption[]
  {(color online) (Left) Coupled spinful ladders with uniform flux attachment; the variable $J$ represents the ladder label and $l=J+1/2$ in the middle of two successive ladders will represent the modes formed by coupling ladders together. (Right) Formation of a two-dimensional quantum Hall system with the filling factor $\nu=1/\widetilde{m}=1/2$. The two chiral edge states $L_1$ and $R_N$ are now diagonal in the K-matrix structure producing an SPT phase of Class A \cite{Thomale}.}
    \label{fig:cspin}
    \vskip -0.5cm
\end{figure}

The general building block for the construction is shown in Fig.~\ref{fig:cspin}. Under the flux constraint (\ref{eq:gfa}), the coupling on the rungs of the $J$-th ladder gives a Sine-Gordon term (\ref{eq:sg}) in the charge sector
  \begin{gather}
     \mathcal{H}_{\perp}^J = - \widetilde{v}\int dx \cos(\sqrt{2} \theta_J^- - \widetilde{m}\sqrt{2} \phi_J^+), \label{eq:il}
  \end{gather}
 where we introduce the charge fields as $\theta_J^- = \Theta_{\rho,J}^-$, $\phi_J^+ = \Phi_{\rho,J}^+$ with $\widetilde{v} = {t_\perp^2}/({\Delta_\sigma \pi a})$ and $\widetilde{m} = 2m = \pm 2$. By analogy with the spinless case, one can build an edge theory on each ladder in  the new basis of Eqs.~(\ref{eq:cm0})-(\ref{eq:cm1}) with the substitution $m \to \widetilde{m}$. Accordingly, the bulk modes ($\theta^J$, $\phi^J$) generate a term
   \begin{gather}
     \mathcal{H}_{\perp}^J = - \widetilde{v} \int dx \cos(2\widetilde{m}\phi^J),
   \end{gather} 
   then pinning each mode $\phi^J$ and producing a gap. The two edge modes on the $J$-th ladder are identified as the chiral fields (see Fig.~\ref{fig:cspin}, right)
  \begin{gather}
    \begin{cases} 
       L_J  = {\phi}^{1, J}_{+1}  = \theta'^J + \phi'^J,   \\
       R_J  = {\phi}^{2, J}_{-1} = \theta'^J - \phi'^J,
     \end{cases} \label{eq:rl}
   \end{gather}
 with ${\phi}_r^{\alpha, J} = {\theta^{\alpha, J}}/{\widetilde{m}} + r \phi^{\alpha, J}$ and $\alpha=1, 2$ representing the wire index inside each ladder. In Fig. \ref{fig:cspin} (right), each ladder behaves as a small quantum Hall system which gives rise to two low-energy chiral edge modes $L_J$ and $R_J$. Neighboring chiral edge modes belonging to two successive ladders can then be coupled through an additional $t_{\perp}'$ term at low energy, smaller than the typical energy scale $\tilde{\Delta}$ at which the term $\tilde{v}$ has flown to strong couplings in each ladder. Here, $\tilde{\Delta}$ can be estimated as $\Lambda_c(\tilde{v}/v)^{1/(2-2|\tilde{m}|K)}$ with $\Lambda_c\sim\Delta_{\sigma}$. As mentioned in Sec. \ref{sec:spinful}, we require the introduction of long-range Coulomb forces (such that $2|\tilde{m}|K<2$) to stabilize the relevance of this energy scale. To couple the left-moving mode $L_{J+1}$ of the ladder $J+1$ with the right-moving mode $R_J$ of the ladder $J$, this requires momentum conservation during the tunnel process, and therefore this requires to apply a magnetic field $\chi'=+\chi$ in the region centered at the position $J+1/2$. 
 
It is then possible to couple two edge modes between ladders taking into account the effect of an additional intra-ladder magnetic flux $\chi' = \chi$ by analogy to the two-dimensional quantum Hall effect, as depicted in Fig.~\ref{fig:cspin} (left). Between the $J$-th and $(J+1)$-th ladders,  the flux constraint then follows
  \begin{align}
     a\left[ (k_F^1 + k_F^2)  - m  \chi' \right] = 0\mod 2\pi.
  \end{align}
 Through a recombination of the fields as in Fig. \ref{fig:cspin}, 
    \begin{gather}
     \begin{cases}
       \  \tilde{\phi}_{J+1/2} = (-R_J + L_{J+1})/2, \\
       \ \tilde{\theta}_{J+1/2} = (R_J + L_{J+1})/2,
     \end{cases}
   \end{gather}
   for $t'_{\perp}\ll t_{\perp}$, the bulk and edge Hamiltonians take the form
  \begin{align}
    \mathcal{H}^{\text{bulk}} &= \tilde{\mathcal{H}}^0 [\tilde{\theta}, \tilde{\phi}] - \sum_{l=J+1/2} \int dx \widetilde{v}' \cos(2\widetilde{m}\tilde{\phi}_{l}), \label{eq:hbulk} \\
    \mathcal{H}^{\text{edge}} &=  \frac{v^e}{8\pi}  \int dx \left[ A_{RR} (\nabla R_N)^2  + A_{LL} (\nabla L_1)^2 \right].
     \label{eq:ht}
  \end{align}
The tunnel process $\tilde{v}'={t'_{\perp}}^2/(\tilde{\Delta}\pi a)$ represents the backscattering process of fractional charges $m(2e)$ from one left-moving chiral edge to another right-moving edge. The bulk quadratic Hamiltonian 
$\tilde{\mathcal{H}}^0$ has the coupled form in $\tilde{\theta}$ and $\tilde{\phi}$
   \begin{align}
      \tilde{\mathcal{H}}^0 [\tilde{\theta}, \tilde{\varphi}] =
    & \frac{v^e}{8\pi} \sum_{l=J+1/2} \int dx (A_{RR} + A_{LL}) [ (\nabla \tilde{\theta}_{l})^2 + (\nabla \tilde{\phi}_l)^2] \notag \\
                  & + A_{RL} (\nabla \tilde{\theta}_{l-1} + \nabla \tilde{\phi}_{l-1})(\nabla \tilde{\theta}_{l} - \nabla \tilde{\phi}_{l}), \label{charge}
   \end{align}
with the non-zero backscattering term $A_{RL} = 2\widetilde{m}^2K^e - 2/K^e$ and the velocity $v_e$ similar to the one introduced in Appendix \ref{app:et}. We also identify $A_{RR} = A_{LL} = \widetilde{m}^2K^e + 1/K^e$ and the new fields satisfy 
  \begin{gather}
    [\tilde{\phi}_l (x), \partial_{x'} \tilde{\theta}_{l'} (x') ]= i(\pi/m) \delta_{l, l'}\delta_{x,x'}.
 \end{gather}
By analogy with the arguments of Sec. \ref{sec:spinful}, once the long-range repulsive interactions are present between the ladders, $\widetilde{v}'$ in $\mathcal{H}^{\text{bulk}}$ becomes relevant. This leads to a gapped bulk, corresponding to the pinning of the phases $\tilde{\phi}_{l=J+1/2}\sim 0$. 

Since $L_J = \tilde{\theta}_{l-1} + \tilde{\phi}_{l-1}$, $R_J = \tilde{\theta}_l - \tilde{\phi}_l$, two edge modes $R_N$ and $L_1$ no longer entangle with the bulk. The coupled wire system now forms a single bulk $\nu = 1/2$ fluid in Fig.~\ref{fig:cspin} (right). Since the edge mode theory for these modes is a chiral Luttinger model with $A_{RL}=0$, the properties of the charge at these edges only depend on the value of $K_e$. If we set $A_{LR}=0$ at an edge with the condition that $\tilde{m}=\pm 2$, then we check from Eq. (\ref{charge}) that the fractional charge $K_e = 1/2$ is now in agreement with the bulk-edge correspondence. This phase which shows a perfectly diagonal structure for the edges can be thought as an SPT phase of Class A in terms of the K-matrix structure \cite{Thomale}. 

Similar arguments could be applied for the reconstruction of the Abelian $\nu=1$ and $\nu=1/3$ quantum Hall states with wires and coupled ladder systems. 

%------------------------------------------------------------------------------------------------------------

\section{Conclusion}
\label{summary}

To summarize, we have studied the effect of orbital magnetic fields on topological superconducting quantum wire systems.  We have shown the occurrence of topological superconducting phases with two or four Majorana fermions per ladder as well as quantum Hall phases by tuning the magnetic flux and densities at the crossing points of the band structure. The adjustment of the chemical potential at the crossing region is a key ingredient to realize the appropriate low-energy physics. When the lowest band is completely filled this can give rise to a charge density wave state.  We have also studied Andreev processes and the induction of topological superconductivity in these quantum wire networks. Then, we have engineered a $p+ip$ topological superconducting state in weakly coupled ladder systems, projecting the low-energy physics on the (partially filled) lowest bonding band for each ladder, {\it i.e.}, or a pair of two-strongly-coupled wires. Chiral effects are introduced here through the application of a space-dependent magnetic field perpendicular to the plane of the wires.  We have also shown how the presence of superconductivity and preformed pairs can stabilize a two-dimensional  $\nu=1/2$ Laughlin quantum Hall phase in hybrid spin-1/2 systems, as a result of Andreev processes in the Luther-Emery description of a superconducting quantum wire. Similar fractional quantum Hall states can be constructed in bosonic coupled wires or quantum spin systems as well \cite{kane2014,alex2015,alex2017}. We have analyzed the phase diagram and the physical observables, discussing the effect of Coulomb interactions. We have suggested implementations of these ideas in solid-state nanowires and cold-atom systems.  Finally, due to the analogy between p-wave superconductors, quantum Ising spin chains \cite{zoller,Levitov} and $\mathbb{Z}_2$ Kitaev quantum spin liquids \cite{Kitaev2D,fan3,fan2,fan1}, one may also anticipate further applicability of these findings in other interacting systems.
\\

%------------------------------------------------------------------------------------------------------------
Acknowledgements:  This work has benefitted from useful discussions with M. Aidelsburger, F. Grusdt, L. Herviou, C. Mora, G. Roux, R. Santachiara, K. Van Houcke and with F. Pollmann, F. Heidrich-Meisner at the DFG meeting FOR2414 in G{\"o}ttingen. This research was funded by the Deutsche Forschungsgemeinschaft (DFG, German Research Foundation) via Research Unit FOR 2414
under project number 277974659 (F.Y. and K.L.H.), and we also acknowledge funding from French ANR BOCA (K.L.H.). This work started as a collaboration between Sherbrooke, CNRS and Ecole Polytechnique via the LIA, LCMQ and the CIFAR. V.P. and I.G. in Sherbrooke were financially supported by the Canada First Research Excellence Fund and by the Natural Sciences and Engineering Research Council of Canada. V.P. acknowledges the CPHT of Ecole Polytechnique, Universit\' e de Sherbrooke, ENS Cachan and ENS Paris for the computing resources and support during his Master internships. A.P. acknowledges support from an Institut Quantique Fellowship at the Universit\' e of Sherbrooke.

\renewcommand{\theequation}{\Alph{section}\arabic{equation}}
\appendix

%-----------------------------------------------------------------------------------------
\section{Perturbative treatment of $\mathcal{H}_\perp$}
\label{app:pert}

Here, we give more detail on the derivations of the Andreev processes in Sec. \ref{sec:hw}. To find the relevant contribution from $t_\perp$, we can develop perturbation theory \cite{rice1994,karyn2001,alex2015} for the hybrid wire Hamiltonian  $\mathcal{H}_0 = \mathcal{H}_1 + \mathcal{H}_2 + \mathcal{H}_\Delta$ with  a weak interwire coupling  
$V = \mathcal{H}_\perp$. Here, $\mathcal{H}_1 + \mathcal{H}_2$ refers to the quadratic part (Luttinger liquid contribution) of the Hamiltonian in each wire $1$ and $2$ respectively, the Hamiltonian density operator $V(x) = \mathcal{H}_\perp (x)$ reads,
  \begin{align}
    V(x) &= -\frac{t_\perp}{2\pi a}  \left\{ \left[ e^{i(\sqrt{2} \theta^-(x) - \chi x)} + e^{-i(\sqrt{2} \theta^-(x) - \chi x)} \right] \right. \notag \\
             &  \times \left[ e^{i((k_F^1 + k_F^2)x - \sqrt{2} \phi^+(x) )} + e^{-i((k_F^1 + k_F^2)x - \sqrt{2} \phi^+(x))} \right] \notag \\
              &  + \left[ e^{i(\sqrt{2} \theta^-(x) - \chi x)} + e^{-i(\sqrt{2} \theta^-(x) - \chi x)} \right] \notag \\
              &  \times \left. \left[ e^{i((k_F^1 - k_F^2)x - \sqrt{2} \phi^-(x) )} + e^{-i((k_F^1 - k_F^2)x - \sqrt{2} \phi^-(x))} \right] \right\}.
    \end{align}
To the $n$-th leading order, the effective Hamiltonian reads
  \begin{gather}
    \mathcal{H}_{\text{eff}} = \mathcal{H}_0 + V^{(n)}.
  \end{gather}
For any observable $A$,  the expectation value under $\mathcal{H}_{\text{eff}}$ takes the form
  \begin{gather}
    \langle A \rangle_{\mathcal{H}_0+V} = \langle A \rangle_{\mathcal{H}_{\text{eff}}} + \mathcal{O}(V^{n}), 
  \end{gather}
where
   \begin{align}
     \langle A \rangle_{\mathcal{H}_0 + V} 
         & = \frac{ \text{Tr} \left[ e^{-\int dx d\tau (\mathcal{H}_0 +V)  } A \right]}{\text{Tr}\left[ e^{-\int dx d\tau (\mathcal{H}_0 +V)} \right]} \notag \\
         & = \left[ \langle A \rangle_{\mathcal{H}_0}  + \sum_{n=1}^\infty \frac{(-1)^n}{n!} \left< \left(\int V\right)^n A \right>_{\mathcal{H}_0} \right] \notag \\
            &  \times \left[ 1  + \sum_{n=1}^\infty \frac{(-1)^n}{n!} \left< \left(\int V\right)^n  \right>_{\mathcal{H}_0} \right]^{-1}
       \end{align}
 with the notation $\int V = \int dx d\tau V(x, \tau)$. As the mode $\theta_1$ is pinned in Eq. (\ref{eq:pd}), its dual field $\phi_1$ oscillates rapidly and renders all terms involving $\phi^+$ and $\phi^-$ irrelevant. Thus
   \begin{gather}
     \left( \int V \right)^n = 0, \qquad n = \text{odd}.
   \end{gather}
We can keep the even order terms
  \begin{gather}
    \langle A \rangle_{\mathcal{H}_0 +V} = \left<  A \right>_{\mathcal{H}_0} + \sum_{k=2,4} \frac{1}{k!} \left< \left( \int V \right)^k A \right>_{\mathcal{H}_0} \notag \\
       - \frac{1}{k!} \left< \left( \int V \right)^k \right>_{\mathcal{H}_0}\left<  A \right>_{\mathcal{H}_0} + \mathcal{O}\left[ \left(\int V \right)^6 \right].
   \end{gather}
  
  %--------------------------------------------------------------------------------------------------------------------------
  \subsection{Second order contribution to the $\pi$-flux}
  
  For the integral at the second order perturbation theory level,
  it is more convenient to switch to the relative and center-of-mass coordinates,
  \begin{gather}
    x = (x_1+x_2)/2, \quad x' = x_1 - x_2, \notag \\
    \tau = (\tau_1+\tau_2)/2, \quad \tau' = \tau_1 - \tau_2. \label{eq:bs}
  \end{gather}
 Further, we introduce polar coordinates for the relative distances $x'$ and $\tau'$,
   \begin{gather}
     d_{12} = \sqrt{(x')^2 +v^2 (\tau')^2}, 
   \end{gather}
 and consider the virtual processes 
   \begin{align}
        |x'|    &= |x_1 - x_2| \le \xi = {v}/{\Delta}, \notag \\
       |\tau'| &= |\tau_1 - \tau_2| \le \Delta^{-1}.
   \end{align}
  Then
   \begin{gather}
     \int dx' d\tau' = \int \frac{2\pi}{v} d_{12} d(d_{12}).
   \end{gather}
   Correspondingly,
   \begin{align}
    &  \int dx_1  d\tau_1 dx_2 d\tau_2 V(x_1, \tau_1) V(x_2, \tau_2) \notag \\
      \simeq &\frac{2\pi a}{v} (\xi - a) \int dx d\tau V(x, \tau) V(x+a, \tau) \notag  \\
      \simeq &\frac{2\pi a}{\Delta} \int dx d\tau V^2(x, \tau) \notag  \\
      = &\frac{4t^2_\perp}{ \pi a \Delta} \int dx d\tau \left[ \cos(2\sqrt{2}\theta^- - 2\chi x) + \cos(2k_F^2x - 2\phi_2) \right. \notag \\
      &\left. + \cos(2\sqrt{2}\theta^- - 2\chi x) \cos(2k_F^2x - 2\phi_2) \right]. \label{eq:ht2}
   \end{align}
  We have dropped out fast-oscillating terms which involve $\phi_1$ fields together with other constant terms.

   For the situation of two wires such that $k_{F}^2\neq \pi/(2a)$, but with the $\pi$-flux gauge choice $\chi a= \pi$, we obtain
   \begin{align}
      \left(\int V\right)^2   
       &=  \frac{4t^2_\perp}{ \pi a \Delta} \int dx d\tau \cos(2\theta_2),  \label{eq:hp2}
    \end{align}
   and in the last equality, $\theta_1$ is pinned to zero and we regard $2\chi x$ as multiples of $2\pi$.   
   
   On the other hand, when $\Delta \gg t_\perp$,
  \begin{align}
     \left< \left( \int V \right)^2 \right>_{\mathcal{H}_0} 
        = &\int dx_1 dx_2 d\tau_1 d\tau_2  \left< V(x_1, \tau_1) V(x_2, \tau_2) \right>_{\mathcal{H}_0} \notag \\
        = &\left( \frac{2\pi }{v} \right) \int dx d\tau  d_{12} d(d_{12}) R({d}_{12})e^{-{d}_{12}/\xi}\notag \\
        \simeq &\left( \frac{2\pi }{v} \right) a (\xi-a) R(a) \int dx d\tau \cdot 1\notag \\
        \simeq  &\left( \frac{2\pi a}{\Delta} \right) R(a) (L \beta) \sim 0,
     \end{align}
  where $R(r)$ denotes a power-law decreasing function. Therefore, the second order contribution reads
  \begin{gather}
     \mathcal{H}_\perp^{(2)} = - \frac{1}{2}\left(\int V\right)^2 =  -\frac{2t^2_\perp}{ \pi a \Delta} \int dx d\tau \cos(2\theta_2).
  \end{gather}
  This reproduces Eq. (\ref{eq:2nd}) in Sec. \ref{sec:hw}, and the proximity effect since $\langle\cos(2\theta_2)\rangle$ now acquires a finite value, implying the pinning of the mode $\theta_2$ in wire $2$ and the opening of a superconducting gap due to the presence of Andreev
  processes, coupling wire $1$ and $2$. 
  
  \subsection{Fourth-order contribution to the arbitrary flux}
  
  Away from half-filling, under our gauge choice (\ref{eq:gfa}) for the flux, the second order term vanishes
  \begin{align}
      2\chi x = \pm 2(k_F^1+k_F^2)x \ne \pm 2\pi n, \quad \left( \int V \right)^2 = 0.
  \end{align} 
 Now if we go to the fourth order, in the same way as in Eq.~(\ref{eq:bs})  by changing the basis twice: (i) from ($x_3$, $\tau_3$, $x_4$, $\tau_4$) to ($\bar{x}$, $\bar{\tau}$, $\bar{x}'$, $\bar{\tau}'$); (ii) from ($x$, $\tau$, $\bar{x}$, $\bar{\tau}$) to ($X$, $\eta$, $X'$, 
 $\eta'$), we reach
   \begin{align}
    & \left( \int V \right)^4 = \left( \frac{2\pi a}{\Delta} \right)^2 \int dx d\tau d\bar{a} d\bar{\tau} V^2(x,\tau) V^2(\bar{x}, \bar{\tau}) \notag \\
                                    &\simeq \left( \frac{2\pi a}{\Delta} \right)^2 \frac{2\pi \cdot 2a}{v} (\xi - 2a) \int dXd\eta V^2(X, \eta) V^2(X+2a, \eta) \notag \\
                                    &\simeq \frac{8t_\perp^4}{\pi a\Delta^3} \int dx d\tau \cos \left[ 2\sqrt{2}( \theta^-(x) - \theta^-(x+2a)) + 4\chi a \right].
   \end{align}
 In the second equality, we notice the relative distance $|X'| = |x - \bar{x}| = |x_1 + x_2 - x_3 - x_4|/2 \in [ 2a, \xi ]$.
Meanwhile,
  \begin{align}
    & \quad \left< \left( \int V \right)^4 \right>_{\mathcal{H}_0} \notag \\
                 &\simeq \left( \frac{2\pi a}{\Delta} \right)^2 \int dx d\tau d\bar{x} d\bar{\tau} 
                 \langle V^2(x, \tau) V^2(\bar{a}, \bar{\tau})\rangle_{\mathcal{H}_0} \notag \\
                 &= 2 \left( \frac{2\pi a}{\Delta} \right)^3 R(2a)(L\beta) \sim 0.
  \end{align}
Taken into account the pinned mode $\theta_1 (x)\sim 0$, we find in the effective Hamiltonian $\mathcal{H}_{\text{eff}}$, that the leading-order contribution from $t_{\perp}$ becomes
  \begin{gather}
    \mathcal{H}_\perp^{(4)} = - \frac{t_\perp^4}{3\pi a \Delta^3} \int dx \cos [ 2 (\theta_2(x+2a) - \theta_2(x)) + 4\chi a].
  \end{gather}

%-----------------------------------------------------------------------------------------
\section{Edge Theory}
\label{app:et}

Here, we study the effective edge theory in the quantum Hall phase found in Sec. \ref{edge}, in the case of two wires or a two-leg ladder implying that the two edges are not fully separated from the bulk. We then perform an integration
on the bulk (gapped) degrees of freedom to build the edge theory. The form of the interwire tunneling Hamiltonian $\mathcal{H}_{\perp}$ (\ref{eq:hp1}) satisfies the classification of the edge theory at the filling $\nu = 1/m$ \cite{alex2015, alex2017}. Here we briefly review the construction approach.

The effective edge Hamiltonian can be built in the Luttinger liquid form
  \begin{gather}
    \mathcal{H} = \mathcal{H}_0^+ + \mathcal{H}_0^- + \mathcal{H}_{\perp},
  \end{gather}
  where $\mathcal{H}_0^\pm$ is given in Eq.~(\ref{eq:h0pm}) and $\mathcal{H}_\perp$ in Eq.~(\ref{eq:hp1}). We are then able to integrate out the other bulk mode $\theta$ in the action   
 \begin{align}
    S [\theta^\pm, \phi^\pm] = \frac{1}{2\pi} \sum_{\alpha = \pm} \int dx d\tau & \left[  vK ( \nabla \theta^\alpha)^2 + \frac{v}{K} (\nabla \phi^\alpha)^2 \right. \notag \\
     & \phantom{=} \left. + 2i \partial_\tau \phi^\alpha \nabla \theta^\alpha  \right].  \label{eq:at}
  \end{align}
 Changing the basis  from $[ \theta^+, \phi^+, \theta^-, \phi^-]$ to $[ \theta, \phi, \theta', \phi']$ via Eqs.~(\ref{eq:cm0}), (\ref{eq:cm}) and (\ref{eq:cm1}), we get
   \begin{align}
        \sum_{\alpha=\pm} (\nabla \theta^\alpha)^2 
              = & \frac{m^2}{2} \left[ (\nabla \theta)^2 + 2\nabla \theta' \nabla \theta + (\nabla \phi)^2 \right. \notag \\
                 & \left. -  2\nabla \phi' \nabla \phi +  (\nabla \theta')^2 + (\nabla \phi')^2  \right],  \notag \\
      \sum_{\alpha=\pm} (\nabla \phi^\alpha)^2 
             = & \frac{1}{2} \left[ (\nabla \theta)^2 - 2\nabla \theta' \nabla \theta + (\nabla \phi)^2\right. \notag \\
                &  \left. +  2\nabla \phi' \nabla \phi +  (\nabla \theta')^2 + (\nabla \phi')^2  \right].   
    \end{align}
   As the bulk mode $\phi$ is pinned, we can safely drop out all terms involving $\nabla \phi$. In the total action (\ref{eq:at}), the bulk $\theta$ mode contributes to
   \begin{gather}
     S[\theta] = \int dx d\tau \frac{v}{2\pi} \left[ G (\nabla \theta)^2 + F \nabla \theta' \nabla \theta\right] + \frac{i}{\pi}\partial_\tau \phi \nabla \theta,
   \end{gather}
   with 
   \begin{gather}
      G = \frac{1}{2}\left( m^2 K + \frac{1}{K}\right), \quad F = m^2 K - \frac{1}{K}. 
   \end{gather}
   We define a general Fourier transform with periodicity on $[0, L]$ ($L$-length of the wire), 
   \begin{gather}
     f(r) = \frac{1}{\Omega} \sum_q f_q e^{iqr},
   \end{gather}
  where $\Omega = \beta L$, $\beta = 1/T$, $r= (x, v\tau)$, $q = (k, \omega_n/v)$ and $qr = kx - \omega_n \tau$. $\omega_n = 2\pi n /\beta (n \in \mathbb{N})$ denote the Matsubara frequencies for bosons. $S[\theta]$ can then be transformed into the momentum space
  \begin{gather}
    S[ \theta ] = \sum_q \left[ \frac{ik\omega_n}{\pi} \phi_q \theta_q^*\right] + \frac{v}{2\pi} \sum_q k^2 (G \theta_q^* \theta_q + F {\theta'_q}^* \theta_q).
  \end{gather}
  Integrating out $\theta$, we get the edge Hamiltonian
    \begin{gather}
      \mathcal{H}^e = \frac{v^e}{2\pi} \int dx \left[ K^e (m\nabla \theta')^2  + \frac{1}{K^e} (\nabla \phi')^2 \right],
    \end{gather}
    with the Luttinger parameters
   \begin{align}
     v^e K^e &= \frac{v}{m^2} \left( G - \frac{F^2}{4G}\right) =  \frac{v}{2} \left[ (K+\frac{1}{m^2K}) - \frac{(K - \frac{1}{m^2 K})^2}{(K+\frac{1}{m^2K})} \right], \notag \\
      \frac{v^e}{K^e} &= vG = \frac{v}{2} \left(m^2 K +\frac{1}{K}\right). \label{veKe}
   \end{align}
   Therefore, when $K = 1$ and $m = \pm 1$,
   \begin{gather}
      K^e = \frac{2}{m^2+1} = 1, \quad v^e = v. 
    \end{gather}
   In terms of edge chiral fields, $L(x) = \phi_{+1}^1(x) = \theta' + \phi', R(x) =  \phi_{-1}^2(x) = \theta' - \phi'$, 
     \begin{gather}
        \mathcal{H}^e = \frac{v}{8\pi} \int dx \left[  A_{RR} (\nabla R)^2 + A_{LL} (\nabla L)^2  + A_{RL} (\nabla R)(\nabla L)\right]. \label{B11}
     \end{gather}
    The backscattering term vanishes: $A_{RL} = 2m^2K^e - 2/K^e = 0$ and $A_{RR} = A_{LL} = m^2 K^e + 1/K^e$. In the end, 
    \begin{gather}
      \mathcal{H}^e = \frac{v}{4\pi} \int dx \left[  (\nabla R)^2 + (\nabla L)^2  \right],
    \end{gather}
    we reach a quantum Hall phase at $\nu  = 1/ m = \pm 1$ in the presence of an arbitrary uniform magnetic flux. 

%----------------------------------------------------------------------------------------
\section{Thouless pump}
\label{app:tp}

Here, we study the bulk response in quantum Hall phases. We address interaction effects and compare the results with those for the CDW phase.

%---------------------------------------------------------------------------------------
\subsection{Thin-torus geometry}
 \begin{figure}[b]
   \begin{center}
      \includegraphics[width=0.9\linewidth]{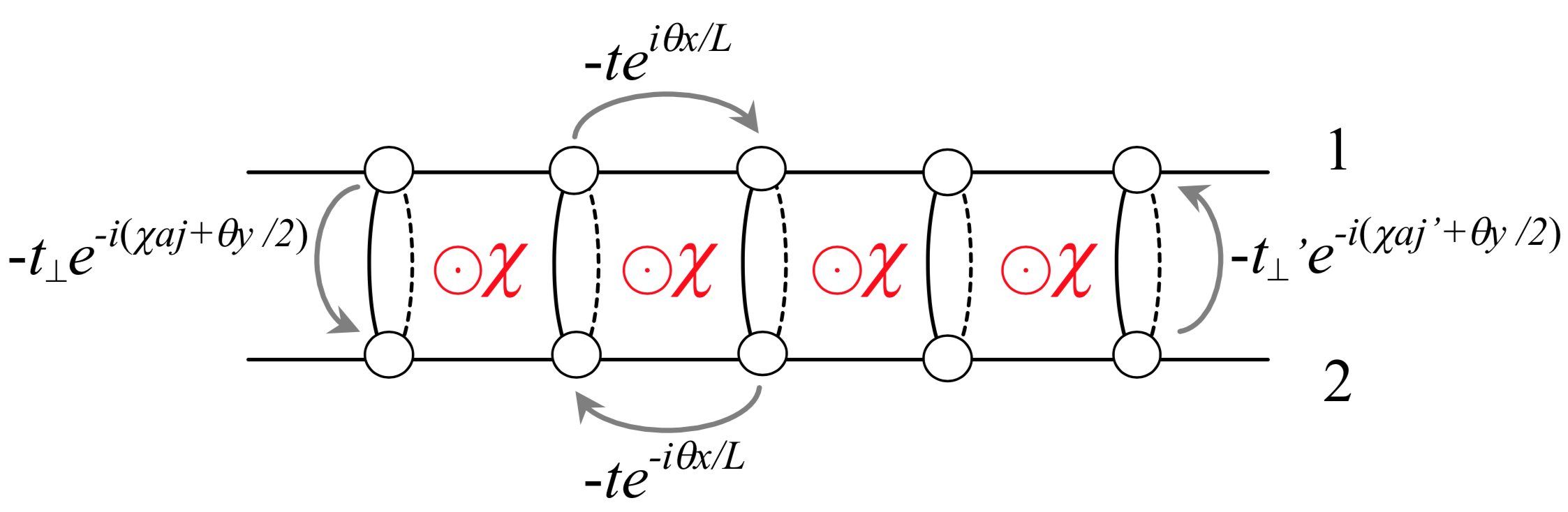}       
    \end{center}
  \vskip -0.5cm \protect\caption[]
  {(color online) Thin-torus lattice with $\theta_{x,y}$.}
    \label{fig:tpump}
    \vskip -0.5cm
\end{figure}
One approach to probe the quantum Hall phase is the detection of the bulk polarization through the Thouless pump \cite{thouless1983, grusdt2014}. We can now gap out the edge mode by mapping the two-wire system to a thin torus, shown in Fig.~\ref{fig:tpump}. We add to the field operators two Aharonov-Bohm phases $\theta_x$, $\theta_y$ along the torus with the periodicity $L_x = L$, $L_y = 2$. The torus Hamiltonian reads
  \begin{gather}
    \mathcal{H} (\theta_y) = -t_\perp \sum_j \left[ c_1^\dagger (j) c_2(j) e^{i(a\chi j + \theta_y/2)} + h.c. \right] \notag \\
      -t'_\perp \sum_j \left[ c_2^\dagger (j) c_1(j) e^{i(a\chi j + \theta_y/2)} + h.c. \right] \notag    \\
      -t \sum_j \left[ c_1^\dagger (j) c_1 (j+1)e^{-i\theta_x/L} +c_2^\dagger (j) c_2 (j+1)e^{-i\theta_x/L}  + h.c. \right] \notag \\
      + \Delta \sum_j \left[ c_1^\dagger (j) c_1^\dagger (j+1) e^{i\theta_x(2j+1)/L } + h.c.\right].
  \end{gather}
The condition $\theta_x = \theta_y = 0$ and $t'_\perp = 0$ gives back the original Hamiltonian. In the following analysis, for simplicity we take $t'_\perp = t_\perp$.
 In the bosonization picture, accordingly, the hopping term between two wires becomes
   \begin{align}
     \mathcal{H}_\perp (\theta_y) 
        = &- \frac{4t_\perp}{\pi a} \int dx \cos(\chi x + {\theta_y}/{2}) \cos(\sqrt{2}\theta^-) \notag \\
           & \times \cos[(k_F^1 + k_F^2)x  - \sqrt{2} \phi^+]. 
     \label{eq:torus}
   \end{align}  
 
%------------------------------------------------------------------------------------------------------------------------------
\subsection{Bulk polarization}

Under the gauge choice (\ref{eq:gfa}) for the arbitrary flux,
   \begin{gather}
     \mathcal{H}_\perp (\theta_y) = - \frac{2t_\perp}{\pi a} \int dx  \cos(\sqrt{2}\theta^-) \cos( \theta_y/2 + m\sqrt{2} \phi^+),
   \end{gather}  
with $m = \pm 1$.
In this geometry, the two modes $\theta^-$ and $\phi^+$ are pinned separately. Therefore, from Eq.~(\ref{eq:cm1}), we obtain that the edge mode $\phi' = (\theta^-/m + \phi^+)/\sqrt{2}$ is indeed gapped out in this geometry. The variation of $ \theta_y$ enters into the original bulk mode $\phi =(-\theta^-/m + \phi^+)/\sqrt{2} = -\theta_y/(4m)$. By changing $\theta_y$ periodically, one can probe the bulk polarization. During the process, one mode $\theta^-$ remains pinned and the magnetization current in the bulk stays fully suppressed: $\langle j_\parallel (x) \rangle \propto \langle \nabla \theta^- \rangle = 0$. For the other mode $\phi^+$, when $\theta_y = 0$, $\phi^+\sim 0$. A charge gap is formed in the total charge response and the system behaves as a charge density wave (CDW).

On an experimental setup \cite{grusdt2014}, in order to induce a variation in $\theta_y$,  one can exert a constant force around the smaller perimeter $\hat{y}$ of the torus (see Fig.~\ref{fig:tpump}) and adiabatically reach twisted boundary conditions: $F \propto \partial_t \theta_y$. When $\theta_y$ goes  from $0$ to $2\pi$, one charge $e$ is transported along the wires,
  \begin{gather}
                    \int_0^L dx  \Delta(n_1(x) + n_2(x)) = \frac{1}{m} \frac{\Delta \theta_y}{2\pi} = \frac{1}{m} = \nu = \pm 1. \label{pumping}
  \end{gather}   
 It gives rise to a quantized Hall current perpendicular to the force
   \begin{gather}
     I_{\text{H}} = e\nu \omega, \label{eq:hcu}
   \end{gather}
  where $\omega$ denotes the frequency of variation.  In Eq.~(\ref{eq:hcu}), one identifies the filling factor with Chern number, which manifests the Zak phase on the thin-torus geometry
  \begin{align}
    \nu  = &- \int_0^{2\pi} d\theta_y \int_{\text{BZ}} \frac{d\theta_x}{2\pi} \Omega_{\theta_x \theta_y} \notag \\
           = &\int_0^{2\pi} d\theta_y \partial \theta_y \int_{\text{BZ}} \frac{d\theta_x}{2\pi} \mathcal{A} ({\theta_x}) \notag \\
           = &\int_0^{2\pi} \frac{d\theta_y}{2\pi} \partial \theta_y \varphi_{\text{Zak}} (\theta_y).
           \label{eq:zak}
      \end{align}
  Here we use the fact that the Berry connection $\mathcal{A} ({\theta_y})$ is periodic in $\theta_x$. In the Berry curvature $\Omega_{\theta_x \theta_y} = \partial_{\theta_x} \mathcal{A} ({\theta_y}) -  \partial_{\theta_y} \mathcal{A} ({\theta_x})$, the first term thus vanishes.
  
Meanwhile, the Zak phase can also be interpreted through the electric polarization \cite{king1993}. 
From macroscopic electrostatics, one relates the polarization density $P(\vec{r})$ to the charge density $\rho(\vec{r})$ through $\nabla \cdot P(\vec{r}) = - \rho (\vec{r})$. Combined with the continuity equation $\partial_t \rho (\vec{r}) + \nabla \cdot \vec{j} (\vec{r}) = 0$, we obtain
  \begin{gather}
    \nabla \cdot \left( \partial_t P(\vec{r}) - \vec{j}(\vec{r}) \right) = 0, \quad \Delta P = \int_0^T dt \vec{j}'.
  \end{gather}
The second equation is valid up to a divergence-free part. It is given by the magnetic current in the bulk, which is identically zero due to the pinned mode $\theta^-$. $\vec{j}'$ represents the adiabatic current induced by the variation in external potentials \cite{xiao2010} and is related to the total velocity by
  \begin{align}
    \vec{j}' &=  en_0 v(\vec{r}) = e\frac{1}{L} \sum_{\vec{r}} v(\vec{r}) = e\int_{\text{BZ}} \frac{dq}{2\pi} v(q), \notag \\
     v(q) &=  \frac{\partial \epsilon(q)}{\hbar \partial q} - \Omega_{qt}.
    \label{eq:av}
  \end{align}
  For completness, in this formula, we have restored the Planck constant $\hbar$. 
 One immediately sees, after the integration, the normal group velocity ${\partial \epsilon(q)}/{(\hbar \partial q)}$ vanishes. The anomalous velocity, on the other hand, comes from the Berry curvature $\Omega_{\mu \nu} = \frac{\partial}{\partial R^{\mu}}\mathcal{A}_\nu (R) -\frac{\partial}{\partial R^{\nu}}\mathcal{A}_\mu (R)$ in the parameter space $R = (q ,t)$. Explicitly, 
   \begin{gather}
     \Omega_{qt} = i \left[ \left< \frac{\partial u}{\partial q} \right| \left.  \frac{\partial u}{\partial t}   \right> - \left< \frac{\partial u}{\partial t} \right| \left.  \frac{\partial u}{\partial q}   \right> \right],
   \end{gather}
 with $|u(q,t)\rangle$  defined in the Bloch form of the instantaneous eigenstates $|\psi_q (x,t) \rangle = e^{iqx} |u(q,t)\rangle$.
 Now $\Omega_{qt}$ gives non-zero contribution to the difference of the polarization
 \begin{gather}
   \Delta P = - e \int_0^T dt \int_{\text{BZ}} \frac{dq}{2\pi} \Omega_{qt}.
  \end{gather}
 On the torus, in one period we can perform a change of variables from $dt dq$ to $d\theta_y d\theta_x$. It follows
  \begin{align}
    \Delta P = & e \int_0^{2\pi} d\theta_y \partial_{\theta_y} \int_{\text{BZ}} \frac{d\theta_x}{2\pi} \mathcal{A} ({\theta_x}) \notag \\
                 = & \frac{e}{2\pi} [\varphi_{\text{Zak}} (\theta_y = 2\pi) - \varphi_{\text{Zak}} (\theta_y = 0)].
    \label{eq:po}
 \end{align}
Along the wires, we conclude $P = l_x e\varphi_{\text{Zak}}/(2\pi) $ where $l_x$ is the number of magnetic unit cells we measure on the $\vec{x}$ direction. Upon the time period $T$, when $\theta_y$ changes from $0$ to $2\pi$, the Zak phase in Eq.~(\ref{eq:zak}) goes continuously from $0$ to $2\pi \nu $ and a quantized change in polarization density $P/e = l_x \nu = \pm l_x$ can be observed. 

Another way to prove Eq.~(\ref{eq:po}) is to re-express the anomalous velocity in Eq.~(\ref{eq:av}) in two dimensions $\vec{\theta} = (\theta_x, \theta_y)$:
 \begin{gather}
  \vec{v} (\theta_x, \theta_y)_{\text{anom.}} = - \frac{\partial \vec{\theta}}{\partial t} \times \left( \partial_{\vec{\theta}} \times \mathcal{A}(\theta_x, \theta_y)\right).
   \end{gather}
Applying  $\mathbf{a} \times (\mathbf{b} \times \mathbf{c}) = \mathbf{b} (\mathbf{a} \cdot \mathbf{c}) - \mathbf{c}(\mathbf{a} \cdot \mathbf{b})$ and $\theta_y = 2\pi t/T$, the anomalous velocity along the $x$ direction reads
  \begin{gather}
   v(\theta_x)_{\text{anom.}} = - \frac{2\pi}{T} \left[ \partial_{\theta_x} \mathcal{A}(\theta_y) - \partial_{\theta_y} \mathcal{A}(\theta_x) \right].
   \end{gather}
 Correspondingly, the difference in polarization 
    \begin{gather}
        \Delta P = - \frac{e}{2\pi} \int_0^T dt \int_{\text{BZ}} {d\theta_x} v(\theta_x)_{\text{anom.}}
    \end{gather}
  gives back Eq.~(\ref{eq:po}) after a change of the variable from $dt$ to $T d\theta_y /(2\pi)$.

%-------------------------------------------------------------------------------------------------------------------
\subsection{Comparison with the $\pi$-flux and stability under Coulomb interactions}

Switching to the $\pi$-flux configuration, the strong tunneling Hamiltonian (\ref{eq:hp_pi}) pins two modes together, $\theta^{-}$, $\phi^+$. The edge mode $\phi' = (\theta^-/m + \phi^+)/\sqrt{2}$ is now gapped out from the beginning.  The system turns into a charge density wave (CDW). If we perform the same Thouless pump measurement in the torus geometry (\ref{eq:torus})
  \begin{align}
     & \mathcal{H}_\perp (\theta_y) =  - \frac{2t_\perp}{\pi a} \int dx  \cos(\sqrt{2}\theta^-) \notag \\
        & \quad \times [ \cos( \theta_y/2 + \sqrt{2} \phi^+) + \cos( \theta_y/2 - \sqrt{2} \phi^+)].            
  \end{align}
As soon as  $\theta_y$ is varied by an external force, the responses in the $\phi^+$ mode differ in signs and cancel with each other. No charge pumping would occur in  the charge density wave state formed by the $\pi$-flux.

Here, we comment briefly on the effects of Coulomb interactions on the Thouless pump at filling factor $\nu = 1$, including both contributions parallel and perpendicular to the wires. The parameters in the Luttinger Hamiltonian are then modified as
$v^\pm K^\pm = vK$, $v^\pm/K^\pm = v/K \pm (aV_\perp)/\pi$. During the charge pumping, however, the velocity of the parallel current remains unaffected:
  \begin{gather}
   \nabla \widetilde{j}_{\parallel}(x) = -\partial_t [ n_1(x) + n_2(x)],  \notag \\
      \widetilde{j}_{\parallel}(x)  = - \frac{v^+K^+\sqrt{2}}{\pi} \nabla \theta^+ (x) = - \frac{vK\sqrt{2}}{\pi} \nabla \theta^+ (x). 
  \end{gather}
 Under Coulomb interactions, we find the bulk is still stable in the Thouless pump measurement, and pumping effects are effectively described through Eq. (\ref{pumping}). 

%---------------------------------------------------------------------------------------------------------------------------------------------------

\bibliographystyle{apsrev4-1}
\bibliography{sample}

\end{document}